\documentclass[11pt]{article}
\usepackage{jheppub}

\usepackage{amsmath,graphicx,amssymb,bbm,subfigure,hyperref,array,slashed,bbm}
\usepackage[all]{xy}

\def\be{\begin{equation}}
\def\ee{\end{equation}}
\def\ba{\begin{eqnarray}}
\def\ea{\end{eqnarray}}

\def\r{\rho}
\def\a{\alpha}

\def\b{\beta}

\def\g{\gamma}
\def\G{\Gamma}
\def\d{\delta}
\def\D{\Delta}

\def\p{\pi}

\def\th{\theta}

\def\m{\mu}
\def\n{\nu}
\def\om{\omega}
\def\Om{\Omega}
\def\l{\lambda}

\def\s{\sigma}

\def\qq{\qquad}

\def\IR{\relax{\rm I\kern-.18em R}}

\def\inv{^{\raise.0ex\hbox{${\scriptscriptstyle -}$}\kern-.05em 1}}

\def \ov {\over}

\title{Mesons from (non) Abelian T-dual backgrounds}

\author{Georgios Itsios$^{1,2}$,}
\author{Carlos N\'u\~nez$^3$}
\author{Dimitrios Zoakos$^4$}

\affiliation{$^1$ Instituto de F\'isica Te\'orica, UNESP-Universidade Estadual Paulista,
R. Dr. Bento T. Ferraz 271, Bl. II, Sao Paulo 01140-070, SP, Brazil}
\affiliation{$^2$ Department of Physics, University of Oviedo, Avda. Calvo Sotelo 18, 33007 Oviedo, Spain}
\affiliation{$^3$ Department of Physics, Swansea University, Swansea SA2 8PP, United Kingdom}
\affiliation{$^4$ Centro de F\'isica do Porto, Universidade do Porto,
Rua do Campo Alegre 687, 4169-007 Porto, Portugal}

\emailAdd{gitsios@gmail.com} 
\emailAdd{c.nunez@swansea.ac.uk} 
\emailAdd{zoakos@gmail.com} 

\abstract{In this work we study mesonic excitations in 
a Quantum Field Theory dual to the non Abelian T-dual of $AdS_5\times S^5$, using a $D6$ brane probe on the Sfetsos-Thompson background. 
Before and after the duality, we observe interesting 
differences between the spectra and
interpret them. The spectrum of masses
and the interactions between mesonic excitations teach valuable lessons about the character of non-Abelian T-duality and its implications for Holography. The case of Abelian T-duality is also studied.
 } 
\keywords{Non Abelian T-duality, Holography, meson excitations.} 

\subheader{}

\begin{document}
\def\Tr{{\textrm{Tr}}}

\maketitle 

\section{Introduction and General Idea of this Paper}

Dualities are a very powerful tool in Theoretical Physics, particularly in the areas of Quantum Field Theory and String Theory. The possibility of calculating observables in the non-perturbative regime of a given physical system 'A' using the perturbative dynamics and degrees of freedom of a system 'B', opened the door to a different appreciation of what Quantum Field Theory or String Theory actually are. Indeed, one of the conceptual changes that the discovery of dualities have catalyzed, is that actually the system A is {\it the same} as the system B, in spite of the possibly very different appearance of  A and B. 

Paradigmatic representations of this idea in the area of String Theory are the web of String dualities \cite{Witten:1995ex}, the Maldacena Conjecture \cite{Maldacena:1997re}, dualities involving topological strings \cite{Gopakumar:1998ki} and many other examples with different physical or mathematical relevance.
There are also numerous examples of dualities purely in Quantum Field Theory, the logic and the conceptual advances they generated are of the same characteristics.

Of course, there are situations in which the two dual descriptions are proposed to be equivalent only in some  regime (for example at low energies)  \cite{Seiberg:1994pq}.  Also, there are examples where a mix of dualities and new parameters are introduced. These typically correspond to a {\it deformation} of the original system 'A' into a new system 'B' whose dynamics is characterized by this new parameter; examples of this are \cite{Lunin:2005jy}, \cite{Maldacena:2009mw}.

But the general lore is that when applying a transformation to dual variables, both systems are the same.
For example, T-duality was shown to be an equivalent way  of rewriting the  sigma model describing the string worldsheet dynamics (as vectorial or axial gaugings of the sigma model) \cite{Rocek:1991ps}. 
This has interesting consequences, like the phenomena of 'supersymmetry without supersymmetry'  \cite{Duff:1998us}. In that case, the T-dual of $AdS_5\times S^5$  (along a $U(1)$ that is fibered over $\mathbb{CP}^2$) is non-susy at the level of  {\it Supergravity}, but  SUSY is recovered when considering the full type IIA String Theory.  

The case of non-Abelian T-duality is particularly interesting for this logic. Indeed, after it was introduced by de la Ossa and Quevedo 
\cite{delaOssa:1992vci}, this generalization of T-duality was subject of careful research. The main puzzling problems concerned the invertibility and global aspects of the duality---see for example the papers
\cite{Alvarez:1993qi}. Interestingly,  Giveon and Rocek  \cite{Giveon:1993ai}, suggested that non-Abelian T-duality is transforming the sigma-model conformal field theory into a {\it different} conformal field theory.  If this were the case, non-Abelian T-duality should be just a {\it transformation} between different sigma models (hence not strictly a duality). Both sigma models give place to different Supergravity backgrounds satisfying the equations of motion, but more importantly different String Theories defined on these backgrounds. The  initial and the 
transformed sigma model would not be identical when including string corrections.

While these puzzling issues with non-Abelian T-duality seemed to have fell off the attention of researchers during various years, it was the work of Sfetsos and Thompson  \cite{Sfetsos:2010uq} that brought them back to our interest. Indeed,  the paper \cite{Sfetsos:2010uq} provided  an algorithmic way to transform Ramond fields, working out interesting examples, like that of $AdS_5\times S^5$, $AdS_3\times S^4\times T^4$, etc. This opened the way to apply it to systems with a  well understood holographic dual. The problem of the 'duality-character' of the non-abelian T-dual transformation gets mapped to the study of the quantum field theory dual of the original and the transformed backgrounds. 

At least at the level of Supergravity the analysis seems to confirm the proposal of 
\cite{Giveon:1993ai}. Indeed, it was understood in a 
variety of papers---see for example \cite{Macpherson:2014eza}
-\cite{Lozano:2016wrs}, that the dual field theory to the non-Abelian T-dual background is {\it different}
from the field theory dual to the seed-background. The holographic approach to the problem also allowed to
resolve some questions regarding global aspects, range of coordinates, etc. Many papers have discussed applications of non-Abelian T-duality to holography \cite{varios1}-\cite{varios4}, where both physical and geometrical aspects of the duality were analyzed.

In this work we consider an observable that combines aspects of Supergravity and String Theory. We study
the mesonic-like excitations of a quenched (in the sense that $\frac{N_f}{N_c}\sim 0$) quantum field theory. 
These excitations will be calculated as fluctuations of 
a probe $D$-brane in a Supergravity background. The mass spectra of these excitations
is a 'clean' observable to learn about the character of the Non-Abelian T-dual 
transformation.

We will consider the example of $AdS_5\times S^5$ and its mesonic excitations represented by the fluctuations of a probe $D7$ brane---the work of Kruczenski, Mateos, Myers and Winters \cite{Kruczenski:2003be}. We will non-Abelian T-dualize the system to obtain the dual to an ${\cal N}=2$ conformal field theory that is probed by a $D6$ brane. This probe brane will have induced charge of $D4$ brane on its worldvolume due to the presence of a Neveu-Schwarz $B_2$ field and will also be affected by a non-trivial dilaton. We will calculate the spectra of masses and interactions for a set of consistent fluctuations in the backgrounds  
before and after the duality and compare these observables. 
We will also perform the comparison with the Abelian T-dual background, again probed by a $D6$/$D4$ system with $B_2$ and dilaton.

{\it \underline{Summary of the Results}}

We will show that for the $D6$ brane fluctuating in 
the non-Abelian T-dual background there are three 
possible consistent fluctuations, types I, II and III. 
The interesting differences between the spectrum of 
masses and interactions  in the 'seed' and T-dualized  backgrounds, happen for the type I excitations. 
These change under 
duality. This points to the fact that
{\it at the Supergravity level} the dual ${\cal N}=2$ 
conformal field theories associated with $D7$ branes probing $AdS_5\times S^5$
and $D6$ branes probing the T-dual backgrounds are {\it different} four dimensional  CFTs.
It might be the case that string corrections rectify these differences, but if
this is the case the mechanism for which it happens 
is unclear to us at present.

{\it \underline{Plan of the paper}}

This is a paper with a high density of long calculations. The main part of the paper contains various sections that summarize  old and well established results together with the new ones. To make the reading more fluid we have written a large number of detailed appendices that the reader may consult. 

The plan of the paper is the following. In Section \ref{sectionsupergravity}, we set conventions and summarize the backgrounds, both  the seed-solution and the ones generated by duality. In Section \ref{sectionKMMW}, we  discuss the results of the important paper   \cite{Kruczenski:2003be}. In Sections \ref{mesonsintduals} and \ref{mesonsinHopftduals}  we calculate the mesonic excitations fluctuating a $D6$ brane both in the non-Abelian and in the Abelian T-dual backgrounds. Spectra of masses are obtained  and compared with the analogous ones in  \cite{Kruczenski:2003be}. It is here that we observe the non-trivial effect of non-Abelian T-duality on the spectrum of excitations.
The interaction between the fluctuations studied in previous sections are discussed in Section \ref{mesoninteractions}. Interesting differences with those in  \cite{Kruczenski:2003be}, that add to the point we are making above are 
discussed there. A summary of the results is given in Section \ref{analysis} and a set of  conclusions and possible future topics of research are given in Section \ref{conclusiones}.  As we mentioned, various detailed appendices complement the presentation.

%%%%%%%%%%%%%%%%%%%%%%%%%%%%%%%%%%%%%%%%%%%%%%%%%%%

\section{The Supergravity Solutions }\label{sectionsupergravity}
In this section, we will summarize the different backgrounds 
in Type II Supergravity that feature in our calculations. 
In Section \ref{mesonsintduals}, these solutions will be probed by $Dp$-branes and we will
compute the spectrum and interactions of the fluctuations of these $D$-branes.

To set our conventions and explain the notation, let us start with our 'seed' solution, the Type IIB $AdS_5\times S^5$ background,
\begin{equation}
 \begin{aligned}
& ds^2=\frac{ R^2}{L^2}dx_{1,3}^2 +\frac{ L^2}{R^2} \Big(   du^2 + u^2 d\phi^2 + d\rho^2 + \rho^2 d\Om_3^2  \Big),
\\[10pt]
& F_5 = dC_4 \ , \qquad C_4 = \frac{R^4}{L^4} dx_{1,3} - \frac{L^4 \ u^2 \ \big(  R^2 + \rho^2  \big)}{R^4} d\phi \wedge \textrm{Vol} \big(  \Om_3 \big) \ ,\label{ads5xs5}
 \end{aligned}
\end{equation}
where we defined,
\begin{equation}
 R^2 = u^2 + \rho^2 \ , \;\;\;  d\Omega^2_3=d\alpha^2+ \sin^2\alpha \, d\beta^2+ \sin^2\alpha \, \sin^2\beta \, d\gamma^2.
\end{equation}
The Ramond five form has a quantized charge\footnote{The integral $\int d \Big[\frac{ \ u^2 \ \big(  R^2 + \rho^2  \big)}{R^4} \Big]=1$, is calculated  moving to polar coordinates $u=R\cos\theta,\; \rho=R\sin\theta$, with $\th \in [0, \pi/2]$. We also use that $2\kappa_{10}^2 T_{Dp}=(2\pi)^{7-p} g_s \alpha'^{\frac{7-p}{2}}$.},
\begin{equation}
\frac{1}{2\kappa_{10}^2 T_{D3}} \int F_5=N_{D3}\;\;\to \,\;\;\frac{L^4}{\alpha'^2}=4\pi g_s N_{D3}.
\end{equation}
The reason to use this particular system of coordinates, introduced in \cite{Kruczenski:2003be}, will be explained in the following sections.  We will now present
the backgrounds generated by non-Abelian T-duality and Abelian T-duality applied on eq.(\ref{ads5xs5}).

\subsection*{The non-Abelian T-dual background }
We will now apply a non-Abelian T-duality transformation on the solution in eq.(\ref{ads5xs5}).
The resulting background was explicitly calculated (and aspects of the dual CFT were studied)
 in various papers \cite{Sfetsos:2010uq}-\cite{Lozano:2016kum}. 
 In Appendix \ref{appendix1}, we re-derive it using the general formulas
 % we will provide a nice algorithmic way of calculating the background using the procedure
 developed in \cite{Itsios:2012dc}.
Here, we will just quote the result for the NS and RR sectors of the supergravity background.

We perform a non-Abelian T-duality on one of the $SU(2)$'s inside the $SO(4)$ of the three-sphere. After the duality  the coordinates on the sphere will be replaced by a set of three
coordinates that we label $(r,\chi,\xi)$. The coordinates  $(\chi,\xi)$ describe a two sphere. The range of the coordinate $r$ is not determined by the duality procedure, but field
theoretic considerations in \cite{Lozano:2016kum} allow to find it. The supergravity background reads, for the Neveu-Schwarz sector,
%\noindent \emph{NS sector}
%
\begin{equation}
 \begin{aligned}
  & {ds^2}=\frac{R^2}{L^2}dx_{1,3}^2 +\frac{L^2}{R^2} \big(   d\rho^2 + du^2 + u^2 d\phi^2  \big) +\frac{4 \ \alpha'^2 R^2}{L^2 \rho^2} dr^2 +\frac{4 \ \alpha'^2 L^2 R^2 \rho^2 \ r^2}{16 \alpha'^2 R^4 r^2 
+ L^4 \rho^4}(d\chi^2 +\sin^2\chi \, d\xi^2),
\\[10pt]
& {B_2}= \frac{16 \ \alpha'^3 r^3 R^4}{16 \ \alpha'^2 r^2 R^4 +L^4 \rho^4}
\textrm{Vol}(S^2)\ , \qquad 
e^{-2{\Phi}}=\frac{ L^2 \rho^2}{64 \ \alpha'^3 R^6} \big(L^4 \rho^4 +16 \ \alpha'^2 r^2 R^4 \big).
\label{NSnatd}
 \end{aligned}
\end{equation}
%\vskip 10pt
For the Ramond-Ramond  sector, we find,
%\noindent \emph{RR sector}
%
\begin{equation}
\begin{aligned}
& {F_2} = d \Bigg[   \frac{L^4 \ u^2 \ \big(   R^2 + \rho^2  \big)}{8 \ \alpha'^{3/2} R^4} \ d\phi  \Bigg], \qquad \ \  {F}_4= {B}_2\wedge {F}_2 \ ,
\\[10pt]
& F_6 = d \Bigg[  \frac{2 \sqrt{\alpha'}}{L^4} r^2 R^3 dx_{1,3} \wedge dR  \Bigg]\ , \qquad
 F_8 = e^{2\Phi} \frac{L^2 r^2 \rho^6}{16 \ \alpha'^{3/2} R^3} \ dx_{1,3} \wedge dR \wedge dr \wedge \textrm{Vol}(S^2) \ .
\label{RRnatd}
\end{aligned}
\end{equation}
As in the papers \cite{Macpherson:2014eza}- \cite{Lozano:2016kum},
the charge of $D6$ branes is quantized and this imposes a relation for the size of the space $L$ after the duality,
\begin{equation}
Q_{D6}=\frac{1}{2\pi g_s \sqrt{\alpha'}}\int F_2=N_{D6}\to \frac{L^4}{\alpha'^2}= 8 g_s N_{D6}.
\end{equation}
The Page charge of $D4$ branes is vanishing, but as suggested 
in \cite{Lozano:2013oma},
 \cite{Lozano:2014ata}  and
 explained in \cite{Macpherson:2014eza}, \cite{Macpherson:2015tka}, \cite{Lozano:2016kum},  the $r$-coordinate
is naturally divided in intervals of size $\pi$. This generates, in the $r$-interval $[q\pi, (q+1)\pi]$, a Page charge for $D4$ branes $Q_{D4}= q N_{D6}$.

In the following sections we consider the dynamics of a probe $D6$ brane that extends along $[x_{1,3}, \rho, \chi, \xi]$. Hence, it is useful to present the expressions for the RR potentials $C_5$ and $C_7$. These potentials are related to the RR forms $F_6$ and $F_8$,
\begin{equation}
 F_6 = dC_5 \ , \qquad F_8 = dC_7 - H \wedge C_5 \ .
\end{equation}
The expressions for $C_1, C_3, C_5$ and $C_7$ are,
\begin{equation}
\label{C1C3C5C7}
 \begin{aligned}
  & C_1 =  \frac{L^4 \ u^2 \ \big(   R^2 + \rho^2  \big)}{8 \ \alpha'^{3/2} R^4} \ d\phi \ , \qquad \,\,\,
  C_3 = \frac{2 \a'^{3/2} L^4}{L^4 \r^4 + 16 \ \a'^2 r^2 R^4} \big(   R^4 - \r^4  \big) r^3 \ d\phi \wedge \textrm{Vol}(S^2) \ , 
 \\[10pt]
  & C_5 = \frac{2 \sqrt{\alpha'}}{L^4} r^2 R^3 dx_{1,3} \wedge dR \ , \qquad 
  C_7 = \frac{4 \alpha'^{3/2} r^3 R^3 \big(   8 \ \alpha'^2 r^2 R^4 - L^4 \rho^4  \big)}{3 L^4 \big(   16 \ \alpha'^2 r^2 R^4 + L^4 \rho^4  \big)} \ dx_{1,3} \wedge dR \wedge \textrm{Vol}(S^2) \ .
 \end{aligned}
\end{equation}

While these expressions are complicated, the combinations $C_7 - B_2 \wedge C_5$, appearing in the Wess-Zumino part of the Lagrangian of the $D6$ brane is simpler,
\begin{equation}
 \begin{aligned}
  & C_7 - B_2 \wedge C_5 = - \frac{4 \alpha'^{3/2}}{3 L^4} R^3 r^3 \ dx_{1,3} \wedge dR \wedge \textrm{Vol}(S^2)\ .
%  \\[10pt]
 % & C_7 + B_2 \wedge C_5 =  \frac{4 \alpha'^{3/2}}{3 L^4} \frac{32 \alpha'^2 r^2 R^4 - L^4 \rho^4}{16 \alpha'^2 r^2 R^4 + L^4 \rho^4} R^3 r^3 \sin\chi \ dx_{1,3} \wedge dR \wedge d\chi \wedge d\xi \ .
 \end{aligned}
\end{equation}
This completes our presentation of the non-Abelian T-dual background.
Let us now present the second background that will be relevant to our calculation, the (Abelian) T-dual of $AdS_5\times S^5$.
\subsection*{The Hopf-T-dual solution}
In order to perform a Hopf-T-duality it is convenient to write the metric of the three-sphere in terms of the left invariant one forms,
\begin{equation}
 \begin{aligned}
  & \om_1 = \frac{1}{\sqrt{2}} \big(  \cos \g \ d\a + \sin \g \  \sin \a \ d \b  \big) \ , \quad
\om_2 = \frac{1}{\sqrt{2}} \big(  - \sin \g \ d\a + \cos \g \  \sin \a \ d \b  \big) \ ,
\\
& \om_3 = \frac{1}{\sqrt{2}}  \big(   d \g + \cos \a \ d \b  \big) \ , \quad  d \Om^2_3 = \frac{1}{2} \big(  \om_1^2 + \om_2^2 + \om_3^2   \big) \ .
 \end{aligned}
\end{equation}
We  calculate the  Hopf-T-duality of the background in eq.(\ref{ads5xs5}), along the direction $\g$. The Neveu-Schwarz sector is,
%\vskip 10pt
%\noindent \emph{NS sector}
%
\begin{equation}
 \begin{aligned}
  & {ds^2}=\frac{R^2}{L^2}dx_{1,3}^2 +\frac{L^2}{R^2} \big(   d\rho^2 + du^2 + u^2 d\phi^2  \big) +\frac{4 \ \alpha'^2 R^2}{L^2 \rho^2} d\g^2 +\frac{L^2 \r^2}{4 R^2}(d\a^2 +\sin^2\a \ d\b^2).
\\[10pt]
& {B_2}= \a' \g \sin \a \ d\a \wedge d \beta \ , \qquad \qquad
e^{-2{\Phi}} = \frac{L^2 \r^2}{ 4 \a' R^2} \ ,
\label{NSatd}
 \end{aligned}
\end{equation}
and the Ramond-Ramond sector is,
%\vskip 10pt
%\noindent \emph{RR sector}
%
\begin{equation}
%\begin{aligned}
F_0 = F_2 = F_8 = 0 \ , \quad F_4 = \frac{L^4 \r^3 u}{2 \sqrt{\a'} R^6} \Big(  \r d u - u d\r  \Big) \wedge d\phi \wedge \textrm{Vol}(S^2) \ , \quad
%\\[10pt]
F_6 = d \Bigg[  \frac{4 \sqrt{\alpha'}}{L^4} \g R^3 dx_{1,3} \wedge dR  \Bigg] \ .
\label{RRatd}
%\end{aligned}
\end{equation}
There is here a quantized charge of $D4$ branes,
\begin{equation}
Q_{D4}=\frac{1}{8\pi^3 g_s \alpha'^{3/2}}\int F_4=N_{D4}\to \frac{L^4}{\alpha'^2}=4\pi g_s  N_{D4}.
\end{equation}
As discussed above, we will consider the dynamics of a $D6$ brane that extends along $[x_{1,3}, \rho, \a, \b]$. The expressions for the RR potentials $C_3$ and $C_5$ will be used in the Wess-Zumino part of the action for the probe brane. These potentials are related to the RR forms $F_4$ and $F_6$ through the relations
$
 F_4 = dC_3,  \;\;F_6 = dC_5 
$
and read,
\begin{equation}
\label{C3C5atd}
 C_3 = \frac{L^4 u \r^3 \phi}{2 \sqrt{\a'} R^6} \big(   u d\r - \r du  \big) \wedge \textrm{Vol}(S^2)  \ , \qquad C_5 = \frac{4 \sqrt{\alpha'}}{L^4} \g R^3 dx_{1,3} \wedge dR \ .
\end{equation}
Moreover, the RR 7-form potential $C_7$, is not trivial in this case. Indeed, we have $F_2 = F_8=0$, but since $F_8 = dC_7 - H \wedge C_5$ ---and in this case $H \wedge C_5 \ne 0$--- we must have a $C_7$ such that $dC_7 = H \wedge C_5$. After a straightforward integration we find,
\begin{equation}
\label{C7atd}
 C_7 = - \frac{\a'^{3/2}}{L^4} R^4 \ \g \ dx_{1,3} \wedge d\g \wedge  \textrm{Vol}(S^2) \ .
\end{equation}
We leave these backgrounds at this point. These will be used below to calculate the action of a probe $D6$-brane. Let us now summarize the results for  the calculation of the fluctuations 
of a probe $D7$ brane in the background of eq.(\ref{ads5xs5}), first discussed in the
seminal paper \cite{Kruczenski:2003be}.

%%%%%%%%%%%%%%%%%%%%%%%%%%%%%%%%%%%%%%%%%%%

\section{Summary of Results for $AdS_5\times S^5$: the work of KMMW}\label{sectionKMMW}
In this section, we discuss the main results of the pioneering  work of Kruczenski, Mateos, Myers and Winters (KMMW) on ${\cal N}=2$ meson spectroscopy  \cite{Kruczenski:2003be}. Our goal 
in this section,  is to provide a self-contained summary of these results. We will write here key-equations for the mass spectra, while
 full details for their derivation will be relegated to the Appendix \ref{appendixb}.

We work in the context of the  original version of the AdS/CFT correspondence \cite{Maldacena:1997re}. The authors of \cite{Kruczenski:2003be} considered the addition of fundamental hypermultiplets (flavor degrees of freedom or \emph{dynamical quarks}) in the $\mathcal{N} = 4$ SYM theory, thus breaking the supersymmetry to $\mathcal{N} = 2$. On the string theory side, the addition of fundamental $N_f$-flavor degrees of freedom (and its associated $SU(N_f)$ global symmetry) corresponds to the introduction of a stack of $N_f$ $D7$ probe branes \cite{Karch:2002sh}. The $D7$ branes and the stack of $N$ $D3$ branes that generate the type IIB supergravity background are taken to be parallel along the Minkowski directions. This configuration can be represented by the following array,
\begin{equation}
\label{D3D7}
\begin{array}{|p{0.8cm}|l|}
\hline
D3 & \quad 0 \;\; 1  \;\; 2  \;\; 3  \;\; - \; - \; - \; - \; - \; -
\\ \hline
D7 & \quad 0  \;\; 1  \;\; 2  \;\; 3 \;\;\; 4 \;\;\; 5 \;\;\; 6 \;\;\; 7 \; - \; -
\\ \hline
\end{array}
\end{equation}
The hypermultiplets in the field theory side arise from the lightest modes of the strings that are stretched between the $D3$ and $D7$ branes (i.e. the $3-7$ and $7-3$ strings) and thus their mass corresponds to the energy of 
those strings, given by $m_q = {l_q}/2 \pi \a'$, where ${l_q}$ 
is the separation of the two stacks of branes in the $89$-directions. 

The analysis that will be presented in this and the following sections is carried out in the probe approximation, which means that we take the number of the flavor branes to be much smaller than that of the color branes (i.e. $N_f \ll N$). In this limit, the flavor branes do not backreact on the geometry generated by the color branes. Equivalently, the dynamics of mesons is
influenced by the dynamics of glueballs, but not vice-versa (the meson dynamics is 'quenched').  For simplicity in our calculation, we  consider $N_f = 1$ probe $D7$-branes.

It should be clarified, that in this particular theory we do not have mesons and  glueballs
like in Yang-Mills or QCD, but it is more appropriate to think about these excitations as bound states. 

In order to study the spectroscopy of mesons (i.e. the quark-antiquark bound states) one has to look at the 
fluctuations of the Born-Infeld (BI) fields on the $D7$-brane \cite{Karch:2002xe}. In the present case, the 
linearized equations of motion for the fluctuations can be solved analytically and as a result one also finds analytical expressions
 for the masses of the scalar and vector mesons. In what follows we present the summary of such mass spectra. We use  a notation that is appropriate to make comparisons with  analogous original  results in the next sections, that we found doing analogous calculations with the Abelian and non-Abelian T-dual backgrounds.

Let us start by considering a $D7$-brane that probes the geometry generated by the $D3$ branes in eq. (\ref{ads5xs5}) according to the array \eqref{D3D7}. Notice that we parametrize the 6-dimensional space, that is transverse to the directions of the $D3$ branes as,
\begin{equation}
 d\vec{Y}^2 = dR^2 + R^2 d\Om_5^2 = du^2 + u^2 d\phi^2 + d\r^2 + \r^2 d\Om_3^2 \ ,
\end{equation}
where $R$ is the holographic coordinate, $d\Om_3^2$ and $d\Om_5^2$ are the line elements for a unit three- and  five-spheres respectively.
 In this parametrization, $\r$ and $\Om_3$ parametrize the directions $(4, 5,6, 7)$ while $u$ and $\phi$ the transverse directions $(8,9)$. Also, $\r$ corresponds to the distance in the directions $(4, 5,6, 7)$ and $u$ is the distance in the directions $(8,9)$. Thus $\r$ and $u$ are related to the usual  holographic coordinate $R$, as $R^2=u^2+\rho^2$. The $D7$ brane is taken in such a way that,
\begin{equation}
\label{originalembedding}
 u = u_0 = \textrm{const} \ , \qquad \phi = \phi_0 = \textrm{const} \ ,
\end{equation}
where $u_0$ measures the distance of the $D7$ and $D3$ branes along their common transverse directions. As a result, the induced metric of the $D7$ brane in our notation takes the form,
\begin{equation}
 ds^2 = \frac{\r^2 + u_0^2}{L^2} dx^2_{1,3} + \frac{L^2}{\r^2 + u_0^2} \Big(   d\r^2 + \r^2 d\Om^2_3  \Big) \ .
 \label{xaxa}
\end{equation}
Notice that in this way, we can have the $D7$ brane ''ending'' at $\rho=0$ without any tadpole problem---see \cite{Karch:2002sh}. 
Also, the quarks acquire a mass proportional to $u_0$, hence 
breaking conformality in  quenched fashion 
and giving a nontrivial dynamics to the system.

In order to study the meson spectrum of this system one has to look at the fluctuations of the DBIWZ action for the $D7$ brane,
\begin{equation}
 \mathcal{L}_{D7} = -T_{D7} \int d^8 \s e^{- \Phi} \sqrt{- \det \big(   P[g] + \mathcal{F}  \big)} + \frac{T_{D7}}{2} \int P[C_4] \wedge \mathcal{F} \wedge \mathcal{F} \ ,
\end{equation}
where $\mathcal{F} \equiv B_2 + 2 \p \a' F$ and F is the field strength of the worldvolume gauge field. Up to quadratic order, it turns out that the fluctuations of the scalars $(u, \phi)$ do not couple to those of the worldvolume gauge field and thus one can study them separately. We will now describe the result for the masses of scalar and vector excitations. Full details of the derivation are relegated to Appendix \ref{appendixb}.

\subsubsection*{Fluctuations of the scalar fields}
Expanding the Lagrangian in eq.(\ref{xaxa}) for  fluctuations of the form
$u=u_0+\delta u$ and $\phi=0+ \delta \phi$, the dynamics of these excitations---to second order-- is given
by a Lagrangian that schematically reads,
\begin{equation}
\mathcal{L}_{2}\sim \frac{\sqrt{- \det g_{ind}}}{\rho^2+u_0^2}\Big( (\partial u)^2 +u_0^2 (\partial \phi)^2   \Big).
\end{equation}
Studying the solutions to the corresponding differential equation,  demanding their regularity, square integrability and a special conspiracy of constants to have a finite series expansion (the full details of this logic  and computation
are discussed in Appendix \ref{appendixb}), 
we obtain the mass spectrum for the fluctuation, labelled by two positive integers $(n,l)$. This reads
\footnote{
Notice that for the scalar and the type II mesons we take $l \ge 1$, i.e. we exclude the modes with $ l = 0 $ in contrast with \cite{Kruczenski:2003be}. The reason we do this is because it turns out that the modes with $ l = 0 $ are not square integrable. For more details about the square integrability of the modes the reader is referred to the subsection \ref{normalizability}.
},
\begin{equation}
 M^2_s(n,l) = \frac{4 u^2_0}{L^4} {\big(   n + l + 1  \big) \big(   n + l + 2  \big)} \ , \qquad n = 0 , 1 , \ldots \ , \qquad l = 1 , 2 , \ldots \ .\label{lnde}
\end{equation}
One also finds that the conformal dimension of the operator corresponding to these fluctuations  is $\Delta = l+3$. Details of 
this derivation are spelled out in Appendix \ref{appendixb}.
Let us now succinctly discuss the case of vectorial fluctuations.

\subsubsection*{Fluctuations of the gauge fields}

Following a similar logic to the one described above, we consider the perturbation,
\begin{equation}
 A_a = 0 + \delta A_a \Rightarrow F_{ab} = 0 + \delta F_{ab} \ .
\end{equation}
According to \cite{Kruczenski:2003be} 
there are three types of consistent fluctuations, each one of them is described below.

\vskip 10pt

\noindent \underline{Type I mesons:}
In this case the proposed fluctuation is,
\begin{equation}
\label{mesonsI}
 \delta A_{\mu} = 0 \ , \qquad \delta A_{\r} = 0 \ , \qquad \delta A_{i} = e^{i k \cdot x} \varphi^{\pm}_{I} (\r) \mathcal{Y}^{l, \pm}_{i} \big( S^3 \big) \ ,
\end{equation}
where $\mathcal{Y}^{l, \pm}_{i} \big(   S^3  \big)$ with $l \ge 1$ are the vector spherical harmonics on $S^3$
The differential equations and the physically acceptable  solutions for this fluctuation are spelled out in Appendix \ref{appendixb}. 
From the solutions we can extract the mass spectra $\tilde{M}=\frac{ML^2}{u_0}$, for the type I fluctuations (mesons) which are  \cite{Kruczenski:2003be} ,
\begin{equation}
 \begin{aligned}
  & M^2_{I,+} = \frac{4 u_0^2}{L^4} \big(  n + l + 2   \big) \big(   n + l + 3  \big) \ , \qquad n = 0 , 1 , \ldots \ , \qquad l = 1 , 2 , \ldots \ ,
  \\[10pt]
  &   M^2_{I,-} = \frac{4 u_0^2}{L^4} \big(  n + l    \big) \big(   n + l + 2  \big) \ , \;\;\;\;\;\; \qquad n = 0 , 1 , \ldots \ , \qquad l = 1 , 2 , \ldots \ .
 \end{aligned}
\end{equation}
Also,  from the asymptotic behavior of the solutions at large values of $\r$ one finds that
the conformal dimensions of the corresponding operators at the UV are $\Delta_{+} = l+5$ and $\Delta_{-} = l+1$.

\vskip 10pt

\noindent \underline{Type II mesons:}
Another consistent  fluctuation is,
\begin{equation}
\label{mesonsII}
 \delta A_{\mu} = \zeta_{\mu} e^{i k \cdot x} \varphi_{II} (\r) \mathcal{Y}^{l} \big( S^3 \big) \ , \qquad k \cdot \zeta = 0 \ , \qquad \delta A_{\r} = 0 \ , \qquad \delta A_{i} = 0 \ .
\end{equation}
and the associated mass spectrum was found to be \cite{Kruczenski:2003be},
\begin{equation}
   M^2_{II} = \frac{4 u_0^2}{L^4} \big(  n + l + 1   \big) \big(   n + l + 2  \big) \ , \qquad n = 0 , 1 , \ldots \ , \qquad l = 1 , 2 , \ldots \ .
\end{equation}
Also, one finds that the conformal dimension of the associated operator is $\Delta = l + 3$. 

\vskip 10pt

\noindent \underline{Type III mesons:}
The fluctuation we consider now is,
\begin{equation}
\label{mesonsIII}
 \delta A_{\mu} =0 \ , \qquad \delta A_{\r} = e^{i k \cdot x} \varphi_{III} (\r) \mathcal{Y}^{l} \big( S^3 \big) \ , \qquad \delta A_{i} = e^{i k \cdot x} \tilde{\varphi}_{III} (\r) \partial_{i} \mathcal{Y}^{l} \big( S^3 \big) \ .
\end{equation}
Things in this case are  slightly more involved (details are explained in Appendix \ref{appendixb}), but following  the same logic as above leads to the  mass spectrum \cite{Kruczenski:2003be},
\begin{equation}
 M^{2}_{III} = \frac{4 u_0^2}{L^4} \big(   n + l + 1  \big) \big(  n + l + 2   \big) \ , \qquad n = 0 , 1 , \ldots \ , \qquad l = 1 , 2 , \ldots \ .
\end{equation}

Finally, from  the behavior of the solution at the boundary we read  the conformal dimension of the associated  operator to be $\D = l+3$.

This completes the summary of the spectrum of masses (the reader unfamiliar with this material should complement it with the explanations in Appendix \ref{appendixb}). The paper \cite{Kruczenski:2003be}  initiated a line of work that  when applied to different gauge-string duals gave numerous results of interest in Physics.
A summary of those results can be found in \cite{Erdmenger:2007cm}.

We will now study and describe in more detail the same results for the fluctuations of a probe $D6$ brane both in the non-Abelian and Abelian T-dual backgrounds of eqs.(\ref{NSnatd})-
(\ref{RRnatd}) and (\ref{NSatd})-(\ref{RRatd}). The comparison between these observable spectra teaches valuable lessons about the character of these dualities.

%%%%%%%%%%%%%%%%%%%%%%%%%%%%%%%%%%%%%%%%%%%

\section{Mesons in the T-dual Backgrounds }
\label{mesonsintduals}

In this section, we study the fluctuations of a 
$D6$ brane, that extends on the Minkowski $R^{1,3}$ 
and $\rho$-directions and wraps the two sphere  
$\Omega_2(\chi,\xi)$ of the Type IIA background 
in eqs.(\ref{NSnatd})-(\ref{RRnatd}). This set-up can be summarized by the following array,
\begin{equation}
\begin{array}{|p{1.2cm}|l|}
\hline
\textrm{NATD} & \;\;\; x_{1,3} \;\;\;\; u \;\;\;\; \phi \;\;\;\; \r \;\;\;\; r \;\;\;\; \Omega_2(\chi,\xi)
\\ \hline
D6 & \;\;\;\; \times \;\;\;\;\ - \;\; - \;\;\; \times \;\; - \;\;\;\;\;\;\;\; \times
\\ \hline
\end{array}
\end{equation}
We will identify these fluctuations with the bosonic part of
 meson superfields in the dual ${\cal N}=2$ CFT. 
The goal of this section is to find the spectra of 
masses for different kinds of consistent fluctuations.

The embedding of the probe $D6$ brane is determined by giving the expression 
of the transverse coordinates of the background $(u,\phi,r)$ as functions
of the worldvolume coordinates,
\begin{eqnarray}
 u=u(x^{\mu},\rho,\chi,\xi),\;\; \phi=\phi(x^{\mu},\rho,\chi,\xi),\;\; r=r(x^{\mu},\rho,\chi,\xi),\nonumber
\end{eqnarray}
and subject these functions to solve the equations of motion coming from the DBIWZ Action,
\begin{equation}
\label{DBIWZ}
 \mathcal{L}_{D6} = -T_{D6} \ \int d^{7}\sigma e^{-\Phi} \sqrt{- \det\big(   g + \mathcal{F}  \big)} + T_{D6} \int \Big(   C_7 - \mathcal{F} \wedge C_5 + \frac{1}{2} C_3 \wedge \mathcal{F}^2 - \frac{1}{6} C_1 \wedge \mathcal{F}^3  \Big) \ ,
\end{equation}
where $\mathcal{F} = B_2 + 2 \pi \alpha' F$ and $F$ is the field strength of the worldvolume gauge field.
The existence of such solutions  is discussed in Appendix \ref{Embedding} and we show  there, that one possible solution is
\begin{equation}
 r = 0 \ , \qquad u, \phi = \textrm{arbitrary constants} \ .\nonumber
\end{equation}
However, such a solution is not useful for the study of the fluctuations 
of the $D6$ brane since, as explained in Appendix
\ref{Embedding}, for $r=0$ the brane action vanishes 
identically (both the DBI and the WZ parts of the action vanish).

At this point, we recall a nice subtlety that appears in 
backgrounds obtained by non-Abelian T-duality. This was first
observed in 
\cite{Lozano:2014ata} and  elaborated upon in \cite{Macpherson:2014eza}, \cite{Macpherson:2015tka}.
Indeed, backgrounds like that in eq.(\ref{NSnatd}), present a shrinking two cycle $\Sigma_2[\chi,\xi]$ that
vanishes at the position $\rho=0$. A Bohm-Aharonov-like 
experiment should not be able to detect a Euclidean 
string that wraps $q$-times the cycle $\Sigma_2$. 
This implies the quantization of the quantity,
\begin{equation}
b_0=\frac{1}{4\pi^2\alpha'}\int_{\Sigma_2}B_2= q, \qquad q = 1,2, \ldots 
\end{equation}
We will impose that $0\leq b_0\leq 1$. Calculating explicitly with the $B_2$ in eq.(\ref{NSnatd}) specialized at $\rho=0$, we find that $b_0=\frac{r}{\pi}$. This implies 
that the coordinate $r$ naturally divides in intervals of size $\pi$. If we are in the interval $q\pi< r< (q+1)\pi$, we should perform a large gauge transformation,
\begin{equation}
\label{B2lg}
 B^{l.g.}_2 = \frac{16 \ \alpha'^3 r^3 R^4}{16 \ \alpha'^2 r^2 R^4 +L^4 \rho^4}
\sin\chi d\chi \wedge d\xi - q \alpha' \pi \sin\chi d\chi \wedge d\xi \ , \qquad q = 1,2, \ldots
\end{equation}
bringing $b_0$ back to the interval $0\leq b_0\leq 1$. It is precisely these large gauge transformations that allow us to find more interesting embeddings, with non-vanishing action.

In fact, in Appendix \ref{Embedding}, we find constant embeddings of the form,
\begin{equation}
\label{straightembedding}
 r = q \ \pi \ , \qquad q = 1, 2, \ldots \qquad u, \phi = \textrm{arbitrary constants} \ .
\end{equation}
Interestingly, the fact that these $D6$ branes 'sit' at the points $r= q\pi$ suggest the 'physical reality' of the intervals.
When discussing interactions between the fluctuations, 
we will find that the couplings between mesons depend on the integer $q$.

In the following we are going to consider the fluctuations of the $D6$ brane around the embedding  in eq.\eqref{straightembedding}. For that purpose we will consider fluctuations of the embedding functions $(r, u, \phi)$ and the worldvolume gauge field $A_M$, following the perturbation scheme below,
\begin{equation}
\label{perturbationScheme}
 r = q \ \pi + \delta r \ , \qquad u = u_0 + \delta u \ , \qquad \phi = 0 + \delta \phi \ , \qquad A_{M} = 0 + \delta A_{M} \ ,
\end{equation}
where the fluctuations depend on all the worldvolume coordinates. The constant $u_0$ must be nontrivial. 
It is $u_0$ the scale that we introduce in our CFT dual to the non-Abelian T-dual background. This scale allows the possibility of having the mesonic bound states.

To study the fluctuations  in eq.(\ref{perturbationScheme}), in Appendix \ref{usefulformulas} we have developed general formulas for a generic perturbation of the form \eqref{genpert}.
%\begin{equation}
% g_{ab} = \bar{g}_{ab} + \delta g_{ab} \ ,  \quad \delta g_{ab} \ll 1 , \;\;\;
%  \mathcal{F}_{ab} = \bar{\mathcal{F}}_{ab} + \delta \mathcal{F}_{ab} \ ,  \quad \delta \mathcal{F}_{ab} \ll 1 ,\;\;\;\Phi = \bar{\Phi} + \delta \Phi \ ,  \quad \delta \Phi \ll 1 \ .\nonumber
% \end{equation}
% where we use bars to denote the unperturbed fields.  
%$\mathcal{F}_{ab}$ is defined as $ \mathcal{F}_{ab} = B_{ab} + 2 \pi \alpha' F_{ab} \ . $
In this case the equations of motion for the fluctuations in eq.(\ref{perturbationScheme}), give a coupled system of second order differential equations that is written in Appendix \ref{equationsfluctuated}. 

A priori, one might expect qualitative differences between 
these fluctuations and those in the paper of KMMW. 
Indeed, not only the isometries of the backgrounds are different, in KMMW the probe is a $D7$ brane, here we have a $D6$ with charge of $D4$ brane induced on its worldvolume due to the large gauge transformation ---see \cite{Macpherson:2014eza}, \cite{Macpherson:2015tka}. Intuitively, we might also expect that the dynamics of our fluctuations to be more of the type studied in \cite{Arean:2006pk}
 for $Dp-D{(p+2)}$ branes. Besides, we have also a $B_2$ NS field and a nontrivial dilaton in our system, that are vanishing in the KMMW case.  Were all these intuitions correct, the effects of the
non-Abelian T-duality would be very noticeable in the spectra. The rest of this section studies this question.

We will make use of the perturbation scheme written in eq.(\ref{perturbationScheme}) in the case in 
which the fluctuations resemble the consistent ones 
classified by KMMW. We will consider fluctuations of Types 
I, II and III. All these will have fluctuations of 
the scalar directions $(u,\phi)$. 
These directions have been 'untouched' by 
the process of non-Abelian T-duality and we expect 
their dynamics to be unaffected too. 
Configurations of Type I will fluctuate the $r$-coordinate 
and the gauge fields on the two-sphere. 
Type II configurations present  a vector meson, 
as the gauge fields on the Minkowski directions will be excited. 
Finally, type III configurations will switch 
on a gauge field on the $\rho$ and sphere directions. Below, we will study the effect of each of these particular consistent fluctuation on the general equations of motion written in Appendix \ref{equationsfluctuated}.
 We will start with what we call Type I fluctuations in our system.

\subsection{Type I solutions}

Let us start with the study of the mesons of type I. For this purpose we use the following ansatz for the
 fluctuations\footnote{The $l=0$ mode for the
fluctuations $\delta A_\chi, \delta A_\xi$, 
does not participate in the following analysis, since $Y^0(\chi,\xi)=1$ 
and its derivative puts to zero the gauge field fluctuations along the directions $\chi$ and $\xi$.},
\begin{equation}
 \begin{aligned}
  & \delta r = F_{(r)}(\r) \ e^{i k \cdot x} \ \mathcal{Y}^l (\chi,\xi) \ ,\quad  
\delta u = F_{(u)}(\r) \ e^{i k \cdot x} \ \mathcal{Y}^l(\chi,\xi) \ , \quad  
\delta \phi = F_{(\phi)}(\r) \ e^{i k \cdot x} \ \mathcal{Y}^l(\chi,\xi)
\\[10pt]
 & \delta A_{\mu} = 0 \ , \;\delta A_{\r} = 0 \ ,\;\delta A_{\chi} =  \phi_{I}(\r) \ e^{i k \cdot x} \ \frac{\partial_{\xi}\mathcal{Y}^l(\chi,\xi)}{\sin\chi} \ ,
\;\delta A_{\xi} = - \phi_{I}(\r) \ e^{i k \cdot x} \ \sin\chi \ \partial_{\chi}\mathcal{Y}^l(\chi,\xi) \ .\label{flucttype1}
 \end{aligned}
\end{equation}
The function $\mathcal{Y}^l$ is the scalar spherical harmonic on a two-sphere and thus it is solution of the eigenvalue problem,
\begin{equation}
\label{eqYl}
 \nabla^2_{S^2} \mathcal{Y}^l = - l (l+1) \ \mathcal{Y}^l \ .
\end{equation}
Moreover, notice that the angular parts in the expressions for $\delta A_{\chi}$ and $\delta A_{\xi}$ above correspond to the vector harmonics defined in terms of the scalar harmonics of $S^2$ as $\sqrt{\tilde{g}} \ \mathcal{Y}^l_i \equiv \tilde{g}_{ij} \ \epsilon^{jk} \nabla_{k} \mathcal{Y}^l$, where $\tilde{g}_{ij}$ are the metric components of the unit two-sphere. Using this ansatz we specialize eqs.(\ref{deltar11})-(\ref{fluctaxi}) for the consistent fluctuation in eq.(\ref{flucttype1}) and rewrite the equations of motion for each fluctuated field as,

\vskip 10pt

\noindent \emph{\underline{Equations for $\delta u$ and $\delta \phi$}}

\vskip 10pt

\begin{equation}
\label{decoupledEqs}
   F''_{(u)} + \frac{3}{\r}F'_{(u)} + \Bigg[   \frac{L^4 M^2}{ (u_0^2 + \r^2)^2 } - \frac{4 l (l+1)}{\r^2}  \Bigg] F_{(u)} = 0 \ ,
  \;\; F''_{(\phi)} + \frac{3}{\r}F'_{(\phi)} + \Bigg[   \frac{L^4 M^2}{ (u_0^2 + \r^2)^2 } - \frac{4 l (l+1)}{\r^2}  \Bigg] F_{(\phi)} = 0 \ .
\end{equation}
\vskip 10pt

\noindent \emph{\underline{Equation for $\delta r$}}

\vskip 10pt

\begin{equation}
\label{rI}
 F''_{(r)} + \frac{u_0^2 + 5 \r^2}{\r \big(   u_0^2 + \r^2  \big)}F'_{(r)} + \Bigg[   \frac{L^4 M^2}{ (u_0^2 + \r^2)^2 } - \frac{4}{\r^2} \ \frac{u_0^2 + 3 \r^2}{u_0^2 + \r^2} - \frac{4 l (l+1)}{\r^2}  \Bigg] F_{(r)} - \frac{8 \pi l (l+1)}{\r^2} \frac{u_0^2 + 3 \r^2}{u_0^2 + \r^2} \phi_{I} = 0 \ .
\end{equation}
\vskip 10pt

\noindent \emph{\underline{Equations for $\delta A_{\chi}$ and $\delta A_\xi$},}

\vskip 10pt

\begin{equation}
\label{AchiI}
  \phi''_{I} + \frac{u_0^2 + 5 \r^2}{\rho \big(   u_0^2 + \r^2  \big)} \phi'_{I} + \Bigg[  \frac{L^4 M^2}{\big(  u_0^2 + \r^2  \big)^2} - \frac{4 l (l+1)}{\rho^2}    \Bigg] \phi_{I} - \frac{2}{\pi \r^2} \frac{u_0^2 + 3 \r^2}{u_0^2 + \r^2} F_{(r)} = 0 \ .
\end{equation}
On the other hand, the equations for $\delta A_{\mu} , \ \mu = t, x_1, x_2, x_3$ and $\delta A_\rho$, are satisfied identically. 
Let us move to find an explicit solution 
for these linear second order differential equations.
\subsubsection{Solving the equations}

We will start first with the solution of the decoupled 
equations \eqref{decoupledEqs}. It is convenient 
to re-write them in terms of the  dimensionless coordinate $z = \frac{\r}{u_0}$,
\begin{equation}
 \begin{aligned}
   & \partial^2_{z}F_{(u)} + \frac{3}{z}\partial_{z}F_{(u)} + \Bigg[   \frac{\bar{M}^2}{ (1 + z^2)^2 } - \frac{4 l (l+1)}{z^2}  \Bigg] F_{(u)} = 0 \ ,
\\[10pt]
   & \partial^2_{z}F_{(\phi)} + \frac{3}{z}\partial_{z}F_{(\phi)} + \Bigg[   \frac{\bar{M}^2}{ (1 + z^2)^2 } - \frac{4 l (l+1)}{z^2}  \Bigg] F_{(\phi)} = 0 \ ,\label{papapa}
 \end{aligned}
\end{equation}
where we have also defined the dimensionless constant $\bar{M}^2 = \frac{L^4 M^2}{u^2_0}$. The solutions are,
\begin{equation}
 \begin{aligned}
    & \textrm{First solution} \qquad z^{2 l} \big(   1 + z^2 \big)^{-\alpha}  {}_2F_1\big( -\alpha, -\alpha + 2 l + 1; 2(l + 1) ; -z^2\big) \ ,
\\[10pt]    
    & \textrm{Second solution} \qquad z^{-2 (l + 1)} \big(   1 + z^2 \big)^{-\alpha} {}_2F_1 
    \big(   -\alpha, -\alpha - 2 l - 1; -2 l; -z^2   \big) \ .\label{mamaza}
 \end{aligned}
\end{equation}
Where we have defined,
\begin{equation}
 \alpha = \frac{-1 + \sqrt{1 + \bar{M}^2}}{2} > 0 \qquad \textrm{or} \qquad \bar{M}^2 = 4 \alpha \big(   \alpha + 1  \big) \ .
\end{equation}
We are interested only in the first solution since it is 
the only one that is regular at $z = 0$. The second solution diverges at $z=0$ 
and cannot be considered to be a small fluctuation.
The mass spectrum comes out if, as discussed in Appendix \ref{appendixb}, we require that,
\begin{equation}
 -\alpha + 2 l + 1 = - n \ , \qquad n = 0 , 1 , 2 , \ldots
\end{equation}
Then we find that,
\begin{equation}
\label{eq:4.15}
 M^2 = \frac{4 u_0^2}{L^4} \big(   n + 2 l + 1   \big) \big(   n + 2 l + 2   \big) \ .
\end{equation}

%\red{How can we compute the mass spectrum here? It seems that the hypergeometric series is always infinite because $\alpha > 0$. Moreover, if we replace $l \rightarrow \frac{l}{2}$ then our differential equation becomes equation (3.10) of Myer's paper. Notice that here we define $\alpha$ in a different way compared to Myer's paper, see eq. (3.12) of that paper.}
Some comments are in order here. 
Notice that the differential equations and the mass spectrum 
for the fluctuated modes $(\delta u, \delta\phi)$, 
are exactly the same as the analogous ones in the KMMW paper if we replace 
\begin{equation}
l_{KMMW}=2l_{here}.
\end{equation}
In this case, the mass spectrum found above coincides with the analogous tower of states in  KMMW---see our eq.(\ref{lnde}). Also, the equations
(\ref{papapa}) coincide with the analogous eqs.(\ref{lns}) in KMMW. 
As expected, the directions that did not participate 
in the non-Abelian T-duality, do not show any change in its equations 
of motion and spectrum (in spite of having 
both backgrounds different isometries, different field 
content and using different probes).

Let us now study the coupled system of differential 
equations \eqref{rI} and \eqref{AchiI}. For our convenience we define the operator $\mathcal{O}_{(I)}$ which acts on functions of $\rho$ as,
\begin{equation}
\label{OperatorO}
 \mathcal{O}_{(I)} f = \partial^2_{\r} f + \frac{u_0^2 + 5 \r^2}{\rho \big(   u_0^2 + \r^2  \big)} \partial_{\r} f + \Bigg[  \frac{L^4 M^2}{\big(  u_0^2 + \r^2  \big)^2} - \frac{4 l (l+1)}{\rho^2}    \Bigg] f \ .
\end{equation}
Then the equations \eqref{rI} and \eqref{AchiI} can be written in matrix form as,
\begin{equation}
 \mathcal{O}_{(I)} \begin{pmatrix}
 			     F_{(r)}
			     \\[5pt]
			     \phi_{I}
 			   \end{pmatrix} = a \ \mathcal{M}_{(I)} \begin{pmatrix}
 			                  			          F_{(r)}
			    						  \\[5pt]
			     						  \phi_{I}
 			  					         \end{pmatrix} \ ,
\end{equation}
where, we defined,
\begin{equation}
 a = \frac{4}{\rho^2} \frac{u_0^2 + 3 \r^2}{u_0^2 + \r^2} \ , \qquad \mathcal{M}_{(I)} = \begin{pmatrix}
			   			      									 1 & \quad 2 \pi l (l+1)
						      									 \\[5pt]
						     									  \frac{1}{2 \pi} & \quad 0			   			      									                 \end{pmatrix} \ .
\end{equation}
To solve this problem, we  diagonalize the matrix $\mathcal{M}_{(I)}$ whose eigenvalues are,
\begin{equation}
 \l_1 = - l \ , \qquad \l_2 = l + 1 \ .
\end{equation}
The matrix that diagonalizes $\mathcal{M}_{(I)}$ is ,
\begin{equation}
 P = \begin{pmatrix}
 	- 2 \pi l & 2 \pi \big(  l + 1  \big)
	\\[5pt]
	     1    &       1
        \end{pmatrix}
\end{equation}
and thus
$
 P^{-1} \mathcal{M}_{(I)} P = D = \textrm{diag}(\l_1 , \l_2) \ .
$
Let us now define the vector,
\begin{equation}
 Y = \begin{pmatrix}
         y_1
         \\[5pt]
         y_2
       \end{pmatrix} = P^{-1} \begin{pmatrix}
                                             F_{(r)}
			    	          \\[5pt]
			     	            \phi_{I}
       					 \end{pmatrix} \ .
\end{equation}
Then the system of differential equations \eqref{rI} and \eqref{AchiI}, can be written as,
\begin{equation}
 \begin{aligned}
  & \partial^2_{\r} y_1 + \frac{u_0^2 + 5 \r^2}{\rho \big(   u_0^2 + \r^2  \big)} \partial_{\r} y_1 + \Bigg[  \frac{L^4 M^2}{\big(  u_0^2 + \r^2  \big)^2} - \frac{4 l (l+1)}{\rho^2} +  \frac{4 l}{\rho^2} \frac{u_0^2 + 3 \r^2}{u_0^2 + \r^2}    \Bigg] y_1 = 0 \ ,
  \\[10pt]
  & \partial^2_{\r} y_2 + \frac{u_0^2 + 5 \r^2}{\rho \big(   u_0^2 + \r^2  \big)} \partial_{\r} y_2 + \Bigg[  \frac{L^4 M^2}{\big(  u_0^2 + \r^2  \big)^2} - \frac{4 l (l+1)}{\rho^2} -  \frac{4 \big(   l + 1  \big) }{\rho^2} \frac{u_0^2 + 3 \r^2}{u_0^2 + \r^2}    \Bigg] y_2 = 0 \ ,
 \end{aligned}
\end{equation}
or, in terms of the coordinate $z$ we have
\begin{equation}
\label{eqx1x2}
 \begin{aligned}
  & \partial^2_{z} y_1 + \frac{1 + 5 z^2}{z \big(   1 + z^2  \big)} \partial_{z} y_1 + \Bigg[  \frac{ \bar{M}^2}{\big(  1 + z^2  \big)^2} - \frac{4 l (l+1)}{z^2} +  \frac{4 l}{z^2} \frac{1 + 3 z^2}{1 + z^2}    \Bigg] y_1 = 0 \ ,
  \\[10pt]
  & \partial^2_{z} y_2 + \frac{1 + 5 z^2}{z \big(   1 + z^2  \big)} \partial_{z} y_2 + \Bigg[  \frac{ \bar{M}^2}{\big(  1 + z^2  \big)^2} - \frac{4 l (l+1)}{z^2} -  \frac{4 \big(   l + 1  \big) }{z^2} \frac{1 + 3 z^2}{1 + z^2}    \Bigg] y_2 = 0 \ .
 \end{aligned}
\end{equation}
Both equations (\ref{eqx1x2}) can be solved in terms of hypergeometric functions. 
The two solutions for $y_1$ are,
\begin{equation}
 \begin{aligned}
    & \textrm{First solution} \qquad z^{2 l} 
\big(   1 + z^2  \big)^{ - \alpha - 1} 
{}_2F_1\big( -\alpha + 2 l - 1, - \alpha + 1 ; 2l + 1 ; -z^2\big) \ ,
\\[10pt]    
    & \textrm{Second solution} \qquad z^{-2 l} 
\big(   1 + z^2 \big)^{ - \alpha - 1} 
{}_2F_1 \big(   -\alpha  - 2 l + 1, -\alpha - 1; -2 l + 1; -z^2   \big) \, ,
 \end{aligned}
\end{equation}
while those for $y_2$ are,
\begin{equation}
 \begin{aligned}
    & \textrm{First solution} 
\qquad z^{2 l + 2} \big(   1 + z^2  \big)^{-\alpha - 1} 
{}_2F_1\big( -\alpha - 1, -\alpha + 2 l + 3; 2l + 3 ; -z^2\big) \ ,
\\[10pt]    
    & \textrm{Second solution} \qquad z^{-2 (l + 1)} 
\big(   1 + z^2 \big)^{-\alpha - 1} 
{}_2F_1 \big(   -\alpha + 1, -\alpha - 2 l - 3; -2 l - 1; -z^2   \big) \ .
 \end{aligned}
\end{equation}
From the expressions above we notice that in both cases the solutions 
that are regular at $z = 0$ are the first ones, 
hence the only ones we will consider here.
Let us now try to obtain the mass spectrum for the mode $y_1$.
In that case we have to require
\begin{equation}
 - \alpha + 2 l - 1 = - n \ , \qquad n = 0 , 1 , 2 , \ldots
\end{equation}
Then for $z\gg 1$ the solution behaves as $z^{- 2 l}$. This vanishes when $z$ approaches the infinity only if $l \ge 1$. The mass spectrum is given by,
\begin{equation}
\label{Mx12}
  M^2 = \frac{4 u_0^2}{L^4} \big(   n + 2 l - 1   \big) \big(   n + 2 l   \big) \ , \qquad l = 1 , 2 , 3 , \ldots \ .
\end{equation}
In order to obtain the mass spectrum for the mode $y_2$  we impose,
\begin{equation}
 - \alpha + 2 l + 3 = - n \ , \qquad n = 0 , 1 , 2 , \ldots
\end{equation}
and the mass spectrum is then given by,
\begin{equation}
\label{eq:4.26}
 M^2 = \frac{4 u_0^2}{L^4} \big(   n + 2 l + 3   \big) \big(   n + 2 l + 4   \big) \ .
\end{equation}
%
%In that case, the behavior of $x_2$ at large z is $z^{-2l - 6}$.

In summary, we have found that the diagonal modes $(y_1,y_2)$ have an equation of motion and a spectrum that
is {\it different} from the analogous one for the modes $\varphi_I^{\pm}$ in KMMW, written in eqs.(\ref{baba}), (\ref{bababa}). 
In other words, we find that for type I fluctuations, the strongly coupled meson dynamics, described by a probe $D6$ brane on the non-Abelian T-dual background in eqs.(\ref{NSnatd})-(\ref{RRnatd}) differs from the analogous one in the seed system, calculated with a $D7$ probe on $AdS_5\times S^5$. This is an interesting contribution of this paper. The non-Abelian T-duality, at least at this stage of the analysis is not just a re-writing of the same physical system. 
We shall now study the square integrability of the solutions of Type I.
\subsubsection{Square integrability of the solutions}
\label{normalizability}

We study now the square integrability of the solutions in this section. We remind the reader that, for a generic fluctuation $\frak{f}(x^{\mu},\rho,\Omega_2)$, we impose,
\begin{equation}
\int d^7x \sqrt{- \det{g_{ind}}} \ \Big| \frak{f}(x^{\mu},\rho,\Omega_2) \Big|^2 <\infty.\label{mamaza}
\end{equation}
We choose to impose this condition, since we are solving a Sturm-Liouville like problem in a curved space. Other possible criteria would be to impose that the modes have finite action. Enforcing  the condition in eq.(\ref{mamaza}) is more stringent than imposing the finiteness of the action.

In the case at hand, using eq.(\ref{NSnatd}), we find,
\begin{equation}
\det g_{ind}= - \frac{R^6}{L^6} 
\ \Bigg[ \frac{4\alpha'^2 L^2 R^2 \rho^2r^2}{16\alpha'^2 R^4 r^2+L^4 \rho^4} \Bigg]^2 
\sin^2\chi.
\end{equation}
We see that the only place where square integrability of the solutions can break down is in the integral in the $\rho$-direction. 
Indeed, since $\sqrt{-\det g_{ind}}\sim \rho^3\sim z^3$ for large values of 
$\rho$ or $z$, we will need that the modes $\frak{f}$ decay 
at least as $\frak{f} \sim \frac{1}{\rho^{2+\epsilon}}$.
Let us first apply this to the solutions for the 
scalar modes $(u,\phi)$ found in eq.(\ref{mamaza}). Since  $\alpha=n+2l+1$ and the hypergeometric asymptotes to ${}_2 F_1\sim z^{2n}$, for large values of the coordinate $z$, the fluctuations behave as,
\begin{equation}
u\sim \phi \sim \frac{z^{2l}}{z^{2n+4l+2}}\times z^{2n}\sim \frac{1}{z^{2l+2}}.
\end{equation}
This implies that the combination
\begin{equation}
\sqrt{\det g} |u|^2\sim\sqrt{\det{g}} |\phi|^2\sim \frac{1}{z^{4l+1}}.
\end{equation}
We then find that the wave functions are square integrable if $l=1,2,3,4....$. The mode with $l=0$ is not square integrable, hence should not be considered part of the spectrum\footnote{The square integrability condition can be thought 
of as a wave function renormalization. All excitations can be square integrable, given
a cutoff $\Lambda$ in the theory. The coupling between a problematic
field $\frak{f}_{NR}$ with an operator made out of square integrable fields $O(\phi_R)$ 
is of the form $V\sim \frak{f}_{NR} O(\phi_R)$. Scaling 
$\frak{f}_{NR}\to \frac{\frak{f}}{\sqrt{\Lambda}}$ to normalize it, implies that when 
removing the cutoff the Hilbert space will split and 
there will be no communication between the square integrable and problematic modes. Thanks to Maurizio Piai for a discussion about this.}. 

We analyze in a similar fashion, the square integrability of the system $(y_1,y_2)$. Let us start with the simpler case of the 
mode $y_2$. As we explained, in this case $\alpha=n+2l+3$ and for large values of $z$ we have,
\begin{equation}
y_2(z)\sim\frac{1}{z^{2l+6}}\to \sqrt{\det g} |y_2|^2\sim \frac{1}{z^{4l+9}},
\end{equation}
rendering all the fluctuations of $y_2$ square integrable.

We move now on to the study of
the fluctuations of $y_1$. We should be more careful here. When $l>1$ we have that $\alpha=n+2l-1$ and $y_1\sim \frac{1}{z^{2l}}$. Hence for $l=2,3,4....$ the fluctuation modes of $y_1$ are square integrable.
For $l=1$ we have that $2l+1-\alpha=1-\alpha$ and $\alpha=n+1$. The $l=1$ modes of $y_1$ decay as $y_1\sim \frac{1}{z^2}$, making the fluctuation not square integrable.

In summary, the scalar modes $(u,\phi)$ are square integrable 
for $l>0$, the $y_2$ modes for $l \ge 1$ and the $y_1$ modes for $l \ge 2$.

%%%%%%%%%%%%%%%%%%%%%%%%%%%%%%%%%%%%%%%%%%%%%%%%%%%

\subsection{Type II solutions}

For the type II solutions we consider the ansatz below,
\begin{equation}
\begin{aligned} & 
\delta r =0 \ ,\quad  
\delta u = F_{(u)}(\r) \ e^{i k \cdot x} \ \mathcal{Y}^l(\chi,\xi) \ , \quad  
\delta \phi = F_{(\phi)}(\r) \ e^{i k \cdot x} \ \mathcal{Y}^l(\chi,\xi)
\\[10pt] & 
\delta A_{\mu} = \zeta_{\mu} \ \phi_{II}(\r) \ e^{i k \cdot x} \ \mathcal{Y}^l(\chi,\xi) \ , \quad k \cdot \zeta = 0 \ , \quad 
\delta A_{\r} = 0 \ ,\quad \delta A_{\chi} = 0 \ , \quad \delta A_{\xi} = 0 \, .
 \end{aligned}
\end{equation}
Using this ansatz in the equations of motion \eqref{deltar11} - \eqref{fluctaxi} we obtain a set of second order differential equations 
for the radial functions $F_{(u)}, F_{(\phi)}$ and $\phi_{II}$. It turns out that the differential equations satisfied by 
$F_{(u)}$ and $F_{(\phi)}$ are given in eq. \eqref{decoupledEqs}---these directions are inert under the duality. 
Hence we focus on the equation for $\phi_{II}$ given below,

\vskip 10pt

\noindent \emph{\underline{Equations for $\delta A_{\mu} , \ \mu = t, x_1, x_2, x_3$}}

\vskip 10pt

\begin{equation}
\label{AmuiII}
  \phi''_{II} + \frac{3}{\rho} \phi'_{II} + \Bigg[  \frac{L^4 M^2}{\big(  u_0^2 + \r^2  \big)^2} - \frac{4 l (l+1)}{\rho^2}    \Bigg] \phi_{II} = 0 \ .
\end{equation}

\vskip 10pt

\noindent Moreover, the equations for $\delta r, \delta A_{\rho}, \delta A_{\chi}$ and $\delta A_{\xi}$ are satisfied identically.
Notice that the eq.\eqref{AmuiII},  is the same as the differential 
equations \eqref{decoupledEqs} and thus it shares the same solutions and the same spectrum. We now study the Type III fluctuations.

%%%%%%%%%%%%%%%%%%%%%%%%%%%%%%%%%%%%%%%%%%%%%%%%%%%

\subsection{Type III solutions}

In this case we adopt the following ansatz for the fluctuations,
\begin{equation}
\begin{aligned}
 & \delta r = 0 \ ,\quad  \delta u = F_{(u)}(\r) \ e^{i k \cdot x} \ \mathcal{Y}^l(\chi,\xi) \ , \quad  
\delta \phi = F_{(\phi)}(\r) \ e^{i k \cdot x} \ \mathcal{Y}^l(\chi,\xi)
\\[10pt]
 & \delta A_{\mu} = 0 \ , \quad \delta A_{\r} = \phi_{III}(\r) \ e^{i k \cdot x} \ \mathcal{Y}^l(\chi,\xi) \ ,\quad 
\delta A_{\chi} = \tilde{\phi}_{III}(\r) \ e^{i k \cdot x} \ \partial_{\chi} \mathcal{Y}^l(\chi,\xi) \ ,
\\[10pt]
& \delta A_{\xi} = \tilde{\phi}_{III}(\r) \ e^{i k \cdot x} \ \partial_{\xi} \mathcal{Y}^l(\chi,\xi) \, .
\end{aligned}
\end{equation}
Notice that here we use a different definition of the vector harmonics that appear in the expressions for $\delta A_{\chi}$ and $\delta A_{\xi}$, 
namely $\mathcal{Y}^l_i \equiv \nabla_{i} \mathcal{Y}^l$. Plugging this ansatz in the equations of motion \eqref{deltar11} - \eqref{fluctaxi} we obtain a set of second order differential equations 
for the radial functions $F_{(u)}, F_{(\phi)}$ and $\phi_{III}, \tilde{\phi}_{III}$. One can verify easily that $F_{(u)}$ and $F_{(\phi)}$ satisfy the equations \eqref{decoupledEqs}. 
For the functions $\phi_{III}$ and $\tilde{\phi}_{III}$ we find,

\vskip 10pt

\noindent \emph{\underline{Equation for $\delta A_{\mu} , \ \mu = t, x_1, x_2, x_3$}}

\vskip 10pt

\begin{equation}
\label{AmuIII}
  \phi'_{III} + \frac{3}{\rho} \phi_{III} - \frac{4 l (l+1)}{\rho^2} \tilde{\phi}_{III} = 0 \ .
\end{equation}

\vskip 10pt

\noindent \emph{\underline{Equation for $\delta A_{\rho}$}}

\vskip 10pt

\begin{equation}
\label{ArIII}
 4 l (l+1) \tilde{\phi}'_{III} + \Bigg[  \frac{L^4 M^2 \rho^2}{(u_0^2 + \rho^2)^2} - 4 l (l+1)  \Bigg] \phi_{III} = 0 \ .
\end{equation}

\vskip 10pt

\noindent \emph{\underline{Equation for $\delta A_{\chi}$ \& $\delta A_{\xi}$}}

\vskip 10pt

\begin{equation}
\label{AchiIII}
 \tilde{\phi}''_{III} - \phi'_{III} + \frac{u_0^2 + 5 \rho^2}{\rho \big(   u_0^2 + \rho^2 \big)} \Big(   \tilde{\phi}'_{III} - \phi_{III}  \Big) 
+ \frac{L^4 M^2}{\big(   u_0^2 + \rho^2 \big)^2} \tilde{\phi}_{III} \, .
\end{equation}

\vskip 10pt

\subsubsection{Solving the equations}

Here we solve the differential equations that we found after imposing the type III constraints to the equations of motion.
This is a system of differential equations between the radial functions $\phi_{III}$ and $\tilde{\phi}_{III}$, namely eqs. \eqref{AmuIII}, \eqref{ArIII} and \eqref{AchiIII}.
Notice that the equation \eqref{AchiIII} can be obtained from \eqref{AmuIII} and \eqref{ArIII}. This means that only two of these equations are independent. 
For convenience we are going to find the solutions for $\phi_{III}$ and $\tilde{\phi}_{III}$ by considering eqs. \eqref{AmuIII} and \eqref{ArIII}. In order to solve this system we differentiate the equation \eqref{AmuIII} and we eliminate $\tilde{\phi}_{III}$ using the equation \eqref{ArIII}. If we do this then we end up with the following differential equation for $\phi_{III}$,
\begin{equation}
\label{eq:4.41}
 \partial_{z} \Bigg[   \frac{1}{z} \partial_{z} \Big(   z^3 \phi_{III}  \Big)  \Bigg]  + \Bigg[   \frac{\bar{M}^2 z^2}{\big(   1 + z^2  \big)^2} - 4 l (l+1)  \Bigg] \phi_{III} = 0 \ .
\end{equation}
The two solutions of this equation are,
\begin{equation}
 \begin{aligned}
    & \textrm{First solution} \qquad z^{2 l - 1} \big(   1 + z^2 \big)^{-\alpha}  {}_2F_1\big( -\alpha, -\alpha + 2 l + 1; 2(l + 1) ; -z^2\big) \ ,
\\[10pt]    
    & \textrm{Second solution} \qquad z^{-2 (l + 1)-1} \big(   1 + z^2 \big)^{-\alpha} {}_2F_1 \big(   -\alpha, -\alpha - 2 l - 1; -2 l ; -z^2   \big) \ .
 \end{aligned}
\end{equation}
Notice that the second solution is always singular at $z = 0$ while the first one is always regular only for $l \ge 1$. In order to compute the mass spectrum we will consider the first solution and we will concentrate on the modes with $l \ge 1$. Now if we require,
\begin{equation}
 - \alpha + 2 l + 1 = - n \ , \qquad n = 0 , 1 , 2 , \ldots
\end{equation}
we see that for large $z$ the solution behaves as $z^{- 2 l - 3}$ and the mass spectrum is given by the formula,
\begin{equation}
  M^2 = \frac{4 u_0^2}{L^4} \big(   n + 2 l + 1   \big) \big(   n + 2 l + 2   \big) \ .
\end{equation}
Obviously this solution is square integrable.
Knowing the solutions for $\phi_{III}$ we can immediately find the solutions for $\tilde{\phi}_{III}$ from the formula \eqref{AmuIII}.

This finishes the analysis of the fluctuations in the background obtained by non-Abelian T-duality. We briefly study now the Abelian T-dual case.
%%%%%%%%%%%%%%%%%%%%%%%%%%%%%%%%%%%%%%%%%%%

\section{Mesons in the Hopf-T-dual background}
\label{mesonsinHopftduals}

In this section we are going to study the dynamics of a $D6$ brane which extends along the directions $x_{1,3}, \rho, \a, \b$ of the 
dual solution in eqs. \eqref{NSatd}, \eqref{RRatd}. Such a configuration can be described schematically with the following array,
\begin{equation}
\begin{array}{|p{1.6cm}|l|}
\hline
\textrm{Hopf-TD} & \;\;\; x_{1,3} \;\;\;\; u \;\;\;\; \phi \;\;\;\; \r \;\;\;\; \g \;\;\;\; \Omega_2(\a,\b)
\\ \hline
D6 & \;\;\;\; \times \;\;\;\;\ - \;\; - \;\;\; \times \;\; - \;\;\;\;\;\;\;\; \times
\\ \hline
\end{array}
\end{equation}
 The DBIWZ action which describes dynamics of the brane now has the following form,
\begin{equation}
\label{DBIWZATD}
 \mathcal{L}_{D6} = -T_{D6} \ \int d^{7}\sigma e^{-\Phi} \sqrt{- \det\big(   g + \mathcal{F}  \big)} + T_{D6} \int \Big(   C_7 - \mathcal{F} \wedge C_5 + \frac{1}{2} \mathcal{F} \wedge \mathcal{F} \wedge C_3  \Big) \ .
\end{equation}

Again,for simplicity we are going to consider constant embeddings, 
%i.e. embeddings of the form
%
\begin{equation}
\label{straightembeddingATDgeneral}
 \g, u, \phi = \textrm{const} \ .
\end{equation}
Before studying those embeddings we must first prove their existence
\footnote{
 For more details the reader is referred to the appendix \ref{EmbeddingATD}.
}. 
The way to do it is to study the equations of motion for the embedding 
functions $\g, u , \phi$,  
allowing a dependence on all the worldvolume coordinates. If we do this we will realize that the only possibility for a constant embedding is,
\begin{equation}
\label{straightembeddingATD}
 \g = 0 \ , \qquad u, \phi = \textrm{arbitrary constants} \ .
\end{equation}

In order to compute the meson spectrum in this case we must examine the fluctuations of the $D6$ brane around the embedding \eqref{straightembeddingATD}. For that purpose we will consider fluctuations of the embedding functions $\g, u, \phi$ and the worldvolume gauge field following the perturbation scheme
\begin{equation}
\label{perturbationSchemeATD}
 \g = 0 + \delta \g \ , \qquad u = u_0 + \delta u \ , \qquad \phi = 0 + \delta \phi \ , \qquad A_{M} = 0 + \delta A_{M} \ ,
\end{equation}
where the fluctuations depend on all the worldvolume coordinates and the constant $u_0$ must be nontrivial. As a result, the equations of motion for the fluctuations give a coupled system of second order differential equations. The coupled system of the equations of motion can be obtained from the equations \eqref{deltar11} - \eqref{fluctaxi} of the NATD case after replacing,
\begin{equation}
\label{NATDtoATD}
 r \rightarrow \g \ , \qquad \chi \rightarrow \a \ , \qquad \xi \rightarrow \b \ .
\end{equation}
Using this correspondence one can map the equations 
%for $\delta r$ of 
for the NATD case to the respective equations in the Abelian T-dual case. 
%for $\delta \g$ and similarly the equations 
%for $\delta A_{\chi}$ and $\delta A_{\xi}$ 
%to those for $\delta A_{\alpha}$ and $\delta A_{\beta}$ respectively. 
%The only terms that are not related through this map are those that are proportional to $\delta r$ and $\delta \g$. It is obvious that the coefficients of these terms are not equal.

The correspondence in eq. \eqref{NATDtoATD} tells us 
that the equations of motion for the fluctuations of the Hopf-T-dual case 
are the same to the equations of motion for the fluctuations in the NATD case. 
This remarkable relation is illustrated
at the level of the quadratic Lagrangians for the fluctuations. 
Indeed, one can prove that the quadratic 
Lagrangians in the Hopf-T-dual and the NATD 
cases differ only by an overall factor of $q \pi$ 
(more details about the quadratic Lagrangians in these two cases 
can be found in the appendices \ref{quadrLagrNATD} and \ref{quadrLagrHopfTD}). 
Due to the existence of such a map, we do not 
have to solve the equations of motion since all of them have already 
been solved in the previous section. We find the same mass spectra.

%%%%%%%%%%%%%%%%%%%%%%%%%%%%%%%%%%%%%%%%%%%%%%%%%%%

%Start with the NATD case - subsection. Describe the brane embedding that we are
%using here (make an array) and say few things about the large gauge transformation
%here. Give the general equations of motion and then focus on each type of mesons -
%three subsubsections. Send the reader to the appendices for the details of the
%fluctuations analysis. 

%Create another small subsection for the Hopf-TD case. Describe the brane embedding
%we consider here (make an array). Mention that it is not needed to present the
%results for the Hopf-TD since these are the same as in the NATD case. Send the
%reader to the appendices for the details of the fluctuations analysis.

%%%%%%%%%%%%%%%%%%%%%%%%%%%%%%%%%%%%%%%%%%%

\section{Interaction of the Fluctuations}
\label{mesoninteractions}

Rather surprisingly, the meson spectrum analysis for both the non-Abelian
T-dual background  in Section \ref{mesonsintduals} 
and the Abelian T dual one in Section \ref{mesonsinHopftduals} cases  turned out to be 
identical. In order to probe the differences between the two distinct {\it rotations}\footnote{We use the word 'rotation' a bit freely here. We just refer to  the different transformations Abelian and non-Abelian, as 'rotations'.} of the archetypical 
$AdS_5\times S^5$ background, we move one step forward in the analysis of the effective meson theory 
and consider the interactions between modes. Following closely \cite{Kruczenski:2003be}, we will expand the 
$D6$-brane action to higher order in the scalar fields. 
We then obtain a (3+1)-dimensional theory through a KK reduction on
the $D6$-brane worldvolume theory,  over the internal $S^2$ and along 
the radial direction $\rho$.

We will restrict our analysis to the interactions of those scalars that are decoupled from the rest, 
namely $u$ and $\phi$, and we will consider cubic and quartic terms in the expansion of the Lagrangian.
As will prove in the following, the analysis of the interaction terms between the decoupled scalar mesons 
in the two effective theories will be enough to distinguish between the Abelian and non-Abelian cases,  a distinction that at the level of the spectrum 
(quadratic expansion of the Lagrangian)  was not possible. 

We expand the brane Lagrangian in the Abelian and the non-Abelian case up to fourth order and,  to perform the 
integration of the radial  $\rho$ and angular coordinates, the following field decomposition is 
used,
\begin{equation} \label{KKansatz}
\delta u \, = \,  \sum_{\alpha} F_{(u)}^{\alpha} (\rho) \, Y^{\alpha}(S^2)\, U_{\alpha} (x^\mu)\ ,
\qquad
\delta \phi \, = \,  \sum_{\beta} F_{(\phi)}^{\beta} (\rho) \, Y^{\beta}(S^2)\, \Phi_{\beta} (x^\mu)\ ,
\end{equation}
where the sum over $\alpha$ \& $\beta$ runs over all values of the quantum numbers $l,m$ \& $n$, denoting the transformation properties under rotations $(l,m)$ and Energy level $n$.
The radial functions in eq. \eqref{KKansatz} satisfy their respective equations
of motion, namely eqs.\eqref{decoupledEqs}.  

Integrating the Lagrangian over $\rho$ and $S^2$ will produce dimensionless constants that 
will depend on the different quantum numbers ($l,m$ \& $n$), which specify each KK mode. 
Due to the orthogonality properties of the angular functions, after 
substituting   eq.\eqref{KKansatz} into the Lagrangian, the corresponding integral of each 
quadratic term will be proportional to $\delta _{l,l'}$, $\delta_{ m,m'}$ and $\delta_{ n,n'}$ .
A simplification is possible for the quadratic terms, which contain a product of two angular
factors, but does not happen for the cubic (and quartic) terms, which contain three (or more) angular functions. 

After the integration, a proper normalization of the fields is required in order to bring the quadratic 
Lagrangian to a canonical form. In this way, the resulting Lagrangian, can be interpreted as a 
(3+1)-dimensional effective field theory for the mesons (a sort of 
'chiral Lagrangian' for this particular system).  Noticing that the required field redefinition is 
different for each set of quantum numbers, with some loss of generality (but not losing qualitative aspects of the dynamics) we present the expressions of the field redefinition 
for the set of quantum numbers $l=2$, \, $m=n=0$,
\begin{align}
&U_{N}(x^{\mu}) \, =\, \frac{24 \sqrt{385 \, \pi}}{u_0^{16} \, \sqrt{q \, N}} \, U(x^{\mu}) 
\quad \& \quad 
\Phi_{N}(x^{\mu}) \, =\, \frac{24 \sqrt{385 \, \pi}}{u_0^{17} \, \sqrt{q \, N}} \, \Phi(x^{\mu})
 \quad \Rightarrow \quad \text{NATD}&
 \\[2pt]
& U_{A}(x^{\mu}) \, =\, \frac{24 \pi \sqrt{385}}{u_0^{16} \, \sqrt{ N}}  \, U(x^{\mu}) 
\quad \& \quad 
\Phi_{A}(x^{\mu}) \, =\, \frac{24 \pi \sqrt{385}}{u_0^{17} \, \sqrt{N}}\, \Phi(x^{\mu})
 \quad \Rightarrow \quad \text{ATD} &
\end{align}
The $U_{N}$ and $\Phi_{N}$ have the same dimensions as $\delta u$ and $\delta \phi$, 
namely [L]$^{-1}$ and [L]$^{-2}$.
Substituting these new fields into the Lagrangian we have the following expression for the quadratic 
part (non-Abelian T-dual --NATD-- and Abelian T-dual-- ATD)\footnote{The mass parameters, for each set of quantum numbers, 
depend on $n$ and $l$ through the expression \eqref{eq:4.15}.} 
\begin{equation} \label{Lag2}
{\cal L}_2  \, = \, - \, \frac{1}{2}\, \left(\partial_\mu U \partial^\mu U +
\partial_\mu \Phi \partial^\mu \Phi\right)  \, - \,  
\frac{1}{2} \, M^2 \, \left(U^2 \, + \,  \Phi^2 \right) \, . 
\end{equation}
If we substitute these redefined fields in the cubic and quartic part of the Lagrangian it is 
possible to read off the dimensionfull cubic and quartic coupling constants. 
From the cubic expansion we have\footnote{In the following expressions $\phi$ is a general scalar, 
either one of the $U$ or $\Phi$. The letters in red-color indicate the factor appears for the non-Abelian case only.}
\begin{equation} \label{cubic}
g_{\phi(\partial \phi)^2} \sim \frac{1}{\sqrt{{\color[rgb]{1,0,0}q} \, N}} \frac{\alpha'}{u_0} 
\sim \frac{1}{\sqrt{{\color[rgb]{1,0,0}q} \, N}} \frac{1}{m_q} 
\quad \& \quad
g_{\phi^3} \sim \, {1 \over \lambda} \, \frac{1}{\sqrt{{\color[rgb]{1,0,0}q} \, N}}  \, m_q \
\quad \Rightarrow \quad \text{{\color[rgb]{1,0,0}N}ATD} 
\end{equation}
while from the quartic terms,
\begin{equation} \label{quartic}
g_{\phi^4} \sim \frac{1}{\lambda}\frac{1}{{\color[rgb]{1,0,0}q} \, N} \, , \quad
g_{\phi^2(\partial \phi)^2} \sim \frac{1}{{\color[rgb]{1,0,0}q} \,  N}\frac{1}{m_q^2}
\quad \& \quad
g_{(\partial \phi)^4} \sim \frac{\lambda}{{\color[rgb]{1,0,0}q} \,N}\frac{1}{m_q^4}
\quad \Rightarrow \quad \text{{\color[rgb]{1,0,0}N}ATD} 
\end{equation}
where $\lambda=g_s N$ is the 't Hooft coupling. Notice that the scaling with the number of colours is the usual one in large-N phenomenology \cite{clasicos}, being meson-decays suppressed in this regime. This approach gives us the coefficient for the interaction terms, their dependence on the mass of the quarks $m_q$ and the quantised parameter $q$, that would have been very hard to compute with field theory methods. 

A few important comments are in order. 
First it is clear, both from eqs. \eqref{cubic} and \eqref{quartic}, that the coupling constants of the non-Abelian T-dual background 
are {\it dressed}---with powers of $q$--- with respect to those of the Abelian case (and the case of $AdS_5\times S^5$). Let us remind the reader that this parameter is related to the large gauge transformation in 
the NATD background and to the massive embedding that we fluctuated to obtain the spectrum.
The different powers of $q$  in equations \eqref{cubic} and \eqref{quartic} are easily understood if one combines 
the $q^{-1/2}$ that comes from the normalization of the scalar fields 
with the fact that in the Lagrangian for the non-Abelian T-dual case,
there is an overall factor of $q$. 

Another point that deserves  notice is that even when the two backgrounds came out of  non trivial 
dualities of the $AdS_5\times S^5$ solution, comparing the expansions 
of the Lagrangian at quadratic, cubic and 
quartic order of the $(u,\phi)$ fluctuations, 
they satisfy the relation  ${\cal L}_{NATD} \, = \, q \, \pi \, {\cal L}_{ATD}$.
A final comment is that four out of the five couplings that appear in eqs. \eqref{cubic} and \eqref{quartic} are also present in the 
$AdS_5\times S^5$ analysis of \cite{Kruczenski:2003be}. The ${\phi^3}$ coupling is not reported in the paper \cite{Kruczenski:2003be}, but our analysis 
of that background shows that it is  present. 

Summarizing the analysis of the interactions of the decoupled scalar sector
($u,\phi$), in the non-Abelian and Abelian backgrounds, 
we arrive to the conclusion that the  couplings calculated with
 the Abelian T-dual background are the same as those calculated with
$AdS_5\times S^5$, while 
those calculated with the  non-Abelian T-dual 
solution are {\it dressed} by factors of $q$.
%%%%%%%%%%%%%%%%%%%%%%%%%%%%%%%%%%%%%%%%%%%
\section{Analysis of the Results}\label{analysis}

In this section, we will summarize the results obtained
by probing the non-abelian T-dual background, we will compare these results
with the analogous ones obtained in $AdS_5\times S^5$. 
Let us start with the quadratic fluctuations and the spectra of masses.

We first deal with the scalar fluctuations $(\delta u, \delta\phi)$. 
These obey the same equations of motion when 
considered as a fluctuation of a $D7$
brane in $AdS_5\times S^5$---see eq.(\ref{lns})--or when studied as fluctuations of a 
$D6$ brane in the Type IIA non-abelian T-dual background--see eq.(\ref{papapa}). 

The mass spectrum is the same as we found in eqs.(\ref{lnde}) and eq.(\ref{eq:4.15}). 
Notice that the $l=0$ mode is not square integrable, 
so it should be excluded from the spectrum.

The profile of the fields $(\delta u, \delta\phi)$ 
vanish asymptotically in the same fashion before and 
after the duality (it is the same solution!), hence they 
correspond to operators of dimensions $\Delta_u=\Delta_\phi=2l+3$. 
The calculation of fluctuations at the quadratic order are 
supposed to match, because the directions $u,\phi$ 
are 'inert/uncharged' under the duality.

Let us now compare the Type II modes. The equation of motion 
for the vector mesons in the Type IIB calculation---see eq.(\ref{eq:B.28}), coincide
with the analogous one in the Type IIA fluctuating $D6$---see eq.(\ref{AmuiII}).
The mass spectrum for the vector mesons coincide before and after the duality
if we replace $l_{KMMW}=2l_{here}$, as already indicated. 
Notice that using the
mapping of angular momentum quantum numbers mentioned above, 
the Type II modes are identical before and after non-Abelian T-duality.

Let us now compare the Type III modes. In the type IIB system
the fluctuations obey the equations (\ref{eqIII1})- (\ref{eq:b.35}), which
after some assumptions appear in the Type IIA calculation---see eqs.(\ref{eq:4.41})
and (\ref{AmuIII}),(\ref{ArIII}). Hence, we will have the same solutions and 
the same spectrum, after the identification $l_{KMMW}=2l_{here}$.

Finally, let us discuss the more interesting case of the Type I excitations.
Aside from the scalar fluctuations $(\delta u, \delta \phi)$ common
to all cases and invariant under duality as we explained, we found 
after diagonalization, the modes $(y_1,y_2)$. They correspond to a fluctuation
along the $r$-direction and the $S^2(\chi,\xi)$ that the $D6$
branes are wrapping. The analogous Type IIB 
$\varphi^{\pm}_{I}$ describe excitations on the three sphere that the $D7$ brane wraps.

The equations of motion for the $D7$ brane 
in Type IIB are eqs.(\ref{baba})-(\ref{eq:B.24})
and the spectrum is given in eq.(\ref{bababa}). The operators dual to $\varphi^{\pm}_{I}$
have dimensions $\Delta_+=l+5$ and $\Delta_-=l+1$. On the Type IIA side, we
find the equations of motion (\ref{eqx1x2}). The solutions show the mass spectrum in
eq.(\ref{Mx12}) for $y_1$ and eq.(\ref{eq:4.26}) for $y_2$. 
We copy them below to ease the reading. In Type IIA we have,
%y_1
\begin{equation}
 \begin{aligned}
  & M^2_{y_1} = \frac{4u_0^2}{L^4}(n+2l-1)(n+2l) \ ,    \qquad l=2,3,4 \ldots \ , \; \qquad \Delta_{y_1} = 2l
   \\[10pt]
 & M^2_{y_2}=\frac{4u_0^2}{L^4}(n+2l+3)(n+2l+4) \ , \;\;\;\;\;   l=1,2 \ldots \ , \; \qquad 	\Delta_{y_2}=2l+6			   
 \end{aligned}
\end{equation}
%
%\begin{eqnarray}
%& & M^2_{x_2}=\frac{4u_0^2}{L^4}(n+2l+3)(n+2l+4),\;\;\;
%\Delta_{x_2}=2l+6,\nonumber\\
%& & M^2_{x_1}=\frac{4u_0^2}{L^4}(n+1)(n+2),\;\;l=0;\nonumber\\  
%& & M^2_{x_1}=\frac{4u_0^2}{L^4}(n+2l-1)(n+2l),\;\; l=2,3,4....\nonumber\\
%& & \Delta_{x_1}= 2l.
%\end{eqnarray}
%
In Type IIB, we have,
\begin{equation}
 \begin{aligned}
  & M^2_{+}=\frac{4u_0^2}{L^4}(n+l+2)(n+l+3) \ , \qquad l = 1,2,3, \ldots \ , \qquad \Delta_{+}=l+5 \ ,
\\[10pt]
  & M^2_{-}=\frac{4u_0^2}{L^4}(n+l)(n+l+2) \ , \;\;\;\;\;\; \qquad l = 1,2,3, \ldots \ , \qquad \Delta_{-}= l+1.
 \end{aligned}
\end{equation}
%
%\begin{eqnarray}
%& & M^2_{+}=\frac{4u_0^2}{L^4}(n+l+2)(n+l+3),\;\;\;
%\Delta_{+}=l+5,\nonumber\\
%& & M^2_{-}=\frac{4u_0^2}{L^4}(n+l)(n+l+2),\;\;l=1,2,3,4....;  
% \Delta_{-}= l+1.\nonumber
%\end{eqnarray}
%
We see that under the scaling $l_{KMMW}=2l_{here}$, 
and for the same excitation level $n_*$, the 
spectrum is not invariant\footnote{Notice that for the 
$y_2, \varphi^{+}_{I}$ fluctuations, we can match the spectra if aside from the scaling of $l$, we also
change the excitation levels $n_*\to n_*+1$.}.

This discrepancy in the result for this observable before and after non-Abelian
T-duality shows that there is some non-trivial effect, at least at the 
level of detail that Supergravity allows us to calculate.

These differences are associated with the two different 
dynamics in the dual conformal field theories. Notice 
that the fields decay with different power law before and after the duality.
The dimensions of the associated operators 
in the dual conformal theories are different.

\begin{figure}[h!]
    \centering
    \includegraphics[width=1\textwidth]{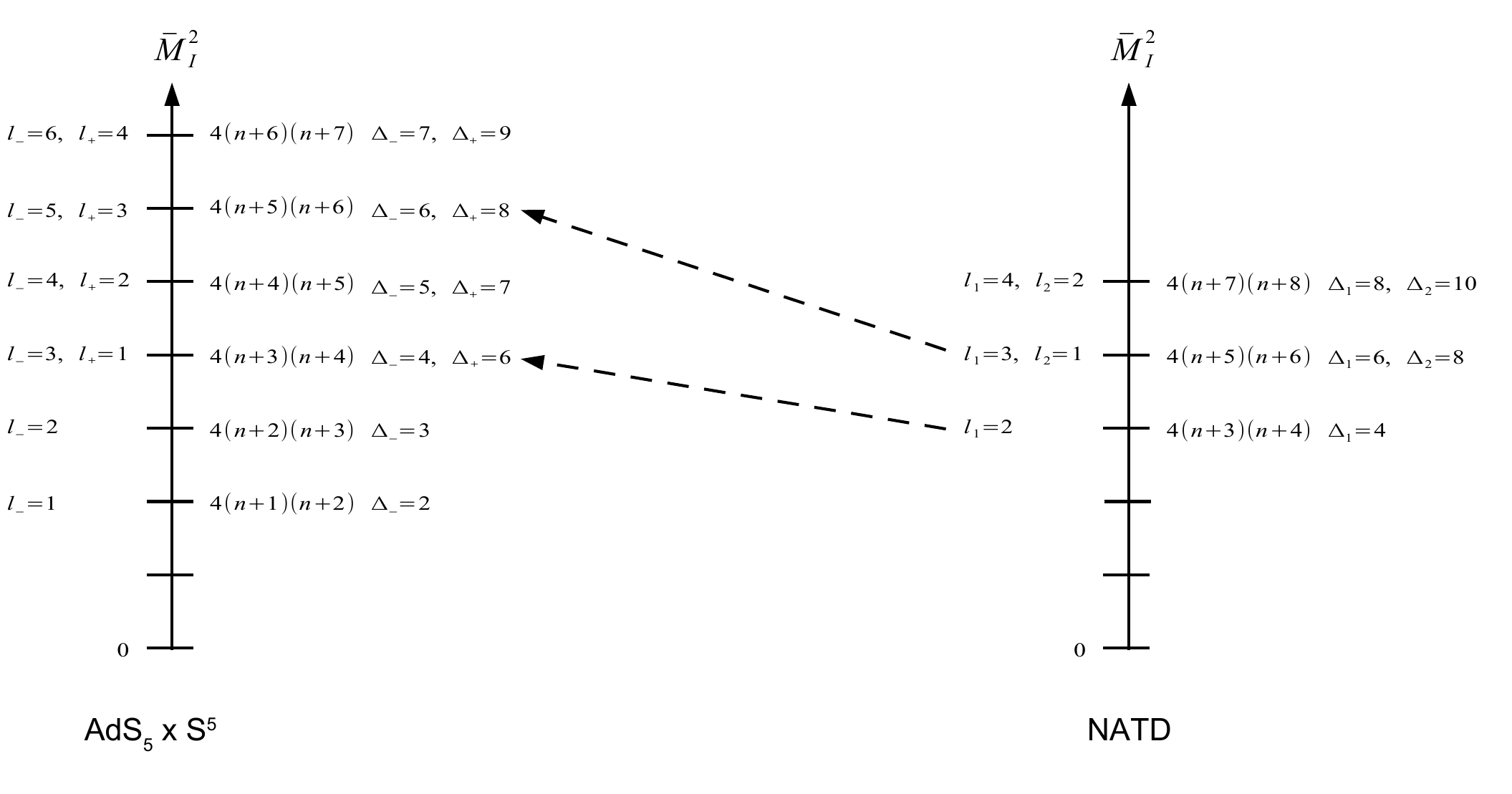}
    \caption{Comparison of the spectra for the type I mesons}
    \label{fig:typeI}
\end{figure}

\begin{figure}[h!]
    \centering
    \includegraphics[width=1\textwidth]{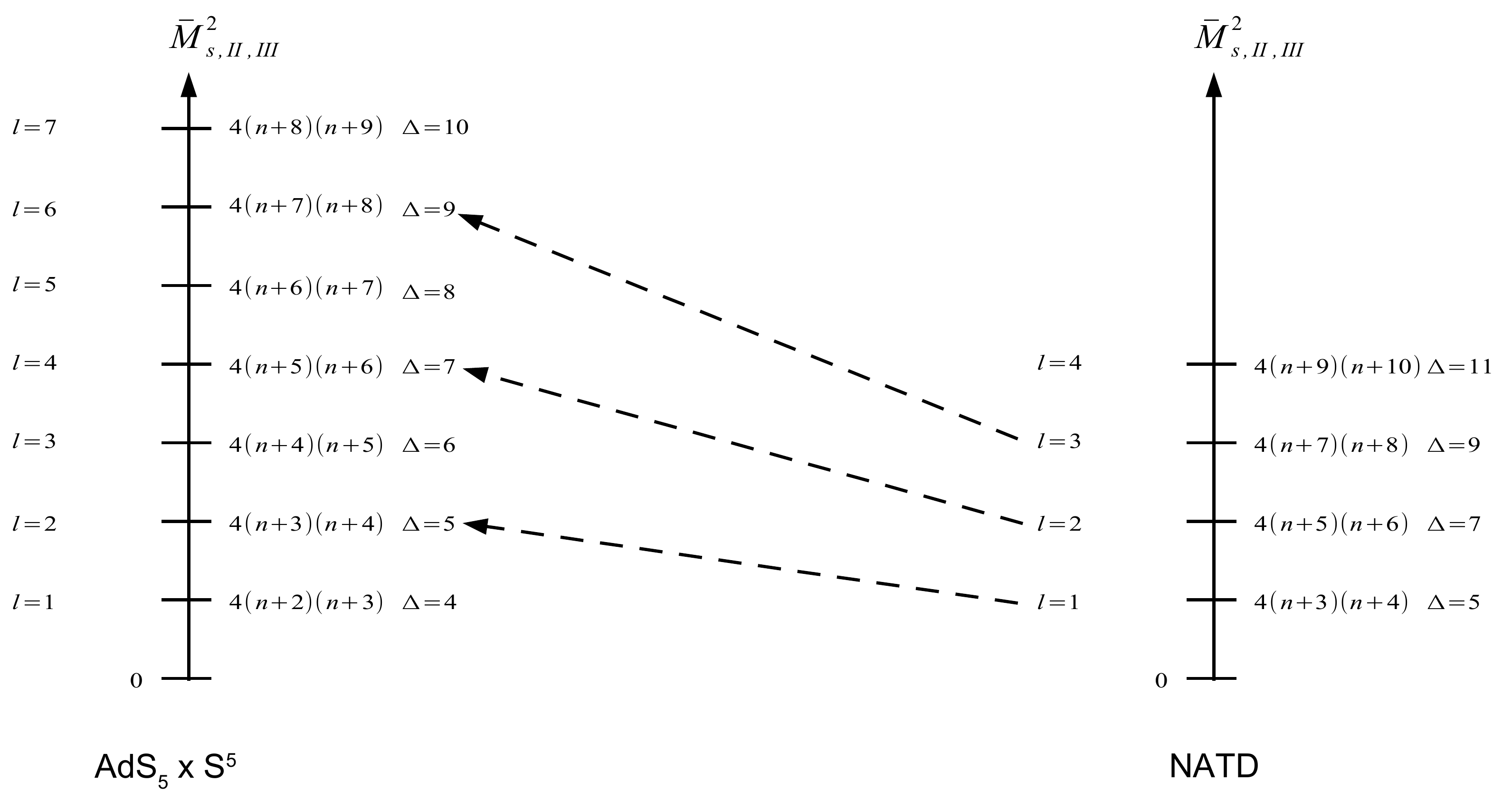}
    \caption{Comparison of the spectra for the scalar and type II, III mesons}
    \label{fig:scalars23}
\end{figure}

{The comparison between the spectra of the original example of \cite{Kruczenski:2003be} and the NATD case can be summarised in the Figures \ref{fig:typeI} \& \ref{fig:scalars23}.}

Regarding the interactions between fluctuations, we found 
that even in the 'inert' sector $(u,\phi)$ thre are differences 
in the coefficients for cubic and quartic interactions calculated
in $AdS_5\times S^5$ and its non-Abelian T-dual.

%%%%%%%%%%%%%%%%%%%%%%%%%%%%%%%%%%%%%%%%%%%

\section{Conclusions and Future Directions}\label{conclusiones}
Since the main results of this work have been summarized in Section \ref{analysis}, here we just limit to observe that {\it at the level of Supergravity}, we find that the fluctuations of a $D6$
brane probing the non-Abelian T-dual background in eqs.(\ref{NSnatd})-(\ref{RRnatd}) are {\it different} from those of a $D7$ brane probing $AdS_5\times S^5$. The same type of effect and differences we have obtained in the Abelian T-dual background.

Inspired by these results, one may wonder
if there exists some stringy mechanism that changes this discrepancy into agreement, with exact matching between the fluctuations and interactions. 

Nevertheless, it may be the case that the sigma model with boundary (corresponding to the supergravity background with the probe brane), are actually different two-dimensional CFTs, not equivalent before and after the duality.

With a holographic perspective, we observe that the ${\cal N}=2$ field theory obtained by adding a quenched flavor hypermultiplet to ${\cal N}=4$ SYM is different from the
field theory dual to the non-Abelian T-dual background with a quenched flavor.

At this level of analysis, we cannot discard that a stringy effect (both $1/N$ and $1/\lambda$ in the field theory) will rectify these differences.

The physical situation we discussed in this paper,
namely a probe $D$-brane embedded in a  background with various RR
and NS $B_2,\Phi$ fields active, is reminiscent (and may have applications)
in other systems. For example, holographic
models of the Chiral Magnetic Effect \cite{Hoyos:2011us}, or models
of String phenomenology with magnetized $D$-branes.

In similar systems, for example in the study of probes in the Polchinski-Strassler system, the effect of the background
$B_2$ field is to introduce a change in the masses of the mesons
$\Delta m\sim B^2$ \cite{Apreda:2006bu}. Similar effects
(a change in the mass spectrum proportional to the deformation parameter) happen when studying
mesons in field theories dual to TsT deformations \cite{Penati:2007vj} of
a given background.

In our case, the non-Abelian T-duality does not
introduce a parameter and should not be thought of as a deformation.
The differences in the observables studied here (masses and interactions) are few and they occur for particular fluctuations. It is also interesting to notice
that even when our system consist of $D4$ and $D6$ branes, the presence of the NS fields
makes it behave more like a $D3$-$D7$ system in the formalism of \cite{Arean:2006pk}.

Some interesting things are left open to be understood with further work. The 
kappa-symmetry preservation of the embeddings used (that should show that 
${\cal N}=2$ SUSY is preserved by the probe).
The reasons for the non-trivial matchings under the $l_{KMMW}=2l_{here}$ change. The fact that the probe $D6$ brane localises at the positions $r= q \pi$ in the non-Abelian T-dual background, suggest that some observables should depend directly upon the parameter $q$ and hence be quantised. Here, we found this behaviour for  the interactions terms mediating meson decay, but it  would be interesting to find other such observables.

More generally, the ''duality-character'' of non-Abelian T-duality remains elusive. It would be good to apply the ideas in this paper and analyze other brane probes and their fluctuations. Not only restricted to mesons, but actually extending this  analysis to other observables in the field theory that rely on probe branes. For example, Baryon vertex, Giant Gravitons,  couplings defined by probe branes, etc. In a variety of papers dealing with the interplay between non-Abelian duality and holography \cite{Macpherson:2014eza}-\cite{varios4}, it 
was seen that 
for each observable in the seed field theory, 
there was some probe brane in the non-Abelian 
T dual background that could play the role of such observable in the corresponding field theory. We suggest that aside from identifying the putative probe, its fluctuations should be studied following the formalism and ideas in this paper.

Finally, to probe the stringy character of non-Abelian T-duality, the best would be to study the full string theory on the non-Abelian T-dualized backgrounds. We have in mind, for example, taking the pp-wave limit and using the formalism initiated in \cite{Berenstein:2002jq}. Other approaches where stringy characteristic of the system play a role in the dynamics, like those represented by the papers \cite{strings} also point to the same conclusions we arrived to in this work.

\section*{Acknowledgments:} Discussions with different colleagues helped us to better understand the way of presenting the material in this paper. We would like to thank specially Carlos Hoyos for reading very 
carefully the preliminary draft and sending us various 
useful comments. We also  gratefully thank Adi Armoni, 
S. Prem Kumar, Yolanda Lozano, David Mateos, Horatiu Nastase,  
Maurizio Piai, Alfonso Ramallo, Konstantinos Sfetsos and Salomon Zacar\'ias.   
G.I. is supported by the FAPESP grants 2016/08972-0 and 2014/18634-9. He also 
acknowledges support by UE-14-GT5LD2013- 618459. The work of D.Z. is partially funded by FCT through the project CERN/FIS-NUC/0045/2015. C.N. is Wolfson Fellow of the Royal Society.

\appendix

%%%%%%%%%%%%%%%%%%%%%%%%%%%%%%%%%%%%%%%%%%%

\section{A nice way to calculate the non-Abelian T-dual of $AdS_5\times S^5$}\label{appendix1}
In this appendix we present a set of general formulas, developed in \cite{Itsios:2012dc}, that are useful to obtain the non-abelian T-dual for a background containing a three sphere
and dualizing on one of the two $SU(2)$'s inside $SO(4)$. Indeed, 
according to \cite{Itsios:2012dc}, if one starts with a type IIB supergravity solution with metric that has the form,
\begin{equation}
ds^2 = ds^2(M_7) + e^{2 A} ds^2(S^3)\ ,\nonumber
\end{equation}
and RR fields that preserve the $SU(2)$ symmetry,
\begin{equation}
\begin{aligned}
& F_5 = G_2 \wedge {\rm Vol}(S^3) - e^{-3 A} \star_7 G_2\ ,
\\[10pt]
& F_3 = G_3 - m {\rm Vol}(S^3)\ ,
\label{IIBfluxansatz}
\\[10pt]
& F_1 = G_1\ ,
\end{aligned}
\end{equation}
then it is very easy to read the fields of the dual type IIA supergravity solution which have the following structure
\footnote{
We have restored the dependence on $\alpha'$ for the dual solution respect to the formulas given in \cite{Itsios:2012dc}.
},
\begin{equation}
 \begin{aligned}
  &  d\hat s^2 =ds^2(M_7) + \alpha'^2 e^{-2 A} dr^2 + {\alpha'^2 r^2 e^{2 A}\ov \alpha'^2 r^2 + e^{4 A}} ds^2(S^2)\ ,
\\[10pt]
& \widehat B =  B+  \widetilde B \ , \qq \widetilde B =  \frac{\alpha'^3 r^3}{\alpha'^2 r^2 + e^{4 A}} {\rm Vol}(S^2)\ , \qquad
\label{NSsec}
e^{-2 \widehat \Phi} = e^{-2 \Phi} e^{2 A} \ \frac{\alpha'^2 r^2 + e^{4 A}}{\alpha'^3}
 \end{aligned}
\end{equation}
and
\begin{equation}
 \begin{aligned}
   \widehat F_0 & =  m \ \alpha'^{-3/2}\ , \qquad \qquad \widehat F_2 = \frac{m \ \alpha'^{3/2} r^3}{\alpha'^2 r^2 + e^{4 A}} \textrm{Vol}(S^2) + \alpha'^{1/2} r dr \wedge G_1 - \alpha'^{-3/2} G_2\ ,
\label{fliIIAmas}
\\[10pt]
  \widehat F_4 & = \frac{\alpha'^{3/2} r^2 e^{4 A} }{\alpha'^2 r^2+e^{4A}} G_1\wedge dr \wedge \textrm{Vol}(S^2)
- \frac{\alpha'^{3/2} r^3}{\alpha'^2 r^2 + e^{4 A}} G_2 \wedge \textrm{Vol}(S^2)
\\[10pt]
& +\alpha'^{1/2} r  dr\wedge  G_3  + \alpha'^{-3/2} e^{3 A} \star_7 G_3 \ .
 \end{aligned}
\end{equation}
In the expressions above $B$ and $\Phi$ are the NS two-form and the dilaton of the 'seed' solution. Moreover, it is important to notice that in order to apply the formulas written above one has to take into account that the three sphere of the original solution must be normalized such that $R^{(S^3)}_{\m \n} = \frac{1}{2} g^{(S^3)}_{\m\n}$. So in our case we have to adjust the line element of the original background so that a three sphere satisfies this condition. This can be achieved by a global rescaling.
For the original seed-background in  eq.\eqref{ads5xs5}, we have
\begin{equation}
 \begin{aligned}
  & e^{2A} = \frac{L^2 \ \rho^2}{4 \ R^2} \ , \qquad B = 0 \ , \qquad \Phi = 0 \ , 
   \\[10pt]
  & G_1 = G_3 = 0 \ , \qquad G_2 = - d \Bigg[   \frac{L^4 \ u^2 \ \big(   R^2 + \rho^2  \big)}{8 \ R^4} \ d\phi  \Bigg] \ .
 \end{aligned}
\end{equation}
From this, using eqs.(\ref{NSsec}), (\ref{fliIIAmas}) we get the expressions for the NS and RR sector of the dual solution written in eqs.(\ref{NSnatd})-(\ref{RRnatd}).

%%%%%%%%%%%%%%%%%%%%%%%%%%%%%%%%%%%%%%%%%%%

\section{Details of the Fluctuation Analysis in KMMW}\label{appendixb}
In this appendix, we present all intermediate steps leading to the derivation of the formulas in Section \ref{sectionKMMW}. Before that, it is useful to make a detour to summarize aspects of the hypergeometric function.
\subsubsection*{The Gaussian hypergeometric function}
The Gaussian hypergeometric function, ${}_2 F_1(a,b;c;x)$, is the solution of the following second order ordinary differential equation,
\begin{equation}
 x (1 - x) y'' + \Big[   c - (1 + a + b) x  \Big] y' - ab y = 0 \ .
\end{equation}
An alternative definition of the hypergeometric function is through power series, namely,
\begin{equation}
 {}_2 F_1(a,b;c;x) = \sum\limits_{n = 0}^{\infty} \frac{(a)_n \ (b)_n}{n! \ (c)_n} x^n = 1 + \frac{ab}{c} x + \frac{a(a+1) b(b+1)}{2! \ c(c+1)} x^2 + \ldots
\end{equation}
where $(a)_n$ is the Pochhammer symbol that is defined as,
\begin{equation}
 (a)_n = \left\{ \begin{array}{ll}
                        a \cdot (a+1) \cdot (a+2) \cdot \ldots \cdot (a + n - 1), & \quad \textrm{for}  \quad n = 1, 2, \ldots
                        \\
                         1 & \quad \textrm{for} \quad n = 0
                        \end{array} \right. \ ,
\end{equation}
or
\begin{equation}
 (a)_n = \frac{\G (a + n)}{\G (a)} \ .
\end{equation}

From the definition in terms of power series we see that the hypergeometric function is symmetric in the first two arguments, i.e.
\begin{equation}
 {}_2 F_1(a,b;c;x) = {}_2 F_1(b,a;c;x) \ .
\end{equation}
Another useful identity is the so called Euler's transformation,
\begin{equation}
 {}_2 F_1(a,b;c;x) = (1 - x)^{c - a - b} {}_2 F_1(c - a, c - b; c; x) \ .
\end{equation}

Moreover, if one of the first two arguments is a negative integer we see that the infinite power series becomes finite, i.e. a polynomial, due to the Pochhammer symbol. To see this let us take $a = - |a| < 0$ and $b>0$. Then,
\begin{equation}
 (a)_n = 0 \quad \textrm{for} \quad n = |a| + 1 = - a + 1
\end{equation}
and the hypergeometric series become a polynomial of degree $|a| = - a$. If it happens that $a<0$ and $b<0$ with $|a|>|b|$ then $(b)_n$ goes to zero "faster" than $(a)_n$ and thus the degree of the polynomial now is $|b| = -b$.
In other words, the hypergeometric 
${}_2 F_1(a,  b; c; -z^2)$ with $b>a$ has a series expansion with a finite number of terms if $b=-n$ (for $n$ a positive integer). In this case the large $z$ asymptotics of the function is
${}_2 F_1(a,  b; c; -z^2)\sim z^{2n}$.

We  impose these conditions and use the identities above when solving the equations of motion for fluctuations
of the probe $D7$ branes in $AdS_5\times S^5$ and for $D6$ branes in the non-Abelian T-dual
of $AdS_5\times S^5$ described in  Section \ref{sectionsupergravity}.

Let us now discuss the details of the dynamics for the fluctuations (at second order) of probe $D7$ branes in $AdS_5\times S^5$.

\subsubsection*{Fluctuations of the scalar fields}

We already mentioned that the directions transverse to the $D7$ brane are $u$ and $\phi$. Hence for the scalar fluctuations around the constant embedding of the equation \eqref{originalembedding} we adopt the following perturbation scheme,
\begin{equation}
 u = u_0 + \delta u \ , \qquad \phi = 0 + \delta \phi \ .
\end{equation}
Then the quadratic Lagrangian of the fluctuations takes the form,
\begin{equation}
 \mathcal{L} \simeq - T_{D7} \sqrt{- \det g} \frac{g^{ab}}{\r^2 + u_0^2} \Big(    \partial_a \delta u \ \partial_b \delta u + u_0^2 \ \partial_a \delta \phi \ \partial_b \delta \phi  \Big) \ ,
\label{lala}\end{equation}
where $g$ here is the induced metric which corresponds to the constant embedding of the equation \eqref{originalembedding} and the indices $a, b$ run over the worldvolume directions of the $D7$ brane. From this expression one can easily see that the linearized equations of motion for $\delta u$ and $\delta \phi$ have the same form, which is,
\begin{equation}
 \frac{L^4}{\big(   \r^2 + u_0^2  \big)^2} \Box f + \frac{1}{\r^3} \partial_{\r} \big( \r^3 \partial_{\r} f \big) + \frac{1}{\r^2} \nabla^2_{S^3} f = 0 \ ,
\end{equation}
where $\Box = \eta^{\mu \nu} \partial_{\mu} \partial_{\nu} \ , \;  \mu, \nu = 0, \ldots, 3$ and $\nabla^2_{S^3}$ is the Laplacian of the unit three-sphere spanned by the directions $\a, \b, \g$.

The above equation of motion can be solved using separation of variables,
\begin{equation}
 f = e^{i k \cdot x} \varphi (\r) \mathcal{Y}^l (S^3) \ ,
\end{equation}
where $\mathcal{Y}^l (S^3)$ with $l \ge 0$ are the scalar spherical harmonics on $S^3$ which satisfy the following eigenvalue problem,
\begin{equation}
 \nabla^2_{S^3} \mathcal{Y}^l = - l (l + 2) \mathcal{Y}^l \ .
\end{equation}
If we plug this ansatz into the equation of motion then we end up with a second order differential equation for the radial function $\varphi$ which has the form,
\begin{equation}
 \partial_z^2 \varphi + \frac{3}{z} \partial_z \varphi + \Bigg[  \frac{\bar{M}^2}{\big(   z^2 + 1  \big)^2} - \frac{ l (l+2) }{z^2}  \Bigg] \varphi = 0,\label{lns}
\end{equation}
and we have defined the dimensionless parameters
\begin{equation}
\label{dimensionlessparams}
 z = \frac{\r}{u_0} \ , \qquad \bar{M}^2 = - \frac{k^2 L^4}{u_0^2} = \frac{M^2 L^4}{u_0^2} \ .
\end{equation}

This equation can be solved analytically in terms of a hypergeometric function. A valid solution must be real-valued, regular and small enough in amplitude in order to justify the use of quadratic Lagrangian. Reality and regularity of the solution at the origin suggest that we must chose the following solution,
\begin{equation}
 \varphi(z) = z^l \big(   z^2 + 1  \big)^{- \a} {}_2 F_1 \big(  -\a, -\a + l + 1; l + 2; - z^2   \big) \ ,
\end{equation}
where
\begin{equation}
\label{parameteralpha}
 2 \a = -1 + \sqrt{1 + \bar{M}^2} \ge 0 \ .
\end{equation}
We have already seen that a hypergeometric function in general can be thought as an infinite series and as a consequence our solution might be divergent as $z \rightarrow \infty$. In such a case the solution becomes problematic since it corresponds to a non-normalizable mode. To avoid this happen we will cause the infinite series to terminate by imposing the following condition,
\begin{equation}
 - \a + l + 1 = - n \ , \qquad n = 0, 1, 2, \ldots \ .
\end{equation}
Then after imposing this condition one easily verifies that at large values of $z$ the solution vanishes as $z^{-l - 2}$. This behavior suggests that the mode with $l = 0$ should not be considered in the spectrum because it is not square integrable. Moreover, this condition together with the equation \eqref{parameteralpha} tells us that the four-dimensional mass spectrum of the scalar mesons is,
\begin{equation}
 M_s(n,l) = \frac{2 u_0}{L^2} \sqrt{\big(   n + l + 1  \big) \big(   n + l + 2  \big)} \ , \qquad n = 0 , 1 , \ldots \ , \qquad l = 1 , 2 , \ldots \ .
\end{equation}
Finally, the conformal dimension of the UV operator is $\Delta = l+3$.

\subsubsection*{Fluctuations of the gauge fields}

We already mentioned that the scalar fluctuations do not couple with the fluctuations of the worldvolume gauge field. This means that in order to study the spectrum of vector mesons we can simply turn off the fluctuations for $u, \phi$ and keep only those for $A_a$. For this reason we adopt the following perturbation scheme,
\begin{equation}
 A_a = 0 + \delta A_a \Rightarrow F_{ab} = 0 + \delta F_{ab} \ .
\end{equation}
Then the linearized equation of motion for $\delta A_a$ takes the simple form,
\begin{equation}
 \partial_{a} \Big(   \sqrt{- \det g} \ \delta F^{ab}  \Big) - \frac{4 \r \big(   \r^2 + u_0^2  \big)}{L^4} \delta^{b}_{i} \varepsilon^{ijk} \partial_{j} \delta A_{k} = 0 \ ,
\end{equation}
where the indices $a,b$ run over the worldvolume coordinates of the $D7$ brane and $i,j,k$ over the $S^3$ directions. Also, $\varepsilon^{ijk}$ is a tensor density (i.e. it takes the values $\pm 1$). According to \cite{Kruczenski:2003be} there are three type of mesons that one can consider here. Each of them is described below.

\vskip 10pt

\noindent \underline{Type I mesons:}

%\begin{equation}
% \delta A_{\mu} = 0 \ , \qquad \delta A_{\r} = 0 \ , \qquad \delta A_{i} = e^{i k \cdot x} \varphi^{\pm}_{I} (\r)\mathcal{Y}^{l, \pm}_{i} \big( S^3 \big) \ ,
%\end{equation}
%
%where $\mathcal{Y}^{l, \pm}_{i} \big(   S^3  \big)$ with $l \ge 1$ are the vector spherical harmonics on $S^3$
The first type of mesons arises when the only non-vanishing fluctuations of the worldvolume gauge field are those along the three-sphere. This kind of fluctuations is described in eq. \eqref{mesonsI} \footnote{
The vector harmonics $\mathcal{Y}^{l, \pm}_{i}$ transform in the $\Big(    \frac{l \mp 1}{2} , \frac{l \pm 1}{2}   \Big)$ of $SO(4)$ and satisfy
\begin{equation}
 \begin{aligned}
   & \nabla_i \nabla^i \mathcal{Y}^{l, \pm}_{j} - R^{k}_{j} \mathcal{Y}^{l, \pm}_{j} = - (l + 1 )^2 \mathcal{Y}^{l, \pm}_{j} \ ,
 \\[5pt]
   & \varepsilon_{ijk} \nabla_j \mathcal{Y}^{l, \pm}_{k} = \pm (l + 1) \mathcal{Y}^{l, \pm}_{i} \ ,
 \\[5pt]
   & \nabla^{i} \mathcal{Y}^{l, \pm}_{i} = 0 \ ,
 \end{aligned}
\end{equation}
where $R^{i}_{j} = 2 \delta^{i}_{j}$ is the Ricci tensor of a unit $S^3$ and $\nabla_{i}$ is the covariant derivative on $S^3$.
}.
 Using this ansatz one finds that $\varphi^{\pm}_{I}$ is given by the following second order differential equation,
\begin{equation}
 \frac{1}{z} \partial_{z} \Big(   z \big(   1 + z^2   \big)^2 \partial_{z} \varphi^{\pm}_{I} (z)   \Big) + \bar{M}^2 \varphi^{\pm}_{I} (z) - \big(   l + 1   \big)^2 \frac{\big(   1 + z^2   \big)^2}{z^2} \varphi^{\pm}_{I} (z) \mp 4 \big(   l + 1   \big) \big(   1 + z^2   \big) \varphi^{\pm}_{I} (z) = 0 \ ,\label{baba}
\end{equation}
where again we expressed everything in terms of the dimensionless quantities written in the equation \eqref{dimensionlessparams}. The solutions that are regular at $z = 0$ are,
\begin{equation}
\label{eq:B.24}
 \begin{aligned}
  & \varphi^{+}_{I} (z) = z^{l + 1} \big(   1 + z^2  \big)^{-1-\a} {}_2 F_1 \big(   l + 2 - \a, -1 -\a; l+2; -z^2  \big) \ ,
 \\[10pt]
 & \varphi^{-}_{I} (z) = z^{l + 1} \big(   1 + z^2  \big)^{-1-\a} {}_2 F_1 \big(   l - \a, 1 -\a; l+2; -z^2  \big) \ ,
 \end{aligned}
\end{equation}
where again the parameter $\a$ is defined in the equation \eqref{parameteralpha}. From the solutions we can extract the mass spectra for the type I mesons which are,
\begin{equation}
 \begin{aligned}
  & M^2_{I,+} = \frac{4u_0^2}{L^4} \big(  n + l + 2   \big) \big(   n + l + 3  \big) \ , \qquad n = 0 , 1 , \ldots \ , \qquad l = 1 , 2 , \ldots \ ,
  \\[10pt]
  &   M^2_{I,-} = \frac{4u_0^2}{L^4} \big(  n + l    \big) \big(   n + l + 2  \big) \ , \;\;\;\;\;\; \qquad n = 0 , 1 , \ldots \ , \qquad l = 1 , 2 , \ldots \ .\label{bababa}
 \end{aligned}
\end{equation}
Also, the asymptotic behavior of the solutions at large values of $z$ is,
\begin{equation}
  \varphi^{+}_{I} (z) \sim z^{-l-5} \ , \qquad  \varphi^{-}_{I} (z) \sim z^{-l-1} 
\end{equation}
and the conformal dimensions of the corresponding operators at the UV are $\Delta_{+} = l+5$ and $\Delta_{-} = l+1$.

\vskip 10pt

\noindent \underline{Type II mesons:}

%\begin{equation}
% \delta A_{\mu} = \zeta_{\mu} e^{i k \cdot x} \varphi_{II} (\r) \mathcal{Y}^{l} \big( S^3 \big) \ , \qquad k \cdot \zeta = 0 \ , \qquad \delta A_{\r} = 0 \ , \qquad \delta A_{i} = 0 \ .
%\end{equation}
%
Another consistent type of fluctuations is the one where the only non-vanishing fluctuations of the worldvolume gauge field are those along the Minkowski directions. This gives rise to the mesons of type II and the corresponding ansatz is given in eq. \eqref{mesonsII}. In this case the equation of motion for the fluctuations of the worldvolume gauge field reduces to,
\begin{equation}
\label{eq:B.28}
  \frac{1}{z^3} \partial_{z} \big(   z^3 \partial_z \varphi_{II} (z)  \big) + \Bigg[ \frac{\bar{M}^2}{\big(   1 + z^2  \big)^2} - \frac{l (l+2)}{z^2} \Bigg] \varphi_{II}(z) = 0 \ .
\end{equation}
The solution of this differential equation which is regular at $z = 0$ is,
\begin{equation}
 \varphi_{II}(z) = z^l \big(  1 + z^2   \big)^{-\a} {}_2 F_1 \big(   l + 1 - \a , - \a ; l + 2 ; - z^2  \big) \ ,
\end{equation}
with $\a$ given in the equation \eqref{parameteralpha}. This solution results to the following mass spectrum,
\begin{equation}
   M^2_{II} = \frac{4u_0^2}{L^4} \big(  n + l + 1   \big) \big(   n + l + 2  \big) \ , \qquad n = 0 , 1 , \ldots \ , \qquad l = 1 , 2 , \ldots \ .
\end{equation}
Also, at $z \rightarrow \infty$ the solution behaves as,
\begin{equation}
 \varphi_{II} (z) \sim z^{-l-2}
\end{equation}
and conformal dimension of the associated UV is $\Delta = l + 3$. Notice that due to this behavior, the $l = 0$ mode is not square integrable and thus we exclude it from the spectrum.

\vskip 10pt

\noindent \underline{Type III mesons:}

%\begin{equation}
% \delta A_{\mu} =0 \ , \qquad \delta A_{\r} = e^{i k \cdot x} \varphi_{III} (\r) \mathcal{Y}^{l} \big( S^3 \big) \ , \qquad \delta A_{i} = e^{i k \cdot x} \tilde{\varphi}_{III} (\r) \partial_{i} \mathcal{Y}^{l} \big( S^3 \big) \ .
%\end{equation}
%
The last category of mesons, which is called type III, corresponds to the perturbation scheme of the equation \eqref{mesonsIII}. In the case of the type III mesons the equation of motion for $b = \mu$ field simplifies to,
\begin{equation}
\label{eqIII1}
 l (l + 2) \tilde{\varphi}_{III} (\r) = \frac{1}{\r} \partial_{\r} \Big(   \r^3 \varphi_{III} (\r)   \Big) \ .
\end{equation}
The solution for $l = 0$ is $\varphi_{III} \sim 1/\r^3$ which is not regular at $\r = 0$ and thus we are not going to consider it.

The equations that arise when we set $b = \r$ and $b = j$  are,
\begin{equation}
\label{eqIII2}
 \begin{aligned}
  & b = \r: \qquad l (l+2) \partial_{\r} \tilde{\varphi}_{III} (\r) = l (l+2) \varphi_{III} (\r) - \frac{M^2 L^4 \r^2}{\big(   \r^2 + u_0^2  \big)^2} \varphi_{III} (\r) \ ,
\\[10pt]
  & b = j: \qquad \partial_{\r} \Bigg[   \frac{\r \big(   \r^2 + u_0^2  \big)^2 }{L^4} \Big(   \partial_{\r} \tilde{\varphi}_{III} - \varphi_{III}   \Big)  \Bigg] + \r M^2 \tilde{\varphi}_{III} = 0 \ .
 \end{aligned}
\end{equation}
We see that we obtain three differential equations for the radial functions $\varphi_{III} (\r)$ and $\tilde{\varphi}_{III} (\r)$. One can easily confirm that these equations are not independent. Indeed, it is easy to recover the equation \eqref{eqIII1} from the formulas in the equation \eqref{eqIII2}. If we now eliminate $\tilde{\varphi}_{III}$ using the first formula in the equation \eqref{eqIII2} and the equation \eqref{eqIII1} then we end up with a second order differential equation for $\varphi_{III}$ which has the following form
\footnote{
The term that is proportional to $\bar{M}^2$ is different from the corresponding term of the equation (3.42) in \cite{Kruczenski:2003be} by an overall sign.
},
\begin{equation}
 \partial_{z} \Bigg(   \frac{1}{z} \partial_{z} \Big(  z^3 \varphi_{III} (z)  \Big)  \Bigg) + \Bigg[ \frac{\bar{M}^2 z^2}{\big(   1 + z^2  \big)^2} - l (l+2) \Bigg] \varphi_{III}(z) = 0 \ ,
\end{equation}
where we expressed everything in terms of the dimensionless parameters \eqref{dimensionlessparams}. The solution of this differential equation that is regular at $z = 0$ is,
\begin{equation}
 \varphi_{III} (z) = z^{l - 1} \big(   1 + z^2  \big)^{-\a} {}_2 F_{1} \big(   - \a + l + 1, - \a ; l+2 ; - z^2  \big) \ .
\end{equation}
From this we obtain the mass spectrum,
\begin{equation}
 M^{2}_{III} = \frac{4u_0^2}{L^4} \big(   n + l + 1  \big) \big(  n + l + 2   \big) \ , \qquad n = 0 , 1 , \ldots \ , \qquad l = 1 , 2 , \ldots \ .
\end{equation}

Finally the behavior of the solution at the boundary is,
\begin{equation}
 \varphi_{III} (z) \sim z^{-l-3}
 \label{eq:b.35}
\end{equation}
and the conformal dimension of the associated UV operator is $\D = l+3$.

All these fluctuations above move in a curved  space as indicated for example in eq.(\ref{lala}). We impose, aside from regularity, a square integrability  condition on these fluctuations. In other words we need that $\int_0^\infty dz \; \sqrt{\det g} \; |\frak{f}(z)|^2$ to be finite. For the case of $AdS_5\times S^5$ we have $\sqrt{\det g}=\rho^3$.

%%%%%%%%%%%%%%%%%%%%%%%%%%%%%%%%%%%%%%%%%%%

\section{The probe embeddings}\label{Embedding}

In this appendix we present in a schematic way the finding of the embeddings of the $D6$ branes that we used in the main text.

\subsection{$D6$ constant embedding in the NATD background}

It turns out that the finding of the embedding of a $D$-brane is a complicated task since in general one has to solve a non-linear system of partial differential equations in order to determine the embedding functions of the brane. Thus here we are going to give the basic arguments which advocate that a constant embedding of the form \eqref{straightembedding} is a good choice for our study.

As a first step we consider that the embedding functions of the $D6$ brane depend on all the worldvolume coordinates and thus we take,
\begin{equation}
 r = r(x^{\m}, \r, \chi, \xi) \ , \qquad u = u(x^{\m}, \r, \chi, \xi) \ , \qquad \phi = \phi(x^{\m}, \r, \chi, \xi) \ .
\end{equation}
Then the dynamics of our $D6$ brane is described by the DBIWZ action given in the equation \eqref{DBIWZ}. In principle the DBIWZ action is a functional of the embedding functions. Thus, knowing the precise form of the DBIWZ action one can compute the equations of motion for the embedding functions $r, u$ and $\phi$.

In order to construct the DBI part of the action we need to find the induced metric on the $D6$ brane which in our case takes the following form,
\begin{equation}
 \begin{aligned}
ds^2_{D6} &= \frac{R^2}{L^2} \eta_{\m\n} dx^\m dx^\n + \frac{L^2}{R^2} d \r^2 + \frac{4 \alpha'^2 L^2 R^2 \r^2 r^2}{16 \alpha'^2 R^4 r^2 + L^4 \r^4} \big(   d\chi^2 + \sin^2\chi \ d\xi^2  \big)
\\[10pt]
& + \Bigg[ \frac{L^2}{R^2} \Big(   \partial_M u \ \partial_N u + u^2 \partial_M \phi \ \partial_N \phi   \Big) + \frac{4 \alpha'^2 R^2}{L^2 \r^2} \partial_M r \ \partial_N r   \Bigg] dx^M dx^N \ ,
 \end{aligned}
\end{equation}
where the indices $\m, \n$ run only over the Minkowski directions and $M,N$ on the worldvolume coordinates of the $D6$ brane.
%
%\begin{equation}
 %\begin{aligned}
%ds^2_{D6} &= \Bigg[   \frac{R^2}{L^2} \eta_{\m\n} + \frac{L^2}{R^2} \Big(   \partial_\m u \ \partial_\n u + u^2 \partial_\m \phi \ \partial_\n \phi   \Big) + \frac{4 \alpha'^2 R^2}{L^2 \r^2} \partial_\m r \ \partial_\n r   \Bigg] dx^\m dx^\n
%\\[10pt]
%& + \Bigg[   \frac{L^2}{R^2} \Big(   1 + (\partial_\r u)^2 + u^2 (\partial_\r \phi)^2  \Big) + \frac{4 \alpha'^2 R^2}{L^2 \r^2} (\partial_r r)^2   \Bigg] d\r^2
%\\[10pt]
%& + \Bigg[   \frac{L^2}{R^2} \Big(    (\partial_\chi u)^2 + u^2 (\partial_\chi \phi)^2  \Big) + \frac{4 \alpha'^2 R^2}{L^2 \r^2} (\partial_\chi r)^2   + \frac{4 \alpha'^2 L^2 R^2 \r^2 r^2}{16 \alpha'^2 R^4 r^2 + L^4 \r^4} \Bigg] d\chi^2
%\\[10pt]
%& + \Bigg[   \frac{L^2}{R^2} \Big(    (\partial_\xi u)^2 + u^2 (\partial_\xi \phi)^2  \Big) + \frac{4 \alpha'^2 R^2}{L^2 \r^2} (\partial_\xi r)^2   + \frac{4 \alpha'^2 L^2 R^2 \r^2 r^2}{16 \alpha'^2 R^4 r^2 + L^4 \r^4} \sin^2 \chi \Bigg] d\xi^2 \ .
% \end{aligned}
%\end{equation}
%
Also we consider the dilaton of the background given in the equation \eqref{NSnatd} and the NS two-form that appears in the equation \eqref{B2lg}. Notice that this expression for the NS two-form is after a large gauge transformation. Here we use this form as it is more generic and we can always go to the case before the large gauge transformation simply by setting $q = 0$.

For the WZ part of the action one needs to evaluate the pullback of the combination
\footnote{
Notice that $\frac{1}{2} C_3 \wedge B^{l.g.}_2 \wedge B^{l.g.}_2 - \frac{1}{6} C_1 \wedge B^{l.g.}_2 \wedge B^{l.g.}_2 \wedge B^{l.g.}_2 = 0 $ since $B^{l.g.}_2 \wedge B^{l.g.}_2 = 0$.
} 
$C_7 - B_2 \wedge C^{l.g.}_5$ which in this case takes the form,
\begin{equation}
 P \Big[ C_7 - B^{l.g.}_2 \wedge C_5 \Big] = \frac{2 \alpha'^{3/2}}{3 L^4} R^2 r^2 \big(   3 q \pi - 2 r  \big) \big(   u \ \partial_{\r} u + \r   \big) \ dx_{1,3} \wedge d\r \wedge \textrm{Vol}(S^2) \ .  
\end{equation}

For the sake of simplicity, instead of presenting the general expressions for the $D6$ brane Lagrangian and the equations of motion that come out of it, we will give the expressions of these quantities only after imposing the following form of the embedding functions,
\begin{equation}
 r = r_0 = \textrm{const} \ ,\qquad u = u_0 = \textrm{const} \ ,\qquad \phi = \phi_0 = \textrm{const} \ .
\end{equation}
In that case the DBI and the WZ parts of the action are,
\begin{equation}
 \begin{aligned}
  & \mathcal{L}_{DBI}^{\textrm{const. emb}} = - \frac{\r \sin\chi}{8 \sqrt{\alpha'} L^2} \sqrt{L^4 q^2 \pi^2 \r^4 + 16 \alpha'^2 r_0^2 \big(   r_0 - q \pi  \big)^2 \big(   u_0^2 + \r^2   \big)^2} \ ,
 \\[10pt]
 &  \mathcal{L}_{WZ}^{\textrm{const. emb}} = \frac{2 \alpha'^{3/2}}{3 L^4} r_0^2 \big(   3 q \pi - 2 r_0  \big) \big(   u_0^2 + \r^2   \big) \r \ \sin\chi  \ ,
 \end{aligned}
\end{equation}
while each of the equations of motion takes the following form,

\vskip 10pt

\noindent \emph{Equation of motion for $r$:}

\vskip 10pt

\begin{equation}
 r_0 \big(   q \pi - r_0  \big) \big(   u_0^2 + \r^2  \big) \frac{  L^2 \big(   q \pi - 2 r_0  \big) \big(   u_0^2 + \r^2  \big) - 2 \sqrt{L^4 q^2 \pi^2 \r^4 + 16 \alpha'^2 r_0^2 \big(   r_0 - q \pi  \big)^2 \big(   u_0^2 + \r^2   \big)^2}  }{\sqrt{L^4 q^2 \pi^2 \r^4 + 16 \alpha'^2 r_0^2 \big(   r_0 - q \pi  \big)^2 \big(   u_0^2 + \r^2   \big)^2}} = 0 \ .
\end{equation}

\vskip 10pt

\noindent \emph{Equation of motion for $u$:}

\vskip 10pt

\begin{equation}
 \frac{u_0 r_0^2 \big(  r_0 - q \pi  \big)^2 \big(   u_0^2 + \r^2   \big)}{\sqrt{L^4 q^2 \pi^2 \r^4 + 16 \alpha'^2 r_0^2 \big(   r_0 - q \pi  \big)^2 \big(   u_0^2 + \r^2   \big)^2}} = 0 \ .
\end{equation}

\vskip 10pt

\noindent \emph{Equation of motion for $\phi$:}

\vskip 10pt

\begin{equation}
 \textrm{Vanishes identically.}
\end{equation}

The relevant expressions for the same quantities that correspond to the NS two-form given in the equation \eqref{NSnatd} can be obtained by setting $q = 0$. In that case we immediately see that the only solution to the equations of motion is if we take $r_0 = 0$ while keeping $u_0$ and $\phi_0$ arbitrary constants. Such a solution is not interesting since for $r_0 = 0$ both the $DBI$ and the $WZ$ actions are zero. The way out to this problem is to consider a large gauge transformation of the NS two-form. Then we see that we can solve the equations of motion for $r_0 = q \pi$ and $u_0, \phi_0$ being arbitrary constants. Obviously this type of embedding does not make the $DBI$ and the $WZ$ actions to vanish.

%%%%%%%%%%%%%%%%%%%%%%%%%%%%%%%%%%%%%%%%%%%

\subsection{$D6$ constant embedding in the Hopf-T-dual background}

\label{EmbeddingATD}

Let us now give a schematic description of the process we followed in order to verify the existence of a constant embedding of the form \eqref{straightembeddingATD}.

Again, we start by considering that the embedding functions of the $D6$ brane depend on all the worldvolume coordinates and thus we take,
\begin{equation}
 \g = \g(x^{\m}, \r, \a, \b) \ , \qquad u = u(x^{\m}, \r, \a, \b) \ , \qquad \phi = \phi(x^{\m}, \r, \a, \b) \ .
\end{equation}
Then the dynamics of our $D6$ brane is described by the DBIWZ action given in the equation \eqref{DBIWZATD}. In order to verify that the $D6$ brane admits a constant embedding of the form \eqref{straightembeddingATDgeneral}, we have to compute the equations of motion of the embedding functions $\g, u$ and $\phi$ and see if \eqref{straightembeddingATDgeneral} is a solution.

In order to construct the DBI part of the action we need to find the induced metric on the $D6$ brane which in our case takes the following form,
\begin{equation}
 \begin{aligned}
ds^2_{D6} & = \frac{R^2}{L^2} \eta_{\m\n} dx^\m dx^\n + \frac{L^2}{R^2} d \r^2 + \frac{L^2 \r^2}{4 R^2} \big(   d\a^2 + \sin^2 \a \ d\b^2  \big)
\\[10pt]
&+ \Bigg[   \frac{L^2}{R^2} \Big(   \partial_M u \ \partial_N u + u^2 \partial_M \phi \ \partial_N \phi   \Big) + \frac{4 \alpha'^2 R^2}{L^2 \r^2} \partial_M \g \ \partial_N \g   \Bigg] dx^M dx^N \ ,
 \end{aligned}
\end{equation}
where the indices $\mu, \nu$ run only over the Minkowski directions and $M,N$ over the $D6$ worldvolume directions. Also we consider the dilaton and the NS two-form given in the equation \eqref{NSatd}.

For the WZ part of the action one needs to evaluate the pullback of the combination $C_7 - C_5 \wedge B_2 + \frac{1}{2} B_2 \wedge B_2 \wedge C_3$ which in this case takes the form,
%
%\begin{equation}
% \begin{aligned}
%  C_7 - C_5 \wedge B_2 + \frac{1}{2} B_2 \wedge B_2 \wedge C_3 & =  - \frac{4 \a'^{3/2}}{L^4} \g^2 R^3 dx_{1,3} \wedge \textrm{Vol}(S^2) \wedge dR
%  \\[10pt]
%  & - \frac{\a'^{3/2}}{L^4} \g \ R^4 \ dx_{1,3} \wedge \textrm{Vol}(S^2) \wedge d\g \ ,
% \end{aligned}  
%\end{equation}
%
%and thus
%
\begin{equation}
 \begin{aligned}
  P \Big[C_7 - C_5 \wedge B_2 + \frac{1}{2} B_2 \wedge B_2 \wedge C_3 \Big] & =  - \frac{4 \a'^{3/2}}{L^4} \big(   u \partial_\r u + \r   \big) \g^2 R^2 dx_{1,3} \wedge \textrm{Vol}(S^2) \wedge d\r
  \\[10pt]
  & - \frac{\a'^{3/2}}{L^4} \g \ \partial_\r \g \ R^4 \ dx_{1,3} \wedge \textrm{Vol}(S^2) \wedge d\r \ .
 \end{aligned}  
\end{equation}

For the sake of simplicity, instead of presenting the general expressions for the $D6$ brane Lagrangian and the equations of motion that come out of it, we will give the expressions of these quantities only after imposing the following form of the embedding functions,
\begin{equation}
 \g = \g_0 = \textrm{const} \ ,\qquad u = u_0 = \textrm{const} \ ,\qquad \phi = \phi_0 = \textrm{const} \ .
\end{equation}
In that case the DBI and the WZ parts of the action are,
\begin{equation}
 \begin{aligned}
  & \mathcal{L}_{DBI}^{\textrm{const. emb}} = - \frac{\r \sin\a}{8 \sqrt{\alpha'} L^2} \sqrt{L^4 \r^4 + 16 \alpha'^2 \g_0^2 \big(   u_0^2 + \r^2   \big)^2} \ ,
 \\[10pt]
 &  \mathcal{L}_{WZ}^{\textrm{const. emb}} = - \frac{4 \alpha'^{3/2}}{L^4} \g_0^2 \big(   u_0^2 + \r^2   \big) \r \ \sin\a  \ ,
 \end{aligned}
\end{equation}
while each of the equations of motion takes the following form,

\vskip 10pt

\noindent \emph{Equation of motion for $\g$:}

\vskip 10pt

\begin{equation}
 \g_0 \big(   u_0^2 + \r^2  \big) \frac{  L^2 \big(   u_0^2 + \r^2  \big) + 2 \sqrt{L^4 \r^4 + 16 \alpha'^2 \g_0^2 \big(   u_0^2 + \r^2   \big)^2}  }{\sqrt{L^4 \r^4 + 16 \alpha'^2 \g_0^2 \big(   u_0^2 + \r^2   \big)^2}} = 0 \ .
\end{equation}

\vskip 10pt

\noindent \emph{Equation of motion for $u$:}

\vskip 10pt

\begin{equation}
 \frac{u_0 \g_0^2 \big(   u_0^2 + \r^2   \big)}{\sqrt{L^4 \r^4 + 16 \alpha'^2 \g_0^2 \big(   u_0^2 + \r^2   \big)^2}} = 0 \ .
\end{equation}

\vskip 10pt

\noindent \emph{Equation of motion for $\phi$:}

\vskip 10pt

\begin{equation}
 \textrm{Vanishes identically.}
\end{equation}

From the equations of motion we immediately see that the only solution to the equations of motion is if we take $\g_0 = 0$ while keeping $u_0$ and $\phi_0$ arbitrary constants. For such a solution we observe that WZ action is trivial while the DBI in non-vanishing.

%%%%%%%%%%%%%%%%%%%%%%%%%%%%%%%%%%%%%%%%%%%

\section{Useful formulas}
\label{usefulformulas}

Here we consider perturbations a) of the induced metric on the probe brane, b) of the worldvolume gauge field and c) of the dilaton. The perturbations have the general form
\footnote{
Here $\mathcal{F}_{ab}$ is defined as: \[ \mathcal{F}_{ab} = B_{ab} + 2 \pi \alpha' F_{ab} \ . \]
},
\begin{equation}
\label{genpert}
 \begin{array}{lll}
  g_{ab} = \bar{g}_{ab} + \delta g_{ab} \ , & \qquad \delta g_{ab} \ll 1 \ ,
\\[5pt]
  \mathcal{F}_{ab} = \bar{\mathcal{F}}_{ab} + \delta \mathcal{F}_{ab} \ , & \qquad \delta \mathcal{F}_{ab} \ll 1 \ ,
\\[5pt]
  \Phi = \bar{\Phi} + \delta \Phi \ , & \qquad \delta \Phi \ll 1 \ .
 \end{array}
\end{equation}
where $\bar{g}_{ab} \ , \bar{\mathcal{F}}$ are the components of the unperturbed metric and worldvolume gauge field respectively, while $\bar{\Phi}$ is the unperturbed dilaton. Under this scheme the perturbed DBI Lagrangian reads
\begin{equation}
 \begin{aligned}
  \mathcal{L}_{DBI} & = - e^{-\Phi} \sqrt{-\det(g +\mathcal{F})} =  - e^{-\bar{\Phi} - \delta \Phi} \sqrt{-\det\Big[ ( \bar{g} +\bar{\mathcal{F}} ) + ( \delta g + \delta \mathcal{F}) \Big]} 
  \\[5pt]
  & = - e^{-\bar{\Phi} - \delta \Phi} \sqrt{-\det( \bar{g} +\bar{\mathcal{F}})} \sqrt{\det (\mathbbm{1} + X)}
  \\[5pt]
  & = \bar{\mathcal{L}}_{DBI} \ e^{ - \delta \Phi} \ \sqrt{\det (\mathbbm{1} + X)} \ ,
 \end{aligned}
\end{equation}
where $\bar{\mathcal{L}}_{DBI}$ is the unperturbed DBI Lagrangian and we have defined the matrix
\begin{equation}
 X \equiv ( \bar{g} + \bar{\mathcal{F}} )^{-1} (\delta g +\delta \mathcal{F}) \ .
\end{equation}
One can decompose the first factor in the definition of $X$ into a symmetric part $\mathcal{G}$ (open string metric) and an antisymmetric part $\mathcal{J}$ as,
\begin{equation}
 ( \bar{g} + \bar{\mathcal{F}} )^{-1} = \mathcal{G}^{-1} + \mathcal{J} \ .
\end{equation}
Taking also into account the expansion
\begin{equation}
 \sqrt{ \det (\mathbbm{1} + X)} = 1 + \frac{1}{2} \Tr X - \frac{1}{4} \Tr(X^2) + \frac{1}{8} (\Tr X)^2 + O(X^3) \ ,
\end{equation}
we can write the perturbed DBI Lagrangian in the following form,
\begin{equation}
  \begin{aligned}
   \mathcal{L}_{DBI} & = \bar{\mathcal{L}}_{DBI} \ e^{ - \delta \Phi} \ \Bigg[  1 + \frac{1}{2} \mathcal{G}^{ab} \delta g_{ab} - \frac{1}{2} \mathcal{J}^{ab} \delta \mathcal{F}_{ab} - \frac{1}{4} \Big(    \mathcal{G}^{ac} \mathcal{G}^{bd} + \mathcal{J}^{ac} \mathcal{J}^{bd}     \Big) \delta g_{bc}  \delta g_{ad} 
  \\[5pt]
  & - \frac{1}{4} \Big(     \mathcal{G}^{ac} \mathcal{G}^{bd} + \mathcal{J}^{ac} \mathcal{J}^{bd} \Big)  \delta \mathcal{F}_{bc}  \delta \mathcal{F}_{ad} - \mathcal{G}^{ac} \mathcal{J}^{bd} \delta g_{cb} \delta \mathcal{F}_{da}
  \\[5pt]
  & + \frac{1}{8} \mathcal{G}^{ab} \mathcal{G}^{cd}  \delta g_{ab}  \delta g_{cd} + \frac{1}{8} \mathcal{J}^{ab} \mathcal{J}^{cd}  \delta \mathcal{F}_{ab}  \delta \mathcal{F}_{cd} - \frac{1}{4} \mathcal{G}^{ab} \mathcal{J}^{cd} \delta g_{ab} \delta \mathcal{F}_{cd} + \ldots \Bigg]
  \end{aligned}
\end{equation}
where also one has to expand $e^{- \delta \Phi}$
\begin{equation}
 e^{- \delta \Phi} = 1 - \delta \Phi + \frac{1}{2} \delta \Phi^2 + \ldots
\end{equation}

Now in order to keep track of the order of the perturbation we find it convenient to express the fluctuations of the metric, the worldvolume gauge field and the dilaton as
\begin{equation}
 \begin{aligned}
  & \delta g_{ab} = \delta g_{ab}^{(1)} + \delta g_{ab}^{(2)} + \ldots \ ,
  \\[5pt]
  & \delta \mathcal{F}_{ab} = \delta \mathcal{F}_{ab}^{(1)} + \delta \mathcal{F}_{ab}^{(2)} + \ldots \ ,
  \\[5pt]
  & \delta \Phi = \delta \Phi^{(1)} + \delta \Phi^{(2)} + \ldots \ ,
 \end{aligned}
\end{equation}
where the index in the parenthesis denotes the order of the perturbation. Then one can split the Lagrangian of the fluctuations in the following way
\begin{equation}
 \mathcal{L}_{DBI} = \mathcal{L}_{DBI}^{(0)} + \mathcal{L}_{DBI}^{(1)} + \mathcal{L}_{DBI}^{(2)} + \ldots \ ,
\end{equation}
where again the index in the  parenthesis denotes the order of the perturbation. More explicitly each order takes the following form

\vskip 10pt

\noindent \emph{ Zeroth order}

\vskip 10pt

\begin{equation}
\label{zerothorderDBI}
 \mathcal{L}_{DBI}^{(0)} = \bar{\mathcal{L}}_{DBI} \ .
\end{equation}

\vskip 10pt

\noindent \emph{ First order}

\vskip 10pt

\begin{equation}
\label{firstorderDBI}
 \mathcal{L}_{DBI}^{(1)} = - \bar{\mathcal{L}}_{DBI} \ \Big(  \delta \Phi^{(1)} - \frac{1}{2} \mathcal{G}^{ab} \delta g_{ab}^{(1)} + \frac{1}{2} \mathcal{J}^{ab} \delta \mathcal{F}_{ab}^{(1)} \Big) \ .
\end{equation}

\vskip 10pt

\noindent \emph{ Second order}

\vskip 10pt

\begin{equation}
\label{secondorderDBI}
 \begin{aligned}
  \mathcal{L}_{DBI}^{(2)} & = \bar{\mathcal{L}}_{DBI} \ \Bigg[   \frac{1}{2} \Big(   \delta \Phi^{(1)}  \Big)^2 - \delta \Phi^{(2)} - \delta \Phi^{(1)} \Bigg(  \frac{1}{2} \mathcal{G}^{ab} \delta g_{ab}^{(1)} - \frac{1}{2} \mathcal{J}^{ab} \delta \mathcal{F}_{ab}^{(1)}   \Bigg)  + \frac{1}{2} \mathcal{G}^{ab} \delta g_{ab} ^{(2)}
  \\[5pt]
  & - \frac{1}{2} \mathcal{J}^{ab} \delta \mathcal{F}_{ab}^{(2)} - \frac{1}{4} \Big(    \mathcal{G}^{ac} \mathcal{G}^{bd} + \mathcal{J}^{ac} \mathcal{J}^{bd}     \Big) \delta g_{bc}^{(1)}  \delta g_{ad}^{(1)}  - \frac{1}{4} \Big(     \mathcal{G}^{ac} \mathcal{G}^{bd} + \mathcal{J}^{ac} \mathcal{J}^{bd} \Big)  \delta \mathcal{F}_{bc}^{(1)}  \delta \mathcal{F}_{ad}^{(1)}
  \\[5pt]
  & - \mathcal{G}^{ac} \mathcal{J}^{bd} \delta g_{cb}^{(1)} \delta \mathcal{F}_{da}^{(1)} + \frac{1}{8} \mathcal{G}^{ab} \mathcal{G}^{cd}  \delta g_{ab}^{(1)}  \delta g_{cd}^{(1)} + \frac{1}{8} \mathcal{J}^{ab} \mathcal{J}^{cd}  \delta \mathcal{F}_{ab}^{(1)}  \delta \mathcal{F}_{cd}^{(1)} - \frac{1}{4} \mathcal{G}^{ab} \mathcal{J}^{cd} \delta g_{ab}^{(1)} \delta \mathcal{F}_{cd}^{(1)}   \Bigg]  \ .
 \end{aligned}
\end{equation}

Obviously the zeroth order part of the DBI Lagrangian does not contribute to the equations of motion of the fluctuations. Similarly one has to prove that also the first order part of the DBI Lagrangian together with the first order part of the WZ term are either constant or can be written as total derivatives so in the end they do not contribute to the equations of motion. In the following we will omit the zeroth and the first order terms and we will focus only to the second order terms.

%%%%%%%%%%%%%%%%%%%%%%%%%%%%%%%%%%%%%%%%%%%%%%%%%%%

\section{Quadratic Lagrangians of the fluctuations}

In this appendix we calculate the quadratic Lagrangians for the fluctuations of the probe $D6$ branes in the NATD and the Hopf-T-dual backgrounds. For this purpose we use the tools developed in the appendix \ref{usefulformulas}. In the end, we will show that the two Lagrangians are the same up to an overall factor of $q \pi$.

\subsection{Fluctuations in the NATD background}
\label{quadrLagrNATD}

Let us start with the study of the fluctuations of the $D6$ probe brane in the NATD background. First we give the entries of the open string metric $\mathcal{G}^{-1}$ and the antisymmetric matrix $\mathcal{J}$. Starting with the open string metric we find,
\begin{equation}
 \begin{array}{ll}
   \mathcal{G}^{\mu \nu} = \frac{L^2}{u_0^2 + \r^2} \ \eta^{\mu\nu} \ , \; \;  \mu,\nu = 0, \ldots, 3 \ , & \qquad \mathcal{G}^{\r\r} =  \frac{u_0^2 + \r^2}{L^2} \ ,
   \\[10pt]
   \mathcal{G}^{\chi \chi} = 4 \frac{u_0^2 + \r^2}{L^2 \r^2} \ , & \qquad    \mathcal{G}^{\xi \xi} =\frac{4}{\sin^2 \chi} \frac{u_0^2 + \r^2}{L^2 \r^2} \ .
 \end{array}
\end{equation}
where $\eta^{\mu\nu}$ is the inverse of the four-dimensional Minkowski metric. For the antisymmetric matrix we find that the only non-vanishing elements are,
\begin{equation}
 \mathcal{J}^{\chi \xi} = -  \mathcal{J}^{\xi \chi} = - \frac{1}{q \alpha' \pi \sin\chi} \ . 
\end{equation}

Let us now consider the fluctuation of the dilaton. Adopting the scheme in eq. \eqref{perturbationScheme} we compute,
\begin{equation}
 \begin{aligned}
   \Phi & = \frac{1}{2} \log \Bigg[   \frac{64 \alpha'^3 \big(   u_0^2 + \r^2  \big)^2}{L^2 \r^2 \D}   \Bigg] - \frac{16 \alpha'^2 q \pi \big(   u_0^2 + \r^2  \big)^2}{\D} \delta r + \frac{u_0 \big(   2 L^4 \r^4 + \D   \big)}{ \big(    u_0^2 + \r^2  \big) \D } \delta u
   \\[10pt]
   & + \frac{8 \alpha'^2 \big(   u_0^2 + \r^2   \big)^2 \big(   \D - 2 L^4 \r^4   \big)}{\D^2} \delta r^2 - \frac{64 L^4 \alpha'^2 q \pi u_0 \r^4 \big(   u_0^2 + \r^2   \big) }{\D^2} \delta u \ \delta r 
   \\[10pt]
   & - \frac{u_0^2 \big(    \D^2 + 10 L^4 \r^4 \D - 8 L^8 \r^8  \big) - \r^2 \big(   \D + 2 L^4 \r^4  \big) \D}{2 \big(   u_0^2 + \r^2  \big)^2 \D^2} \delta u^2 + \dots
 \end{aligned}
\end{equation}
where for convenience we have defined the function $\D$ to be,
\begin{equation}
 \D = L^4 \r^4 + 16 \alpha'^2 q^2 \pi^2 \big(   u_0^2 + \r^2  \big)^2 \ .
\end{equation}

Let us do the same with the metric. We will compute the line element for each order in the expansion,

\vskip 5pt

\noindent \emph{ Zeroth order}

\begin{equation}
 ds_{(0)}^2 =\frac{u_0^2 + \r^2}{L^2}dx_{1,3}^2 + \frac{L^2}{u_0^2 + \r^2} d\r^2 +\frac{4 L^2 \alpha'^2 q^2 \pi^2 \r^2 \big(   u_0^2 + \r^2  \big)}{\D}(d\chi^2 +\sin^2\chi \ d\xi^2).
\end{equation}

\vskip 5pt

\noindent \emph{ First order}

\begin{equation}
 \begin{aligned}
  ds_{(1)}^2 & = \frac{2 u_0}{L^2} \ \delta u \ dx_{1,3}^2 - \frac{2 L^2 u_0}{\big(   u_0^2 + \r^2  \big)^2} \ \delta u \ d \r^2 + 8 L^2 \alpha'^2 q \pi \r^2 \Bigg[     \frac{L^4 \r^4 \big(   u_0^2 + \r^2  \big)}{\D^2} \delta r
  \\[5pt]
  & - \frac{q \pi u_0 \big(   \D - 2 L^4 \r^4  \big)}{\D^2} \delta u \Bigg] \ \big( d\chi^2 +\sin^2\chi d\xi^2 \big).
 \end{aligned}
\end{equation}

\vskip 5pt

\noindent \emph{ Second order}

\begin{equation}
 \begin{aligned}
  ds_{(2)}^2 & =  \frac{4 \alpha'^2 \big(   u_0^2 + \r^2  \big)}{L^2 \r^2} \partial_{M} \delta r \ \partial_{N} \delta r \  dx^{M} dx^{N} + \frac{L^2}{u_0^2 + \r^2} \Big[   \partial_{M} \delta u \ \partial_{N} \delta u + u_0^2 \ \partial_{M} \delta \phi \ \partial_{N} \delta \phi  \Big] dx^{M} dx^{N}
  \\[10pt]
  &  + \frac{4 L^2 \alpha'^2 \r^2}{\D^3} \Bigg[   q^2 \pi^2 \frac{u_0^2 \big(  3 \D^2 -18 L^4 \r^4 \D + 16 L^8 \r^8  \big) - \D \r^2 \big(   \D - 2 L^4 \r^4  \big)}{u_0^2 + \r^2} \delta u^2
  \\[10pt]
  & -  4 L^4 q \pi u_0 \r^4 \big(   3 \D - 4 L^4 \r^4  \big) \delta r \ \delta u - L^4 \r^4 \big(   u_0^2 + \r^2  \big) \big(   3 \D - 4 L^4 \r^4  \big) \delta r^2  \Bigg] \ \big( d\chi^2 +\sin^2\chi d\xi^2 \big) 
  \\[10pt]
  & + \frac{\delta u^2}{L^2} dx_{1,3}^2 + L^2 \frac{3 u_0^2 - \r^2}{ \big(   u_0^2 + \r^2  \big)^3 } \delta u^2 \ d\r^2 \ ,
 \end{aligned}
\end{equation}
where the indices $M,N$ run over $x_{1,3},\r, \chi, \xi$.

Similarly we compute the fluctuation of the NS two-form $B_2$,
\begin{equation}
 \begin{aligned}
  B_2 &  = - \frac{L^4 \alpha' q \pi \r^4}{\D} \sin\chi \ d\chi \wedge d\xi + \frac{16 \alpha'^3 q^2 \pi^2}{\D^2}\Bigg[  \Big(  u_0^2 + \r^2   \Big)^2 \Big(   2 L^4 \r^4 + \D  \Big) \delta r
  \\[5pt]
  & + 4 L^4 q \pi u_0 \r^4 \Big(   u_0^2 + \r^2  \Big) \delta u   \Bigg] \ \sin \chi \ d\chi \wedge d\xi - \frac{16 L^4 \alpha'^3 q \pi \r^4}{\D^3} \Bigg[   \Big(   u_0^2 + \r^2  \Big)^2 \Big(   \D - 4 L^4 \r^4  \Big) \delta r^2  
  \\[5pt]
  & + 4 q \pi u_0 \Big(   u_0^2 + \r^2 \Big) \Big(   \D - 4 L^4 \r^4   \Big) \delta r \ \delta u - 2 q^2 \pi^2 \Bigg(   8 u_0^2 L^4 \r^4 - \Big(   5 u_0^2 - \r^2  \Big) \D  \Bigg) \delta u^2 \Bigg]  \ \sin \chi \ d\chi \wedge d\xi 
  \\[5pt]
  &  + \ldots
 \end{aligned}
\end{equation}

Now we have all the ingredients to compute the DBI Lagrangian up to second order in the fluctuations using the formulas \eqref{zerothorderDBI}, \eqref{firstorderDBI} and \eqref{secondorderDBI}. It turns out that the zeroth and first order terms take the following simple form,
\begin{equation}
 \mathcal{L}_{DBI}^{(0)} = - \frac{q \pi \r^3}{8 \sqrt{\alpha'}} \sin\chi \ , \qquad \mathcal{L}_{DBI}^{(1)} = \frac{\pi \r^3}{4 \sqrt{\alpha'}} \delta F_{\chi \xi} \ ,
\end{equation}
while the second order term is given by the following expression,
\begin{equation}
 \begin{aligned}
  \mathcal{L}_{DBI}^{(2)} & = - \frac{\alpha'^{3/2} q \pi}{4} \r \sin\chi \ \eta^{\mu \nu} \partial_{\m} \delta r \ \partial_{\n} \delta r - \frac{L^4 q \pi \r^3 \sin\chi}{16 \sqrt{\alpha'} \big(   u_0^2 + \r^2  \big)^2} \eta^{\mu\nu} \Big(   \partial_{\mu} \delta u \ \partial_{\nu} \delta u + u_0^2 \partial_{\mu} \delta \phi \ \partial_{\nu} \delta \phi   \Big)
  \\[5pt]
  & - \frac{\alpha^{3/2} q \pi \big(    u_0^2 + \r^2  \big)^2 \sin\chi }{L^4 \r} \Bigg[    \delta r^2 + \frac{\r^2}{4} \big(   \partial_\r \delta r  \big)^2 + \big(   \partial_{\chi} \delta r   \big)^2 + \frac{1}{\sin^2\chi} \big(   \partial_\xi \delta r   \big)^2   \Bigg] 
  \\[5pt]
  & - \frac{q \pi \r \sin\chi}{4 \sqrt{\alpha'}} \Bigg[  \frac{\r^2}{4} \big(   \partial_\r \delta u  \big)^2 + \big(   \partial_{\chi} \delta u   \big)^2 + \frac{1}{\sin^2\chi} \big(   \partial_\xi \delta u   \big)^2 + \frac{u_0^2 \r^2}{4} \big(   \partial_\r \delta \phi  \big)^2 + u_0^2 \big(   \partial_{\chi} \delta \phi   \big)^2 
  \\[5pt]
  &+ \frac{u_0^2}{\sin^2\chi} \big(   \partial_\xi \delta \phi   \big)^2   \Bigg] - \frac{4 \alpha'^{3/2} q \pi^2 \big(   u_0^2 + \r^2  \big)^2 }{L^4 \r} \delta F_{\chi \xi} \ \delta r - \frac{\alpha'^{3/2} L^4 q \pi^3 \r^3 \sin\chi}{8 \big(   u_0^2 + \r^2  \big)^2} \eta^{\m\n} \eta^{\kappa \lambda} \delta F_{\mu \kappa} \delta F_{\nu \lambda} 
  \\[5pt]
  & - \alpha'^{3/2} q \pi^3 \r \sin\chi \ \eta^{\mu \nu} \Bigg[   \frac{\r^2}{4} \delta F_{\mu \r} \delta F_{\nu \r} + \delta F_{\mu \chi} \delta F_{\nu \chi} + \frac{1}{\sin^2 \chi} \delta F_{\mu \xi} \delta F_{\nu \xi}   \Bigg]
  \\[5pt]
  &- \frac{\alpha'^{3/2} q \pi^3 \r \big(  u_0^2 + \r^2  \big)^2 \sin\chi}{L^4} \Bigg[   \delta F_{\r \chi}^2 + \frac{1}{\sin^2 \chi} \delta F_{\r \xi}^2 + \frac{4}{\r^2 \sin^2 \chi} \delta F_{\chi \xi}^2   \Bigg]\ ,
 \end{aligned} 
\end{equation}
where the indices $\mu, \nu, \kappa, \lambda$ run over the Minkowski directions $x_{1,3}$.

For the WZ term one has to perturb also the RR potentials $C_5$ and $C_7$ which are given in the equation \eqref{C1C3C5C7}. In this case, before perturbing the RR potentials we notice that the combination $C_7 - B^{l.g.}_2 \wedge C_5$ simplifies to,
\begin{equation}
 C_7 - B_2^{l.g.} \wedge C_5 = \frac{2 \alpha'^{3/2} r^2 R^3}{3L^4} \big(   3 q \pi - 2 r   \big) \sin \chi \ dx_{1,3} \wedge dR \wedge d\chi \wedge d \xi \ .
\end{equation}

Now in the expansion of $\mathcal{L}_{WZ}$ one has also to consider the fluctuation of the worldvolume field strength that is inside $\mathcal{F}$. So for the zeroth and first order we find the simple expressions below,
\begin{equation}
 \begin{aligned}
   \mathcal{L}_{WZ}^{(0)} & = \frac{2 \alpha'^{3/2} q^3 \pi^3 \r \big(   u_0^2 + \r^2 \big)}{3L^4} \sin \chi \ ,
   \\[10pt]
   \mathcal{L}_{WZ}^{(1)} & =  \frac{2 \alpha'^{3/2} q^3 \pi^3 u_0}{3L^4} \sin \chi \ \partial_{\r} \Big[   \big(   u_0^2 + \r^2 \big) \delta u   \Big] - \frac{4 \alpha'^{3/2} q^2 \pi^3 \r \big(   u_0^2 + \r^2   \big)}{L^4} \delta F_{\chi \xi} \ ,
 \end{aligned}
\end{equation}
while for the second order term we get,
\begin{equation}
 \begin{aligned}
   \mathcal{L}_{WZ}^{(2)} & = - \frac{2 \alpha'^{3/2} q \pi \r \big(   u_0^2 + \r^2  \big)}{L^4} \sin\chi \ \delta r^2 + \frac{\alpha'^{3/2} q^3 \pi^3}{3 L^4} \sin \chi \ \partial_{\r} \Big[  \big(   3 u_0^2 + \r^2  \big) \delta u^2  \Big]
   \\[10pt]
   & - \frac{4 \alpha'^{3/2} q^2 \pi^3 u_0 \big(   u_0^2 + \r^2  \big)}{L^4} \Big[   \delta F_{\r \chi} \ \partial_{\xi} \delta u - \delta F_{\r \xi} \ \partial_{\chi} \delta u +\delta F_{\chi \xi} \ \partial_{\r} \delta u   \Big]
   \\[10pt]
   & - \frac{8 \alpha'^{3/2} q \pi^2 \r}{L^4} \Big[   \big(   u_0^2 + \r^2  \big) \delta r + q \pi u_0 \ \delta u   \Big] \delta F_{\chi \xi}  \ .
 \end{aligned}
\end{equation}
Notice that the second term in the first line of the formula above is a total derivative and thus we can drop it. Moreover, the last two lines can be simplified if we integrate by parts the three terms of the second line and combine them with the terms of the third line. After doing that we obtain,
\begin{equation}
 \begin{aligned}
  & - \frac{4 \alpha'^{3/2} q^2 \pi^3 u_0 \big(   u_0^2 + \r^2  \big)}{L^4} \Big[   \delta F_{\r \chi} \ \partial_{\xi} \delta u - \delta F_{\r \xi} \ \partial_{\chi} \delta u +\delta F_{\chi \xi} \ \partial_{\r} \delta u   \Big]
   \\[10pt]
   & - \frac{8 \alpha'^{3/2} q \pi^2 \r}{L^4} \Big[   \big(   u_0^2 + \r^2  \big) \delta r + q \pi u_0 \ \delta u   \Big] \delta F_{\chi \xi} = - \frac{8 \a'^{3/2} q \pi^2 (u_0^2 + \r^2) \r}{L^4} \d r \d F_{\chi \xi} + \textrm{tot. der.}
 \end{aligned}
\end{equation}

After dropping the total derivative terms, we can rewrite the second order WZ term as,
\begin{equation}
 \mathcal{L}_{WZ}^{(2)} = - \frac{2 \alpha'^{3/2} q \pi \r \big(   u_0^2 + \r^2  \big)}{L^4} \sin\chi \ \delta r^2 - \frac{8 \a'^{3/2} q \pi^2 (u_0^2 + \r^2) \r}{L^4} \d r \d F_{\chi \xi} \ .
\end{equation}

Obviously the zeroth order terms of the DBI and WZ parts do not contribute to the equations of motion for the fluctuations. Looking at the first order terms we notice that they can be written as total derivatives and thus we do not consider them in the calculation for the equations of motion. Finally, the total second order Lagrangian for the fluctuations is given by the sum of $\mathcal{L}_{DBI}^{(2)}$ and $\mathcal{L}_{WZ}^{(2)}$.

\subsection{Fluctuations in the Hopf-T-dual background}
\label{quadrLagrHopfTD}

We continue with the study of the fluctuations of the $D6$ probe brane in the Hopf-T-dual background. First we give the entries of the open string metric $\mathcal{G}^{-1}$ and the antisymmetric matrix $\mathcal{J}$. Starting with the open string metric we find,
\begin{equation}
 \begin{array}{ll}
   \mathcal{G}^{\mu \nu} = \frac{L^2}{u_0^2 + \r^2} \ \eta^{\mu\nu} \ , \; \;  \mu,\nu = 0, \ldots, 3 \ , & \qquad \mathcal{G}^{\r\r} =  \frac{u_0^2 + \r^2}{L^2} \ ,
   \\[10pt]
   \mathcal{G}^{\chi \chi} = 4 \frac{u_0^2 + \r^2}{L^2 \r^2} \ , & \qquad    \mathcal{G}^{\xi \xi} =\frac{4}{\sin^2 \a} \frac{u_0^2 + \r^2}{L^2 \r^2} \ .
 \end{array}
\end{equation}
where $\eta^{\mu\nu}$ is the inverse of the four-dimensional Minkowski metric. For the antisymmetric matrix $\mathcal{J}$ we conclude that all of its entries are zero. This is because the pullback of the NS two-form on the $D6$-brane worldvolume is trivial.

Let us now consider the fluctuation of the dilaton. Adopting the scheme in eq. \eqref{straightembeddingATD} we compute,
\begin{equation}
 \begin{aligned}
   \Phi = \ln \frac{2 \sqrt{\a' \big(   u_0^2 + \r^2  \big)}}{L \r} + \frac{u_0}{u_0^2 + \r^2} \d u - \frac{u_0^2 - \r^2}{2 \big(   u_0^2 + \r^2  \big)^2} \d u^2 + \dots
 \end{aligned}
\end{equation}

Let us do the same with the metric. We will compute the line element for each order in the expansion,

\vskip 5pt

\noindent \emph{ Zeroth order}

\begin{equation}
 ds_{(0)}^2 =\frac{u_0^2 + \r^2}{L^2}dx_{1,3}^2 + \frac{L^2}{u_0^2 + \r^2} d\r^2 +\frac{L^2 \r^2}{4 \big(   u_0^2 + \r^2  \big)}(d\a^2 +\sin^2\a \ d\b^2).
\end{equation}

\vskip 5pt

\noindent \emph{ First order}

\begin{equation}
 \begin{aligned}
  ds_{(1)}^2 & = \frac{2 u_0}{L^2} \ \delta u \ dx_{1,3}^2 - \frac{2 L^2 u_0}{\big(   u_0^2 + \r^2  \big)^2} \ \delta u \ d \r^2 - \frac{L^2 \r^2 u_0}{2 (u_0^2 + \r^2)^2} \d u \big( d\a^2 +\sin^2\a \ d\b^2 \big).
 \end{aligned}
\end{equation}

\vskip 5pt

\noindent \emph{ Second order}

\begin{equation}
 \begin{aligned}
  ds_{(2)}^2 & =  \frac{4 \alpha'^2 \big(   u_0^2 + \r^2  \big)}{L^2 \r^2} \partial_{M} \delta \g \ \partial_{N} \delta \g \  dx^{M} dx^{N} + \frac{L^2}{u_0^2 + \r^2} \Big[   \partial_{M} \delta u \ \partial_{N} \delta u + u_0^2 \ \partial_{M} \delta \phi \ \partial_{N} \delta \phi  \Big] dx^{M} dx^{N}
  \\[10pt]
  & + \frac{L^2 \r^2 (3 u_0^2 - \r^2)}{4 (u_0^2 + \r^2)^3} \d u^2 \big( d\a^2 +\sin^2\a \ d\b^2 \big) 
  + \frac{\delta u^2}{L^2} dx_{1,3}^2 + L^2 \frac{3 u_0^2 - \r^2}{ \big(   u_0^2 + \r^2  \big)^3 } \delta u^2 \ d\r^2 \ ,
 \end{aligned}
\end{equation}
where the indices $M,N$ run over the worldvolume directions of the $D6$ brane $x_{1,3},\r, \a, \b$.

Similarly we compute the fluctuation of the NS two-form $B_2$,
\begin{equation}
 \begin{aligned}
  B_2  =  \a' \d \g \ \sin \a \ d\a \wedge d\b+ \ldots
 \end{aligned}
\end{equation}

Now we have all the ingredients to compute the DBI Lagrangian up to second order in the fluctuations using the formulas \eqref{zerothorderDBI}, \eqref{firstorderDBI} and \eqref{secondorderDBI}. It turns out that the zeroth and first order terms take the following simple form,
\begin{equation}
 \mathcal{L}_{DBI}^{(0)} = - \frac{\r^3}{8 \sqrt{\alpha'}} \sin\a \ , \qquad \mathcal{L}_{DBI}^{(1)} = 0 \ ,
\end{equation}
while the second order term is given by the following expression,
\begin{equation}
 \begin{aligned}
  \mathcal{L}_{DBI}^{(2)} & = - \frac{\alpha'^{3/2}}{4} \r \sin\a \ \eta^{\mu \nu} \partial_{\m} \delta \g \ \partial_{\n} \delta \g - \frac{L^4 \r^3 \sin\a}{16 \sqrt{\alpha'} \big(   u_0^2 + \r^2  \big)^2} \eta^{\mu\nu} \Big(   \partial_{\mu} \delta u \ \partial_{\nu} \delta u + u_0^2 \partial_{\mu} \delta \phi \ \partial_{\nu} \delta \phi   \Big)
  \\[5pt]
  & - \frac{\alpha^{3/2} \big(    u_0^2 + \r^2  \big)^2 \sin\a }{L^4 \r} \Bigg[    \delta \g^2 + \frac{\r^2}{4} \big(   \partial_\r \delta \g  \big)^2 + \big(   \partial_{\a} \delta \g   \big)^2 + \frac{1}{\sin^2\a} \big(   \partial_\b \delta \g   \big)^2   \Bigg] 
  \\[5pt]
  & - \frac{ \r \sin\a}{4 \sqrt{\alpha'}} \Bigg[  \frac{\r^2}{4} \big(   \partial_\r \delta u  \big)^2 + \big(   \partial_{\a} \delta u   \big)^2 + \frac{1}{\sin^2\a} \big(   \partial_\b \delta u   \big)^2 + \frac{u_0^2 \r^2}{4} \big(   \partial_\r \delta \phi  \big)^2 + u_0^2 \big(   \partial_{\a} \delta \phi   \big)^2 
  \\[5pt]
  &+ \frac{u_0^2}{\sin^2\a} \big(   \partial_\b \delta \phi   \big)^2   \Bigg] - \frac{4 \alpha'^{3/2} \pi \big(   u_0^2 + \r^2  \big)^2 }{L^4 \r} \delta F_{\a \b} \ \delta \g - \frac{\alpha'^{3/2} L^4 \pi^2 \r^3 \sin\a}{8 \big(   u_0^2 + \r^2  \big)^2} \eta^{\m\n} \eta^{\kappa \lambda} \delta F_{\mu \kappa} \delta F_{\nu \lambda} 
  \\[5pt]
  & - \alpha'^{3/2} \pi^2 \r \sin\a \ \eta^{\mu \nu} \Bigg[   \frac{\r^2}{4} \delta F_{\mu \r} \delta F_{\nu \r} + \delta F_{\mu \a} \delta F_{\nu \a} + \frac{1}{\sin^2 \a} \delta F_{\mu \b} \delta F_{\nu \b}   \Bigg]
  \\[5pt]
  &- \frac{\alpha'^{3/2} \pi^2 \r \big(  u_0^2 + \r^2  \big)^2 \sin\a}{L^4} \Bigg[   \delta F_{\r \a}^2 + \frac{1}{\sin^2 \a} \delta F_{\r \b}^2 + \frac{4}{\r^2 \sin^2 \a} \delta F_{\a \b}^2   \Bigg]\ ,
 \end{aligned} 
\end{equation}
where the indices $\mu, \nu, \kappa, \lambda$ run over the Minkowski directions $x_{1,3}$.

For the WZ term one has to perturb also the RR potentials $C_3, C_5$ and $C_7$ which are given in the equations \eqref{C3C5atd} and \eqref{C7atd} respectively. Moreover, one has also to consider the fluctuation of the worldvolume field strength that is inside $\mathcal{F}$. A simple computation tells us that the zeroth and the first order terms of the WZ action vanish, i.e.
\begin{equation}
   \mathcal{L}_{WZ}^{(0)} = \mathcal{L}_{WZ}^{(1)} =0 \ ,
\end{equation}
while for the second order term we get,
\begin{equation}
 \mathcal{L}_{WZ}^{(2)} = - \frac{4 \alpha'^{3/2} \r \big(   u_0^2 + \r^2  \big)}{L^4} \sin\a \ \delta \g^2
 - \frac{\a'^{3/2}}{L^4} \big(   u_0^2 + \r^2  \big)^2 \sin \a \ \delta \g \ \partial_\r \delta \g
 - \frac{8 \a'^{3/2} \pi (u_0^2 + \r^2) \r}{L^4} \ \d \g \ \d F_{\a \b} \ .
\end{equation}
Notice that we can integrate by parts the second term in the expression above and combine it with the first one. After dropping the total derivative term that comes from the integration by parts we get,
\begin{equation}
 \mathcal{L}_{WZ}^{(2)} = - \frac{2 \alpha'^{3/2} \r \big(   u_0^2 + \r^2  \big)}{L^4} \sin\a \ \delta \g^2
 - \frac{8 \a'^{3/2} \pi (u_0^2 + \r^2) \r}{L^4} \ \d \g \ \d F_{\a \b} \ .
\end{equation}

Obviously the zeroth order term of the DBI part does not contribute to the equations of motion for the fluctuations.  Finally, the total second order Lagrangian for the fluctuations is given by the sum of $\mathcal{L}_{DBI}^{(2)}$ and $\mathcal{L}_{WZ}^{(2)}$. 

Notice that the total action of the NATD case is \emph{almost} identical to the total action here under the map in equation \eqref{NATDtoATD}. The only difference between the two actions is an overall factor of $q \pi$. This will also imply that the equations of motion of the fluctuations here will be mapped to the equations of motion of the NATD case using the relation \eqref{NATDtoATD}.

%%%%%%%%%%%%%%%%%%%%%%%%%%%%%%%%%%%%%%%%%%%%%%%%%%%

\section{General Equations for the Fluctuations}
\label{equationsfluctuated}

In this section we study the general equations for a fluctuation of the form of eq.(\ref{perturbationScheme}).

\noindent \emph{Equation for $\delta r$}

\vskip 10pt

\begin{equation}
 \begin{aligned}
  & L^4 \rho^2 \ \Box \delta r + \rho^2 \big(  u_0^2 + \rho^2  \big)^2 \partial_{\rho}^2 \delta r + \rho \big(  u_0^2 + \rho^2  \big) \big(  u_0^2 + 5 \rho^2  \big) \partial_{\rho} \delta r - 4 \big(  u_0^2 + \rho^2  \big) \big(   u_0^2 + 3 \rho^2  \big) \delta r
  \\[10pt]
  & + 4 \big(  u_0^2 + \rho^2  \big)^2 \nabla_{S^2}^2 \delta r - 8 \pi \frac{\big(  u_0^2 + \rho^2  \big) \big(   u_0^2 + 3 \rho^2  \big)}{\sin\chi} \delta F_{\chi \xi} = 0 \ .\label{deltar11}
 \end{aligned}
\end{equation}

\vskip 10pt

\noindent \emph{Equation for $\delta u$}

\vskip 10pt

\begin{equation}
 \begin{aligned}
  L^4 \rho^2 \Box \delta u + \rho^2 \big(  u_0^2 + \rho^2  \big)^2 \partial_{\rho}^2 \delta u + 3 \rho \big(  u_0^2 + \rho^2  \big)^2 \partial_{\rho} \delta u + 4 \big(  u_0^2 + \rho^2  \big)^2 \nabla_{S^2}^2 \delta u = 0 \ .
 \end{aligned}
\end{equation}

\vskip 10pt

\noindent \emph{Equation for $\delta \phi$}

\vskip 10pt

The equation for $\delta\phi$ is exactly the same as the one for $\delta u$.

\vskip 10pt

\noindent \emph{Equation for $\delta A_{C} , \;\; C = t, x_1 , x_2 , x_3 , \r$}

\vskip 10pt

\begin{equation}
 \begin{aligned}
  & L^4 \rho^2 \Box \delta A_{C} - L^4 \rho^2 \partial_{C} \big(   \eta^{\mu\nu} \partial_{\mu} \delta A_{\nu}  \big) - \rho^2 \big(   u_0^2 + \rho^2  \big)^2 \partial_{\rho} \delta F_{C \rho} - 3 \rho \big(   u_0^2 + \rho^2  \big)^2 \delta F_{C \rho}
  \\[10pt]
  &+  4 \big(  u_0^2 + \rho^2  \big)^2 \nabla_{S^2}^2 \delta A_{C} - 4 \big(  u_0^2 + \rho^2  \big)^2 \partial_{C} \partial_{\chi} \delta A_{\chi} - 4 \big(  u_0^2 + \rho^2  \big)^2 \cot\chi \ \partial_{C} \delta A_{\chi}  
  \\[10pt]
  & - 4 \frac{\big(  u_0^2 + \rho^2  \big)^2} {\sin^2\chi} \partial_{C} \partial_{\xi} \delta A_{\xi} = 0 \ .  
\end{aligned} 
\end{equation}
%
% The last three terms maybe can be written in terms of the Laplace-Beltrami operator for vectors.

% \vskip 10pt

% \noindent \emph{Equation for $\delta A_{\rho}$}

% \vskip 10pt

% \begin{equation}
% \begin{aligned}
%   & L^4 \rho^2 \Box \delta A_{\rho} - L^4 \rho^2 \partial_{\rho} \big(   \eta^{\mu\nu} \partial_{\mu} \delta A_{\nu}  \big) +  4 \big(  u_0^2 + \rho^2  \big)^2 \nabla_{S^2}^2 \delta A_{\rho} - 4 \big(  u_0^2 + \rho^2  \big)^2 \partial_{\rho} \partial_{\chi} \delta A_{\chi}  
%  \\[10pt]
%  & - 4 \big(  u_0^2 + \rho^2  \big)^2 \cot\chi \partial_{\rho} \delta A_{\chi} - 4 \frac{\big(  u_0^2 + \rho^2  \big)^2} {\sin^2\chi} \partial_{\rho} \partial_{\xi} \delta A_{\xi} = 0 \ .
% \end{aligned} 
% \end{equation}

\vskip 10pt

\noindent \emph{Equation for $\delta A_{\chi}$}

\vskip 10pt

\begin{equation}
 \begin{aligned}
   & - 2 \frac{\big(  u_0^2 + \rho^2  \big) \big(   u_0^2 + 3 \rho^2  \big)}{\sin\chi} \partial_{\xi} \delta r + \pi L^4 \rho^2 \Box \delta A_{\chi} - \pi L^4 \rho^2 \partial_{\chi} \big(   \eta^{\mu\nu} \partial_{\mu} \delta A_{\nu}  \big) + \pi \rho^2 \big(  u_0^2 + \rho^2  \big)^2 \partial_{\rho} \delta F_{\rho \chi}
   \\[10pt]
   & + \pi \rho \big(   u_0^2 + \rho^2  \big) \big(  u_0^2 + 5 \rho^2   \big) \delta F_{\rho \chi} - 4 \pi \frac{
   \big(  u_0^2 + \rho^2  \big)^2}{\sin^2 \chi}\partial_{\xi} \delta F_{\chi \xi} = 0 \ .
\end{aligned} 
\end{equation}

\vskip 10pt

\noindent \emph{Equation for $\delta A_{\xi}$}

\vskip 10pt

\begin{equation}
 \begin{aligned}
   & - 2 \frac{\big(  u_0^2 + \rho^2  \big) \big(   u_0^2 + 3 \rho^2  \big)}{\sin\chi} \partial_{\chi} \delta r - \pi L^4 \frac{\rho^2}{\sin^2 \chi} \Box \delta A_{\xi} + \pi L^4 \frac{\rho^2}{\sin^2 \chi} \partial_{\xi} \big(   \eta^{\mu\nu} \partial_{\mu} \delta A_{\nu}  \big) - \pi \frac{\rho^2}{\sin^2 \chi} \big(  u_0^2 + \rho^2  \big)^2 \partial_{\rho} \delta F_{\rho \xi}
   \\[10pt]
   & - \pi \rho \frac{\big(   u_0^2 + \rho^2  \big) \big(  u_0^2 + 5 \rho^2   \big)}{\sin^2 \chi} \delta F_{\rho \xi} - 4 \pi \frac{
   \big(  u_0^2 + \rho^2  \big)^2}{\sin^2 \chi}\partial_{\chi} \delta F_{\chi \xi} + 4 \pi \big(  u_0^2 + \rho^2  \big)^2 \frac{
   \cot \chi}{\sin^2 \chi} \delta F_{\chi \xi} = 0 \ .\label{fluctaxi}
\end{aligned} 
\end{equation}
For simplicity we have defined the operators $\Box$ and $\nabla_{S^2}^2$ as,
\begin{equation}
 \Box f_1 = \eta^{\mu \nu} \partial_{\mu} \partial_{\nu} f_1 \ , \qquad \nabla_{S^2}^2 f_2 = \frac{1}{\sin\chi} \partial_{\chi} \big(   \sin\chi \ \partial_{\chi} f_2 \big) + \frac{1}{\sin^2\chi} \partial_{\xi}^2 f_2 \ ,
\end{equation}
where $f_1 = f_1(x^{\mu})$ and $f_2 =  f_2(\chi, \xi)$.

In the main part of the paper we  solve this coupled system of differential equations for three different consistent fluctuations.

%%%%%%%%%%%%%%%%%%%%%%%%%%%%%%%%%%%%%%%%%%%%%%%%%%%%%%%%%%%%%%%%%%%%%%%%%%%%%%%%%


\begin{thebibliography}{99}

%\cite{Witten:1995ex}
\bibitem{Witten:1995ex} 
For a very representative paper see  E.~Witten,
  {\it String theory dynamics in various dimensions},
  Nucl.\ Phys.\ B {\bf 443}, 85 (1995),
  %doi:10.1016/0550-3213(95)00158-O,
  \href{http://arxiv.org/abs/hep-th/9503124}{{\tt hep-th/9503124}}.
  %%CITATION = doi:10.1016/0550-3213(95)00158-O;%%
  %2073 citations counted in INSPIRE as of 31 Oct 2016
For a review see
%\cite{Obers:1998fb}
%\bibitem{Obers:1998fb} 
  N.~A.~Obers and B.~Pioline,
  {\it U duality and M theory},
  Phys.\ Rept.\  {\bf 318}, 113 (1999),
  %doi:10.1016/S0370-1573(99)00004-6,
  \href{http://arxiv.org/abs/hep-th/9809039}{{\tt hep-th/9809039}}.
  %%CITATION = doi:10.1016/S0370-1573(99)00004-6;%%
  %253 citations counted in INSPIRE as of 31 Oct 2016


%\cite{Maldacena:1997re}
\bibitem{Maldacena:1997re}
  J.~M.~Maldacena,
  {\it The Large N limit of superconformal field theories and supergravity},
  Int.\ J.\ Theor.\ Phys.\  {\bf 38} (1999) 1113,
   [Adv.\ Theor.\ Math.\ Phys.\  {\bf 2} (1998) 231],
  %doi:10.1023/A:1026654312961,
  \href{http://arxiv.org/abs/hep-th/9711200}{{\tt hep-th/9711200}}.
  %%CITATION = doi:10.1023/A:1026654312961;%%
  %12201 citations counted in INSPIRE as of 24 Oct 2016


%
\bibitem{Gopakumar:1998ki} 
  R.~Gopakumar and C.~Vafa,
  {\it On the gauge theory / geometry correspondence},
  Adv.\ Theor.\ Math.\ Phys.\  {\bf 3}, 1415 (1999),
  \href{http://arxiv.org/abs/hep-th/9811131}{{\tt hep-th/9811131}}.
  %%CITATION = HEP-TH/9811131;%%
  %523 citations counted in INSPIRE as of 31 Oct 2016



\bibitem{Seiberg:1994pq} 
  N.~Seiberg,
  {\it Electric - magnetic duality in supersymmetric nonAbelian gauge theories},
  Nucl.\ Phys.\ B {\bf 435}, 129 (1995),
  %doi:10.1016/0550-3213(94)00023-8,
  \href{http://arxiv.org/abs/hep-th/9411149}{{\tt hep-th/9411149}}.
  %%CITATION = doi:10.1016/0550-3213(94)00023-8;%%
  %1367 citations counted in INSPIRE as of 31 Oct 2016


%\cite{Lunin:2005jy}
\bibitem{Lunin:2005jy} 
  O.~Lunin and J.~M.~Maldacena,
  {\it Deforming field theories with U(1) x U(1) global symmetry and their gravity duals},
  JHEP {\bf 0505}, 033 (2005),
  %doi:10.1088/1126-6708/2005/05/033,
  \href{http://arxiv.org/abs/hep-th/0502086}{{\tt hep-th/0502086}}.
  %%CITATION = doi:10.1088/1126-6708/2005/05/033;%%
  %435 citations counted in INSPIRE as of 31 Oct 2016




\bibitem{Maldacena:2009mw} 
  J.~Maldacena and D.~Martelli,
  {\it The Unwarped, resolved, deformed conifold: Fivebranes and the baryonic branch of the Klebanov-Strassler theory},
  JHEP {\bf 1001}, 104 (2010),
  %doi:10.1007/JHEP01(2010)104,
  \href{http://arxiv.org/abs/arXiv:0906.0591}{{\tt arXiv:0906.0591}}.
  %%CITATION = doi:10.1007/JHEP01(2010)104;%%
  %73 citations counted in INSPIRE as of 31 Oct 2016%\cite{Gaillard:2010qg}
%\bibitem{Gaillard:2010qg} 
  J.~Gaillard, D.~Martelli, C.~Nunez and I.~Papadimitriou,
  {\it The warped, resolved, deformed conifold gets flavoured},
  Nucl.\ Phys.\ B {\bf 843}, 1 (2011),
  %doi:10.1016/j.nuclphysb.2010.09.011,
  \href{http://arxiv.org/abs/arXiv:1004.4638}{{\tt arXiv:1004.4638}}.
  %%CITATION = doi:10.1016/j.nuclphysb.2010.09.011;%%
  %54 citations counted in INSPIRE as of 31 Oct 2016%\cite{Elander:2011mh}
%\bibitem{Elander:2011mh} 
  D.~Elander, J.~Gaillard, C.~Nunez and M.~Piai,
  {\it Towards multi-scale dynamics on the baryonic branch of Klebanov-Strassler},
  JHEP {\bf 1107}, 056 (2011),
  %doi:10.1007/JHEP07(2011)056,
  \href{http://arxiv.org/abs/arXiv:1104.3963}{{\tt arXiv:1104.3963}}.
  %%CITATION = doi:10.1007/JHEP07(2011)056;%%
  %48 citations counted in INSPIRE as of 31 Oct 2016


\bibitem{Rocek:1991ps} 
  M.~Rocek and E.~P.~Verlinde,
  {\it Duality, quotients, and currents},
  Nucl.\ Phys.\ B {\bf 373}, 630 (1992),
  %doi:10.1016/0550-3213(92)90269-H,
  \href{http://arxiv.org/abs/hep-th/9110053}{{\tt hep-th/9110053}}.
  %%CITATION = doi:10.1016/0550-3213(92)90269-H;%%
  %386 citations counted in INSPIRE as of 31 Oct 2016
%\cite{Duff:1998us}
\bibitem{Duff:1998us} 
  M.~J.~Duff, H.~Lu and C.~N.~Pope,
  %``AdS(5) x S**5 untwisted,''
  Nucl.\ Phys.\ B {\bf 532}, 181 (1998)
  doi:10.1016/S0550-3213(98)00464-7
  [hep-th/9803061].
  %%CITATION = doi:10.1016/S0550-3213(98)00464-7;%%
  %114 citations counted in INSPIRE as of 02 Nov 2016%\cite{Duff:1997qz}
%\bibitem{Duff:1997qz} 
  M.~J.~Duff, H.~Lu and C.~N.~Pope,
  %``Supersymmetry without supersymmetry,''
  Phys.\ Lett.\ B {\bf 409}, 136 (1997)
  doi:10.1016/S0370-2693(97)00687-4
  [hep-th/9704186].
  %%CITATION = doi:10.1016/S0370-2693(97)00687-4;%%
  %72 citations counted in INSPIRE as of 02 Nov 2016

%\cite{delaOssa:1992vci}
\bibitem{delaOssa:1992vci} 
  X.~C.~de la Ossa and F.~Quevedo,
  {\it Duality symmetries from nonAbelian isometries in string theory},
  Nucl.\ Phys.\ B {\bf 403}, 377 (1993),
  %doi:10.1016/0550-3213(93)90041-M,
  \href{http://arxiv.org/abs/hep-th/9210021}{{\tt hep-th/9210021}}.
  %%CITATION = doi:10.1016/0550-3213(93)90041-M;%%
  %204 citations counted in INSPIRE as of 31 Oct 2016

%\cite{Alvarez:1993qi}
\bibitem{Alvarez:1993qi} 
  E.~Alvarez, L.~Alvarez-Gaume, J.~L.~F.~Barbon and Y.~Lozano,
  {\it Some global aspects of duality in string theory},
  Nucl.\ Phys.\ B {\bf 415}, 71 (1994),
  %doi:10.1016/0550-3213(94)90067-1,
  \href{http://arxiv.org/abs/hep-th/9309039}{{\tt hep-th/9309039}}.
  %%CITATION = doi:10.1016/0550-3213(94)90067-1;%%
  %174 citations counted in INSPIRE as of 31 Oct 2016%\cite{Alvarez:1994np}
%\bibitem{Alvarez:1994np} 
  E.~Alvarez, L.~Alvarez-Gaume and Y.~Lozano,
  {\it On nonAbelian duality},
  Nucl.\ Phys.\ B {\bf 424}, 155 (1994),
  %doi:10.1016/0550-3213(94)90093-0,
  \href{http://arxiv.org/abs/hep-th/9403155}{{\tt hep-th/9403155}}.
  %%CITATION = doi:10.1016/0550-3213(94)90093-0;%%
  %121 citations counted in INSPIRE as of 31 Oct 2016%\cite{Alvarez:1994wj}
%\bibitem{Alvarez:1994wj} 
  E.~Alvarez, L.~Alvarez-Gaume and Y.~Lozano,
  {\it A Canonical approach to duality transformations},
  Phys.\ Lett.\ B {\bf 336}, 183 (1994),
  %doi:10.1016/0370-2693(94)00982-1,
  \href{http://arxiv.org/abs/hep-th/9406206}{{\tt hep-th/9406206}}.
  %%CITATION = doi:10.1016/0370-2693(94)00982-1;%%
  %125 citations counted in INSPIRE as of 31 Oct 2016%\cite{Elitzur:1994ri}
%\bibitem{Elitzur:1994ri} 
  S.~Elitzur, A.~Giveon, E.~Rabinovici, A.~Schwimmer and G.~Veneziano,
  {\it Remarks on nonAbelian duality},
  Nucl.\ Phys.\ B {\bf 435}, 147 (1995),
  %doi:10.1016/0550-3213(94)00426-F,
  \href{http://arxiv.org/abs/hep-th/9409011}{{\tt hep-th/9409011}}.
  %%CITATION = doi:10.1016/0550-3213(94)00426-F;%%
  %43 citations counted in INSPIRE as of 31 Oct 2016%\cite{Klimcik:1995ux}
%\bibitem{Klimcik:1995ux} 
  C.~Klimcik and P.~Severa,
  {\it Dual nonAbelian duality and the Drinfeld double},
  Phys.\ Lett.\ B {\bf 351}, 455 (1995),
  %doi:10.1016/0370-2693(95)00451-P,
  \href{http://arxiv.org/abs/hep-th/9502122}{{\tt hep-th/9502122}}.
  %%CITATION = doi:10.1016/0370-2693(95)00451-P;%%
  %157 citations counted in INSPIRE as of 31 Oct 2016%\cite{Lozano:1995jx}
%\bibitem{Lozano:1995jx} 
  Y.~Lozano,
  {\it NonAbelian duality and canonical transformations},
  Phys.\ Lett.\ B {\bf 355}, 165 (1995),
  %doi:10.1016/0370-2693(95)00777-I,
  \href{http://arxiv.org/abs/hep-th/9503045}{{\tt hep-th/9503045}}.
  %%CITATION = doi:10.1016/0370-2693(95)00777-I;%%
  %68 citations counted in INSPIRE as of 31 Oct 2016



%\cite{Giveon:1993ai}
\bibitem{Giveon:1993ai} 
  A.~Giveon and M.~Rocek,
  {\it On nonAbelian duality},
  Nucl.\ Phys.\ B {\bf 421}, 173 (1994),
  %doi:10.1016/0550-3213(94)90230-5,
  \href{http://arxiv.org/abs/hep-th/9308154}{{\tt hep-th/9308154}}.
  %%CITATION = doi:10.1016/0550-3213(94)90230-5;%%
  %130 citations counted in INSPIRE as of 31 Oct 2016%\cite{Giveon:1994mw}
%\bibitem{Giveon:1994mw} 
  A.~Giveon and M.~Rocek,
  {\it Introduction to duality},
  In *Greene, B. (ed.), Yau, S.T. (ed.): Mirror symmetry II* 413-426,
  \href{http://arxiv.org/abs/hep-th/9406178}{{\tt hep-th/9406178}}.
  %%CITATION = HEP-TH/9406178;%%
  %22 citations counted in INSPIRE as of 31 Oct 2016



%\cite{Sfetsos:2010uq}
\bibitem{Sfetsos:2010uq} 
  K.~Sfetsos and D.~C.~Thompson,
  {\it On non-abelian T-dual geometries with Ramond fluxes},
  Nucl.\ Phys.\ B {\bf 846}, 21 (2011),
  %doi:10.1016/j.nuclphysb.2010.12.013,
  \href{http://arxiv.org/abs/arXiv:1012.1320}{{\tt arXiv:1012.1320}}.
  %%CITATION = doi:10.1016/j.nuclphysb.2010.12.013;%%
  %62 citations counted in INSPIRE as of 25 Oct 2016




  %
\bibitem{Macpherson:2014eza} 
  N.~T.~Macpherson, C.~Nunez, L.~A.~Pando Zayas, V.~G.~J.~Rodgers and C.~A.~Whiting,
  {\it Type IIB supergravity solutions with AdS$_{5}$ from Abelian and non-Abelian T dualities},
  JHEP {\bf 1502}, 040 (2015),
  %doi:10.1007/JHEP02(2015)040,
  \href{http://arxiv.org/abs/arXiv:1410.2650}{{\tt arXiv:1410.2650}}.
  %%CITATION = doi:10.1007/JHEP02(2015)040;%%
  %17 citations counted in INSPIRE as of 25 Oct 2016
  
  %\cite{Itsios:2013wd}
\bibitem{Itsios:2013wd} 
  G.~Itsios, C.~Nunez, K.~Sfetsos and D.~C.~Thompson,
  {\it Non-Abelian T-duality and the AdS/CFT correspondence:new N=1 backgrounds},
  Nucl.\ Phys.\ B {\bf 873}, 1 (2013),
  %doi:10.1016/j.nuclphysb.2013.04.004,
  \href{http://arxiv.org/abs/arXiv:1301.6755}{{\tt arXiv:1301.6755}}.
  %%CITATION = doi:10.1016/j.nuclphysb.2013.04.004;%%
  %47 citations counted in INSPIRE as of 31 Oct 2016
  
  
  
  %\cite{Macpherson:2015tka}
\bibitem{Macpherson:2015tka}
  N.~T.~Macpherson, C.~Nunez, D.~C.~Thompson and S.~Zacarias,
  {\it Holographic Flows in non-Abelian T-dual Geometries},
  JHEP {\bf 1511} (2015) 212
  %doi:10.1007/JHEP11(2015)212,
  \href{http://arxiv.org/abs/arXiv:1509.04286}{{\tt arXiv:1509.04286}}.
  %%CITATION = doi:10.1007/JHEP11(2015)212;%%
  %4 citations counted in INSPIRE as of 14 Apr 2016

  
%\cite{Lozano:2016kum}
\bibitem{Lozano:2016kum} 
  Y.~Lozano and C.~Nunez,
  {\it Field theory aspects of non-Abelian T-duality and $ \mathcal{N}  =$ 2 linear quivers},
  JHEP {\bf 1605}, 107 (2016),
  %doi:10.1007/JHEP05(2016)107,
  \href{http://arxiv.org/abs/arXiv:1603.04440}{{\tt arXiv:1603.04440}}.
  %%CITATION = doi:10.1007/JHEP05(2016)107;%%
  %2 citations counted in INSPIRE as of 25 Oct 2016

%\cite{Lozano:2016wrs}
\bibitem{Lozano:2016wrs} 
  Y.~Lozano, N.~T.~Macpherson, J.~Montero and C.~Nunez,
  {\it Three-dimensional N=4 Linear Quivers and non-Abelian T-duals},
  \href{http://arxiv.org/abs/arXiv:1609.09061}{{\tt arXiv:1609.09061}}.
  %%CITATION = ARXIV:1609.09061;%%
  
  

\bibitem{varios1}
 %\cite{Lozano:2011kb}
%\bibitem{Lozano:2011kb}
  Y.~Lozano, E.~O Colgain, K.~Sfetsos and D.~C.~Thompson,
  {\it Non-abelian T-duality, Ramond Fields and Coset Geometries},
  JHEP {\bf 1106} (2011) 106,
  \href{http://arxiv.org/abs/arXiv:1104.5196}{{\tt arXiv:1104.5196}};
  %\cite{Itsios:2012dc}
%\bibitem{Itsios:2012dc}
  G.~Itsios, Y.~Lozano, E.~O Colgain and K.~Sfetsos,
  {\it Non-Abelian T-duality and consistent truncations in type-II supergravity},
  JHEP {\bf 1208} (2012) 132,
  \href{http://arxiv.org/abs/arXiv:1205.2274}{{\tt arXiv:1205.2274}};
  %%CITATION = doi:10.1007/JHEP08(2012)132;%%
  %35 citations counted in INSPIRE as of 10 Mar 2016
  %%CITATION = doi:10.1007/JHEP06(2011)106;%%
  %46 citations counted in INSPIRE as of 10 Mar 2016
 %\cite{Itsios:2013wd}
%\bibitem{Itsios:2013wd} 
  G.~Itsios, C.~Nunez, K.~Sfetsos and D.~C.~Thompson,
  {\it Non-Abelian T-duality and the AdS/CFT correspondence:new N=1 backgrounds},
  Nucl.\ Phys.\ B {\bf 873}, 1 (2013),
  \href{http://arxiv.org/abs/arXiv:1301.6755}{{\tt arXiv:1301.6755}};
  %%CITATION = doi:10.1016/j.nuclphysb.2013.04.004;%%
  %43 citations counted in INSPIRE as of 16 Feb 2016%\cite{Lozano:2012au}
%\bibitem{Lozano:2012au} 
  Y.~Lozano, E.~� Colg�in, D.~Rodr�guez-G�mez and K.~Sfetsos,
  {\it Supersymmetric $AdS_6$ via T Duality},
  Phys.\ Rev.\ Lett.\  {\bf 110}, no. 23, 231601 (2013),
  \href{http://arxiv.org/abs/arXiv:1212.1043}{{\tt arXiv:1212.1043}};
  %%CITATION = doi:10.1103/PhysRevLett.110.231601;%%
  %36 citations counted in INSPIRE as of 16 Feb 2016%\cite{Itsios:2012zv}
%\bibitem{Itsios:2012zv} 
  G.~Itsios, C.~Nunez, K.~Sfetsos and D.~C.~Thompson,
  {\it On Non-Abelian T-Duality and new N=1 backgrounds},
  Phys.\ Lett.\ B {\bf 721}, 342 (2013),
  \href{http://arxiv.org/abs/arXiv:1212.4840}{{\tt arXiv:1212.4840}};
  %%CITATION = doi:10.1016/j.physletb.2013.03.033;%%
  %31 citations counted in INSPIRE as of 16 Feb 2016
%\cite{Jeong:2013jfc}
%\bibitem{Jeong:2013jfc} 
  J.~Jeong, O.~Kelekci and E.~O Colgain,
  {\it An alternative IIB embedding of F(4) gauged supergravity},
  JHEP {\bf 1305}, 079 (2013),
  \href{http://arxiv.org/abs/arXiv:1302.2105}{{\tt arXiv:1302.2105}}.
  %%CITATION = doi:10.1007/JHEP05(2013)079;%%
  %27 citations counted in INSPIRE as of 16 Feb 2016
  
  \bibitem{varios2}
%\bibitem{Barranco:2013fza} 
  A.~Barranco, J.~Gaillard, N.~T.~Macpherson, C.~Nunez and D.~C.~Thompson,
  {\it G-structures and Flavouring non-Abelian T-duality},
  JHEP {\bf 1308}, 018 (2013),
  %doi:10.1007/JHEP08(2013)018,
  \href{http://arxiv.org/abs/arXiv:1305.7229}{{\tt arXiv:1305.7229}}.
  %%CITATION = doi:10.1007/JHEP08(2013)018;%%
  %26 citations counted in INSPIRE as of 16 Feb 2016%\cite{Macpherson:2013zba}
%\bibitem{Macpherson:2013zba} 
  N.~T.~Macpherson,
  {\it Non-Abelian T-duality, $G_2$-structure rotation and holographic duals of $N=1$ Chern-Simons theories},
  JHEP {\bf 1311}, 137 (2013),
  %doi:10.1007/JHEP11(2013)137,
  \href{http://arxiv.org/abs/arXiv:1310.1609}{{\tt arXiv:1310.1609}}.
  %%CITATION = doi:10.1007/JHEP11(2013)137;%%
  %18 citations counted in INSPIRE as of 16 Feb 2016%\cite{Gaillard:2013vsa}
  
  
 % \cite{Macpherson:2014eza}
%  \cite{Lozano:2013oma} %\cite{Lozano:2014ata}
\bibitem{Lozano:2013oma} 
  Y.~Lozano, E.~O.~Colgain and D.~Rodriguez-Gomez,
  {\it Hints of 5d Fixed Point Theories from Non-Abelian T-duality},
  JHEP {\bf 1405}, 009 (2014),
  %doi:10.1007/JHEP05(2014)009,
  \href{http://arxiv.org/abs/arXiv:1311.4842}{{\tt arXiv:1311.4842}}.
  %%CITATION = doi:10.1007/JHEP05(2014)009;%%
  %23 citations counted in INSPIRE as of 14 Dec 2015


  
\bibitem{varios3} 
  J.~Gaillard, N.~T.~Macpherson, C.~Nunez and D.~C.~Thompson,
  {\it Dualising the Baryonic Branch: Dynamic SU(2) and confining backgrounds in IIA},
  Nucl.\ Phys.\ B {\bf 884}, 696 (2014),
  %doi:10.1016/j.nuclphysb.2014.05.004,
  \href{http://arxiv.org/abs/arXiv:1312.4945}{{\tt arXiv:1312.4945}}.
  %%CITATION = doi:10.1016/j.nuclphysb.2014.05.004;%%
  %18 citations counted in INSPIRE as of 16 Feb 2016
%\cite{Elander:2013jqa}
%bibitem{Elander:2013jqa} 
  D.~Elander, A.~F.~Faedo, C.~Hoyos, D.~Mateos and M.~Piai,
  {\it Multiscale confining dynamics from holographic RG flows},
  JHEP {\bf 1405}, 003 (2014),
 % doi:10.1007/JHEP05(2014)003
  \href{http://arxiv.org/abs/arXiv:1312.7160}{{\tt arXiv:1312.7160}}.
  %%CITATION = doi:10.1007/JHEP05(2014)003;%%
  %16 citations counted in INSPIRE as of 16 Feb 2016%\cite{Caceres:2014uoa}
%\bibitem{Caceres:2014uoa} 
  E.~Caceres, N.~T.~Macpherson and C.~Nunez,
  {\it New Type IIB Backgrounds and Aspects of Their Field Theory Duals},
  JHEP {\bf 1408}, 107 (2014),
  %doi:10.1007/JHEP08(2014)107
  \href{http://arxiv.org/abs/arXiv:1402.3294}{{\tt arXiv:1402.3294}}.
  %%CITATION = doi:10.1007/JHEP08(2014)107;%%
  %17 citations counted in INSPIRE as of 16 Feb 2016%\cite{Pradhan:2014zqa}
%\bibitem{Pradhan:2014zqa} 
  P.~M.~Pradhan,
  {\it Oscillating Strings and Non-Abelian T-dual Klebanov-Witten Background},
  Phys.\ Rev.\ D {\bf 90}, no. 4, 046003 (2014),
  %doi:10.1103/PhysRevD.90.046003
  \href{http://arxiv.org/abs/arXiv:1406.2152}{{\tt arXiv:1406.2152}}.
  %%CITATION = doi:10.1103/PhysRevD.90.046003;%%
  %9 citations counted in INSPIRE as of 16 Feb 2016%\cite{Sfetsos:2014tza}
%\bibitem{Sfetsos:2014tza} 
  K.~Sfetsos and D.~C.~Thompson,
  {\it New ${\cal N} = 1$ supersymmetric $AdS_5$ backgrounds in Type IIA supergravity},
  JHEP {\bf 1411}, 006 (2014),
  %doi:10.1007/JHEP11(2014)006
  \href{http://arxiv.org/abs/arXiv:1408.6545}{{\tt arXiv:1408.6545}}.
  %%CITATION = doi:10.1007/JHEP11(2014)006;%%
  %11 citations counted in INSPIRE as of 16 Feb 2016
  
  %\cite{Lozano:2014ata}
\bibitem{Lozano:2014ata}
  Y.~Lozano and N.~T.~Macpherson,
  {\it A new AdS$_{4}$/CFT$_{3}$ dual with extended SUSY and a spectral flow},
  JHEP {\bf 1411} (2014) 115,
 % doi:10.1007/JHEP11(2014)115
  \href{http://arxiv.org/abs/arXiv:1408.0912}{{\tt arXiv:1408.0912}}.
  %%CITATION = doi:10.1007/JHEP11(2014)115;%%
  %12 citations counted in INSPIRE as of 10 Mar 2016
  
  %\cite{Kelekci:2014ima}
\bibitem{Kelekci:2014ima}
  O.~Kelekci, Y.~Lozano, N.~T.~Macpherson and E.~O.~Colg�in,
  {\it Supersymmetry and non-Abelian T-duality in type II supergravity},
  Class.\ Quant.\ Grav.\  {\bf 32} (2015) no.3,  035014,
  %doi:10.1088/0264-9381/32/3/035014
  \href{http://arxiv.org/abs/arXiv:1409.7406}{{\tt arXiv:1409.7406}}.
  %%CITATION = doi:10.1088/0264-9381/32/3/035014;%%
  %15 citations counted in INSPIRE as of 25 Mar 2016
  
 
   
 
\bibitem{varios4} 
%\bibitem{Kooner:2014cqa} 
  K.~S.~Kooner and S.~Zacarias,
  {\it Non-Abelian T-Dualizing the Resolved Conifold with Regular and Fractional D3-Branes},
  JHEP {\bf 1508}, 143 (2015),
  %doi:10.1007/JHEP08(2015)143
  \href{http://arxiv.org/abs/arXiv:1411.7433}{{\tt arXiv:1411.7433}}.
  %%CITATION = doi:10.1007/JHEP08(2015)143;%%
  %6 citations counted in INSPIRE as of 16 Feb 2016
%\cite{Araujo:2015npa}
%\bibitem{Araujo:2015npa} 
  T.~R.~Araujo and H.~Nastase,
  {\it $\mathcal{N}=1$ SUSY backgrounds with an AdS factor from non-Abelian T duality},
  Phys.\ Rev.\ D {\bf 91}, no. 12, 126015 (2015),
  %doi:10.1103/PhysRevD.91.126015
  \href{http://arxiv.org/abs/arXiv:1503.00553}{{\tt arXiv:1503.00553}}.
  %%CITATION = doi:10.1103/PhysRevD.91.126015;%%
  %8 citations counted in INSPIRE as of 16 Feb 2016%\bibitem{Lozano:2015bra} 
  Y.~Lozano, N.~T.~Macpherson, J.~Montero and E.~O.~Colgain,
  {\it New $AdS_3 \times S^2$ T-duals with $ \mathcal{N}=\left(0,4\right) $ supersymmetry},
  JHEP {\bf 1508}, 121 (2015),
  %doi:10.1007/JHEP08(2015)121
  \href{http://arxiv.org/abs/arXiv:1507.02659}{{\tt arXiv:1507.02659}}.
  %%CITATION = doi:10.1007/JHEP08(2015)121;%%
  %7 citations counted in INSPIRE as of 16 Feb 2016
  %\cite{Lozano:2015cra}
  Y.~Lozano, N.~T.~Macpherson and J.~Montero,
  {\it A $ \mathcal{N}=2 $ supersymmetric AdS$_{4}$ solution in M-theory with purely magnetic flux},
  JHEP {\bf 1510} (2015) 004,
  %doi:10.1007/JHEP10(2015)004
  \href{http://arxiv.org/abs/arXiv:1507.02660}{{\tt [arXiv:1507.02660}}.
  %%CITATION = doi:10.1007/JHEP10(2015)004;%%
  %9 citations counted in INSPIRE as of 25 Sep 2016
%\bibitem{Araujo:2015dba} 
  T.~R.~Araujo and H.~Nastase,
  {\it Non-Abelian T-duality for nonrelativistic holographic duals},
  JHEP {\bf 1511}, 203 (2015),
  %doi:10.1007/JHEP11(2015)203
  \href{http://arxiv.org/abs/arXiv:1508.06568}{{\tt arXiv:1508.06568}}.
  %%CITATION = doi:10.1007/JHEP11(2015)203;%%
  %3 citations counted in INSPIRE as of 16 Feb 2016
%\cite{Zayas:2015azn}
%\bibitem{Zayas:2015azn} 
  L.~A.~P.~Zayas, V.~G.~J.~Rodgers and C.~A.~Whiting,
  {\it Supergravity solutions with AdS$_{4}$ from non-Abelian T-dualities},
  JHEP {\bf 1602}, 061 (2016),
  %doi:10.1007/JHEP02(2016)061
  \href{http://arxiv.org/abs/arXiv:1511.05991}{{\tt arXiv:1511.05991}}.
  %%CITATION = doi:10.1007/JHEP02(2016)061;%%%\cite{Dimov:2015rie}
  
  
  %\cite{Kruczenski:2003be}
\bibitem{Kruczenski:2003be} 
  M.~Kruczenski, D.~Mateos, R.~C.~Myers and D.~J.~Winters,
  {\it Meson spectroscopy in AdS / CFT with flavor},
  JHEP {\bf 0307}, 049 (2003),
  %doi:10.1088/1126-6708/2003/07/049
  \href{http://arxiv.org/abs/hep-th/0304032}{{\tt hep-th/0304032}}.
  %%CITATION = doi:10.1088/1126-6708/2003/07/049;%%
  %442 citations counted in INSPIRE as of 25 Oct 2016

%\cite{Itsios:2012dc}
\bibitem{Itsios:2012dc}
  G.~Itsios, Y.~Lozano, E.~O Colgain and K.~Sfetsos,
  {\it Non-Abelian T-duality and consistent truncations in type-II supergravity},
  JHEP {\bf 1208} (2012) 132
  %doi:10.1007/JHEP08(2012)132,
  \href{http://arxiv.org/abs/arXiv:1205.2274}{{\tt arXiv:1205.2274}}.
  %%CITATION = doi:10.1007/JHEP08(2012)132;%%
  %36 citations counted in INSPIRE as of 05 Jul 2016

%\cite{Karch:2002sh}
\bibitem{Karch:2002sh}
  A.~Karch and E.~Katz,
  {\it Adding flavor to AdS / CFT},
  JHEP {\bf 0206} (2002) 043,
  %doi:10.1088/1126-6708/2002/06/043,
  \href{http://arxiv.org/abs/hep-th/0205236}{{\tt hep-th/0205236}}.
  %%CITATION = doi:10.1088/1126-6708/2002/06/043;%%
  %811 citations counted in INSPIRE as of 24 Oct 2016 
  
%\cite{Karch:2002xe}
\bibitem{Karch:2002xe}
  A.~Karch, E.~Katz and N.~Weiner,
  {\it Hadron masses and screening from AdS Wilson loops},
  Phys.\ Rev.\ Lett.\  {\bf 90} (2003) 091601,
  %doi:10.1103/PhysRevLett.90.091601,
  \href{http://arxiv.org/abs/hep-th/0211107}{{\tt hep-th/0211107}}.
  %%CITATION = doi:10.1103/PhysRevLett.90.091601;%%
  %78 citations counted in INSPIRE as of 24 Oct 2016


%\cite{Erdmenger:2007cm}
\bibitem{Erdmenger:2007cm} 
  J.~Erdmenger, N.~Evans, I.~Kirsch and E.~Threlfall,
  {\it Mesons in Gauge/Gravity Duals - A Review},
  Eur.\ Phys.\ J.\ A {\bf 35}, 81 (2008),
  %doi:10.1140/epja/i2007-10540-1
  \href{http://arxiv.org/abs/arXiv:0711.4467}{{\tt arXiv:0711.4467}}.
  %%CITATION = doi:10.1140/epja/i2007-10540-1;%%
  %349 citations counted in INSPIRE as of 27 Oct 2016
  %\cite{Lozano:2014ata}
  


%\cite{Arean:2006pk}
\bibitem{Arean:2006pk} 
  D.~Arean and A.~V.~Ramallo,
  {\it Open string modes at brane intersections},
  JHEP {\bf 0604}, 037 (2006),
  %doi:10.1088/1126-6708/2006/04/037
  \href{http://arxiv.org/abs/hep-th/0602174}{{\tt hep-th/0602174}}.
  %%CITATION = doi:10.1088/1126-6708/2006/04/037;%%
  %52 citations counted in INSPIRE as of 29 Oct 2016%\cite{Myers:2006qr}
%\bibitem{Myers:2006qr} 
  R.~C.~Myers and R.~M.~Thomson,
  {\it Holographic mesons in various dimensions},
  JHEP {\bf 0609}, 066 (2006),
  %doi:10.1088/1126-6708/2006/09/066
  \href{http://arxiv.org/abs/hep-th/0605017}{{\tt hep-th/0605017}}.
  %%CITATION = doi:10.1088/1126-6708/2006/09/066;%%
  %61 citations counted in INSPIRE as of 29 Oct 2016
%
%\cite{Witten:1979kh}
\bibitem{clasicos} 
  E.~Witten,
  {\it Baryons in the 1/n Expansion},
  Nucl.\ Phys.\ B {\bf 160}, 57 (1979).
%  doi:10.1016/0550-3213(79)90232-3
  %%CITATION = doi:10.1016/0550-3213(79)90232-3;%%
  %2315 citations counted in INSPIRE as of 04 Nov 2016%\cite{'tHooft:1973jz}
%\bibitem{'tHooft:1973jz} 
  G.~'t Hooft,
  {\it A Planar Diagram Theory for Strong Interactions},
  Nucl.\ Phys.\ B {\bf 72}, 461 (1974).
 % doi:10.1016/0550-3213(74)90154-0
  %%CITATION = doi:10.1016/0550-3213(74)90154-0;%%
  %4193 citations counted in INSPIRE as of 04 Nov 2016
  
%\cite{Hoyos:2011us}
\bibitem{Hoyos:2011us}
 C.~Hoyos, T.~Nishioka and A.~O'Bannon,
 {\it A Chiral Magnetic Effect from AdS/CFT with Flavor},
 JHEP {\bf 1110}, 084 (2011),
 %doi:10.1007/JHEP10(2011)084,
 \href{http://arxiv.org/abs/arXiv:1106.4030}{{\tt arXiv:1106.4030}}.
 %%CITATION = doi:10.1007/JHEP10(2011)084;%%
 %51 citations counted in INSPIRE as of 07 Nov 2016

%\cite{Apreda:2006bu}
\bibitem{Apreda:2006bu}
 R.~Apreda, J.~Erdmenger, D.~Lust and C.~Sieg,
 {\it Adding flavour to the Polchinski-Strassler background},
 JHEP {\bf 0701}, 079 (2007),
 %doi:10.1088/1126-6708/2007/01/079,
 \href{http://arxiv.org/abs/hep-th/0610276}{{\tt hep-th/0610276}}.
 %%CITATION = doi:10.1088/1126-6708/2007/01/079;%%
 %27 citations counted in INSPIRE as of 07 Nov 2016

%\cite{Penati:2007vj}
\bibitem{Penati:2007vj}
 S.~Penati, M.~Pirrone and C.~Ratti,
 {\it Mesons in marginally deformed AdS/CFT},
 JHEP {\bf 0804}, 037 (2008),
 %doi:10.1088/1126-6708/2008/04/037,
 \href{http://arxiv.org/abs/arXiv:0710.4292}{{\tt arXiv:0710.4292}}.
 %%CITATION = doi:10.1088/1126-6708/2008/04/037;%%
 %16 citations counted in INSPIRE as of 07 Nov 2016
%\cite{Arean:2005ar}
%\bibitem{Arean:2005ar}
 D.~Arean, A.~Paredes and A.~V.~Ramallo,
 {\it Adding flavor to the gravity dual of non-commutative gauge theories},
 JHEP {\bf 0508}, 017 (2005),
 %doi:10.1088/1126-6708/2005/08/017,
 \href{http://arxiv.org/abs/arXiv:0710.4292}{{\tt hep-th/0505181}}.
 %%CITATION = doi:10.1088/1126-6708/2005/08/017;%%
 %39 citations counted in INSPIRE as of 07 Nov 2016  
  
\bibitem{Berenstein:2002jq} 
  D.~E.~Berenstein, J.~M.~Maldacena and H.~S.~Nastase,
  {\it Strings in flat space and pp waves from N=4 superYang-Mills},
  JHEP {\bf 0204}, 013 (2002),
  %doi:10.1088/1126-6708/2002/04/013
  \href{http://arxiv.org/abs/hep-th/0202021}{{\tt hep-th/0202021}}.
  %%CITATION = doi:10.1088/1126-6708/2002/04/013;%%
  %1604 citations counted in INSPIRE as of 02 Nov 2016




\bibitem{strings}
%\cite{Sfetsos:1994vz}
%\bibitem{Sfetsos:1994vz} 
  K.~Sfetsos,
  {\it Gauged WZW models and nonAbelian duality},
  Phys.\ Rev.\ D {\bf 50}, 2784 (1994),
  %doi:10.1103/PhysRevD.50.2784
  \href{http://arxiv.org/abs/hep-th/9402031}{{\tt hep-th/9402031}}.
  %%CITATION = doi:10.1103/PhysRevD.50.2784;%%
  %93 citations counted in INSPIRE as of 10 Nov 2016%\cite{Polychronakos:2010hd}
%\bibitem{Polychronakos:2010hd} 
  A.~P.~Polychronakos and K.~Sfetsos,
  {\it High spin limits and non-abelian T-duality},
  Nucl.\ Phys.\ B {\bf 843}, 344 (2011),
  %doi:10.1016/j.nuclphysb.2010.09.006
  \href{http://arxiv.org/abs/arXiv:1008.3909}{{\tt arXiv:1008.3909}}.
  %%CITATION = doi:10.1016/j.nuclphysb.2010.09.006;%%
  %15 citations counted in INSPIRE as of 10 Nov 2016%\cite{Zacarias:2014wta}
%\bibitem{Zacarias:2014wta} 
  S.~Zacarias,
  {\it Semiclassical strings and Non-Abelian T-duality},
  Phys.\ Lett.\ B {\bf 737}, 90 (2014),
  %doi:10.1016/j.physletb.2014.08.016
  \href{http://arxiv.org/abs/arXiv:1401.7618}{{\tt arXiv:1401.7618}}.
  %%CITATION = doi:10.1016/j.physletb.2014.08.016;%%
  %10 citations counted in INSPIRE as of 16 Feb 2016

\end{thebibliography}
\end{document}